\documentclass[aps,physrev,preprint,nofootinbib]{revtex4-2}
\makeatletter
\def\l@subsubsection#1#2{}
\makeatother
\usepackage[english]{babel}
\usepackage[utf8]{inputenc}
\usepackage[T1]{fontenc}
\usepackage{amsmath,amssymb,braket,slashed,mathrsfs}
\usepackage{pifont}

\usepackage[squaren,Gray,cdot,binary]{SIunits}
\usepackage{graphicx}
\graphicspath{{./figures/}}
\usepackage{color,hyperref}
\usepackage{minibox}
\usepackage{todonotes}
\usepackage{cleveref}
\usepackage{marginnote}

\crefformat{section}{Sec.~#2#1#3}
\crefmultiformat{section}{Secs.~#2#1#3}
    { and~#2#1#3}{, #2#1#3}{ and~#2#1#3}
\crefrangeformat{section}{Secs.~#3#1#4\,--\,#5#2#6}
\crefformat{subsection}{Sec.~#2#1#3}
\crefmultiformat{subsection}{Secs.~#2#1#3}
    { and~#2#1#3}{, #2#1#3}{ and~#2#1#3}
\crefrangeformat{subsection}{Secs.~#3#1#4\,--\,#5#2#6}
\crefformat{subsubsection}{Sec.~#2#1#3}
\crefmultiformat{subsubsection}{Secs.~#2#1#3}
    { and~#2#1#3}{, #2#1#3}{ and~#2#1#3}
\crefrangeformat{subsubsection}{Secs.~#3#1#4\,--\,#5#2#6}
\crefformat{figure}{Fig.~#2#1#3}
\crefmultiformat{figure}{Figs.~#2#1#3}
    { and~#2#1#3}{, #2#1#3}{ and~#2#1#3}
\crefrangeformat{figure}{Figs.~#3#1#4\,--\,#5#2#6}
\crefformat{table}{Table~#2#1#3}
\crefmultiformat{table}{Tables~#2#1#3}
    { and~#2#1#3}{, #2#1#3}{ and~#2#1#3}
\crefrangeformat{table}{Tables~#3#1#4\,--\,#5#2#6}
\crefformat{equation}{Eq.~(#2#1#3)}
\crefmultiformat{equation}{Eqs.~(#2#1#3)}{ and~(#2#1#3)}{, (#2#1#3)}{ and~(#2#1#3)}
\crefrangeformat{equation}{Eqs.~(#3#1#4)--(#5#2#6)}
\crefformat{appendix}{App.~#2#1#3}
\crefmultiformat{appendix}{Apps.~#2#1#3}
    { and~#2#1#3}{, #2#1#3}{ and~#2#1#3}
\crefrangeformat{appendix}{Apps.~#3#1#4\,--\,#5#2#6}

\definecolor{linkcolor}{rgb}{.17578125,.1875,.5703125}
\hypersetup{
    pdfstartview={FitH},                         
    pdftitle={TITLE},                            
    pdfauthor={AUTHOR},                         
    pdfsubject={Physics article},                
    pdfcreator={Antonin Portelli},               
    pdfkeywords={KEW1} {KEW2},                   
    colorlinks=true,                             
    linkcolor=linkcolor,                         
    citecolor=linkcolor,                         
    filecolor=black,                             
    urlcolor=linkcolor      						 
}

\newcommand{\ie}{\textit{i.e.}~}

\DeclareMathOperator{\arctanh}{arctanh}
\DeclareMathOperator{\bigo}{\mathcal{O}}

\newcommand{\diff}{\mathrm{d}}

\renewcommand{\epsilon}{\varepsilon}

\newcommand{\R}{\mathbb{R}}

\newcommand{\T}{\mathbb{T}}
\renewcommand{\tilde}{\widetilde}

\DeclareMathOperator*{\sump}{\sideset{}{^\prime}\sum}

\newcommand{\Z}{\mathbb{Z}}

\newcommand{\nsp}{\mathbf{n}}

\newcommand{\ksp}{\mathbf{k}}
\newcommand{\kspu}{\hat{\mathbf{k}}}
\newcommand{\psp}{\mathbf{p}}
\newcommand{\pspl}{\mathbf{p}_{\ell}}
\newcommand{\vel}{\mathbf{v}}
\newcommand{\vell}{\mathbf{v}_{\ell}}
\newcommand{\ml}{m_{\ell}}

\newcommand{\oml}{\omega_{\ell}}
\newcommand{\omg}{\omega_{\lambda}}
\newcommand{\rl}{r_{\ell}}

\newcommand{\qedl}{\mathrm{QED}_{\mathrm{L}}}

\newcommand{\uoe}{School of Physics and Astronomy, 
The University of Edinburgh, Edinburgh EH9 3FD, United Kingdom}
\newcommand{\bern}{Albert Einstein Center for Fundamental Physics, Institute for Theoretical Physics, Universität Bern, Sidlerstrasse 5, CH-3012 Bern}

\begin{document}
  \title{Relativistic, model-independent determination of electromagnetic finite-size effects beyond the point-like approximation}
 
  \author{M. Di Carlo}\affiliation{\uoe}
  \author{M. T. Hansen}\affiliation{\uoe}
  \author{N. Hermansson-Truedsson}\email[corresponding author, ]{nils@itp.unibe.ch}\affiliation{\bern}
  \author{A. Portelli}\affiliation{\uoe}
  
  \begin{abstract}
    We present a relativistic and model-independent method to derive
    structure-dependent electromagnetic finite-size effects. This is a
    systematic procedure, particularly well-suited for automatization, which 
    works at arbitrarily high orders in the large-volume expansion.
    Structure-dependent coefficients appear as zero-momentum derivatives of
    physical form factors which can be obtained through experimental
    measurements or auxiliary lattice calculations. As an application we derive
    the electromagnetic finite-size effects on the pseudoscalar meson mass and
    leptonic decay amplitude, through orders $\bigo(1/L^3)$ and $\bigo(1/L^2)$, respectively.
    The structure dependence appears at this order
    through the meson charge radius and the real radiative leptonic amplitude, which
    are known experimentally. 
  \end{abstract}
  \maketitle
  \pagebreak
  \section{Introduction}
  Lattice quantum chromodynamics (QCD) makes it possible to perform precision tests of the Standard Model
(SM) using observables for which non-perturbative physics plays an important role. In recent years, for example, it has been used to determine hadronic
corrections to the muon anomalous magnetic moment~\cite{Aoyama:2020ynm}
and, in the flavour physics sector, decay rates
needed for the extraction of Cabibbo-Kobayashi-Maskawa (CKM) matrix elements~\cite{Aoki:2019cca}, in
particular $\left| V_{us}\right| $ and $\left| V_{ud}\right|$, including
radiative corrections~\cite{Carrasco:2015xwa,DiCarlo:2019thl,Desiderio:2020oej,Frezzotti:2020bfa}.

Among other sources of systematic uncertainty in lattice QCD calculations, it is important to quantify the role of the finite volume (FV). This is particularly important when quantum electrodynamics (QED) is included, since the long-range nature of the interaction leads to power-like rather than exponentially suppressed finite-volume effects (FVEs), even in simple quantities like masses and leptonic decay rates.
The power-like FVEs can either be estimated numerically by fitting functional forms to
simulation results at various volumes, or by deriving the volume scaling using
analytic techniques, see
Refs.~\cite{Borsanyi:2014jba,Davoudi:2014qua,Lubicz:2016xro,Davoudi:2018qpl,Bijnens:2019ejw}.

In order to reach sub-percent precision in lattice calculations, isospin
breaking (IB) effects are essential. This means including strong effects coming
from the quark mass difference $m_{u}-m_{d}\neq 0$ as well as electromagnetic
(EM) effects by considering QCD coupled to QED. The latter effects are
particularly complicated for several reasons. Because QED does not have a mass gap, zero-momentum photon modes lead to new infrared divergences and difficulties
in defining charged particles in a FV. The problem can also be understood via Gauss' law, which predicts a flux through a surface containing a charged particle that contradicts naive periodic boundary
conditions~\cite{Hayakawa:2008an,Borsanyi:2014jba,Davoudi:2018qpl}.
However, it is still
possible to define QED in a finite volume in ways that remove or modify the problematic zero
modes. Many prescriptions have been defined, including QED$_{\mathrm{L}}$~\cite{Hayakawa:2008an}, the most commonly used
approach nowadays, but also
QED$_{\mathcal{C}}$~\cite{WIESE199245,Kronfeld:1992ae,KRONFELD1991521,Polley:1993bn,Lucini:2015hfa},
QED$_{\mathrm{M}}$~\cite{Endres:2015gda,Bussone:2017xkb}, QED$_{\mathrm{TL}}$~\cite{Duncan:1996xy,Duncan:1996be} and the infinite-volume reconstruction method~\cite{Feng:2018qpx,Christ:2020jlp,Feng:2021zek}. 

In QED$_{\mathrm{L}}$
the photon zero modes are subtracted on each energy slice, providing a
straightforward regularization of zero-mode singularities in finite-volume QED.
This approach breaks the locality of the theory but still admits a transfer
matrix, preserving its quantum mechanical
interpretation~\cite{Borsanyi:2014jba,Davoudi:2018qpl}. In this paper, we
consider QED$_{\mathrm{L}}$ on a space-time with an infinite time direction, but
compact, periodic space directions of length $L$. We expect that the formalism developed here can be generalized to different formulations of finite-volume QED.

As mentioned above, electromagnetic FVEs are particularly significant as they can scale with inverse
powers of the spatial extent $L$. These are potentially larger than the
exponentially suppressed effects from QCD alone, so an analytic knowledge of EM
FVEs is of great interest for precision calculations in lattice QCD+QED. In a
given hadronic process, the EM FVEs will in principle depend on the structure of the hadrons involved, although it was proven~\citep{Borsanyi:2014jba,Lubicz:2016xro} that, due to gauge invariance, some of the leading-order coefficients are universal.\footnote{Universal is understood as independent of the hadron's structure, i.e.~equivalent with the point-like limit.} For example, in the pseudoscalar finite-volume mass shift, the first non-universal (structure dependent) contribution occurs at order
$\bigo(1/L^3)$~\cite{Davoudi:2014qua}, and is encoded in the EM charge radius of the particle as well as a contribution dictated by the branch-cut of the forward Compton amplitude, described below. At higher orders, other physical quantities appear, e.g.~the EM polarisabilities. For leptonic decays, it is known that structure-dependence occurs at
order $1/L^2$. The point-like EM FVEs for these decays were derived through order $1/L$ in
Ref.~\cite{Lubicz:2016xro} and the point-like limit through order $1/L^3$ was also considered in Ref.~\cite{Tantalo:2016vxk}. 

In this paper, we develop a relativistic and model-independent approach to
derive EM FVEs beyond the point-like approximation. Strongly inspired by
pioneering work on multi-hadron states in a finite
volume~\cite{Luscher:1986pf,Kim:2005gf,Hansen:2015zga}, and following the general proofs of universality of EM FVEs~\citep{Borsanyi:2014jba,Lubicz:2016xro}, the main approach
here is to relate the $1/L$ expansion of amplitudes to momentum singularities of
FV Feynman integrands. The hadronic structure is introduced through generic, relativistic expansions of the vertex functions into hadronic form factors. Our method for deriving FVEs  is systematic and
well-suited for automation. In this vein, most of the analytic results presented in this work are collected in a supplementary \textsc{Mathematica}
notebook that we have also made available~\cite{klfv-zenodo}. Together with the notebook \texttt{FVE\_calculation.nb}
we also provide the package \texttt{fvtools.wl}, that allows the user to compute a variety of different finite-volume coefficients entering EM FVEs.
This more automated approach differs from similar calculations,
e.g.~Ref.~\cite{Lubicz:2016xro}, where focus is put on separately studying master
integrals specific to a given process. We demonstrate the efficiency of our
approach by computing the leading structure-dependent EM FVEs on the
pseudoscalar mass and leptonic decay rate. An important result herein is the
derivation of summation formulae, which generalize those of
Refs.~\cite{Davoudi:2014qua,Davoudi:2018qpl,Bijnens:2019ejw} by also including
infrared (IR) divergent cases needed for leptonic decays. 

In Sec~\ref{sec:sum} we derive the summation formulae which act in later sections
as generic building blocks to calculate EM FVEs. Following this, in
Sec.~\ref{sec:se} we study the finite-volume effects on pseudoscalar masses up to and including
order $\bigo(1/L^3)$. Next, leptonic decays are studied in Sec.~\ref{sec:kl2}.
In particular, we first introduce the structure-dependent matrix elements needed
and derive the finite-size scaling up to and including order $\bigo(1/L^2)$.
After this, we numerically study the effects in Sec.~\ref{sec:num}. Conclusions
and an outlook are given in Sec.~\ref{sec:conclusions}. In
App.~\ref{sec:cj} we provide further mathematical
details on summation formulae, and an exponentially fast method to evaluate
FV coefficients numerically, generalizing the algorithm proposed
in Ref.~\cite{Davoudi:2018qpl} to the case of IR-divergent coefficients.

  \section{Summation formulae}
  \label{sec:sum}
In this section we summarise the derivation of the core mathematical identities used in the calculations of EM FVEs presented in the next sections. These identities allow one to compute the asymptotic behavior in the spatial extent $L$ of general classes of sums over quantized momenta, converging to momentum integrals in the $L\to+\infty$ limit. This asymptotic behavior is known to be deeply related to the regularity of the integrand, and the formulae presented here can be seen as a direct generalization for IR-divergent integrals of similar identities presented in Ref.~\citep{Davoudi:2018qpl}.

As derived in the next sections, electromagnetic first-order corrections to
QCD+QED correlation functions are related to generic sums/integrals of functions of the form
  \begin{equation}
      g_r(k_{\lambda};\{\psp\})=\frac{f(k_{\lambda};\{\psp\})}{\omg(\ksp)^r}\,,
  \end{equation}
 where $\ksp$ is the photon 3-momentum to be summed/integrated,  $\{\psp \}$ is an arbitrary set of external momenta, and $\lambda$ a photon mass IR regulator. Additionally we define the 4-vector $k_{\lambda}=(\lambda,\ksp)$ and the energy function $\omg(\ksp)=\sqrt{\ksp^2+\lambda^2}$. 
  We also consider the spherical coordinates associated to~$k_{\lambda}$
  \begin{equation}
      |k_{\lambda}|=\omg(\ksp)\quad\text{and}\quad
      \hat{k}_{\lambda}=\frac{k_{\lambda}}{|k_{\lambda}|},\qquad\text{with the spatial part}\qquad
      \kspu_{\lambda}=\frac{\ksp}{\omg(\ksp)}=\frac{|\ksp|}{\omg(\ksp)}\kspu\,.
  \end{equation}
  In these coordinates, we assume that $f(k_{\lambda};\{\psp\})$ is analytic in $\omg(\ksp)$ in the vicinity of
  ${\omg(\ksp)=0}$ and non-zero in the $\omg(\ksp)\to 0$ limit.
  EM FVEs are then given by the sum-integral difference
  \begin{equation}
    F_r[f(k_{\lambda};\{\psp\})]=
    \left(\frac{1}{L^3}\sump_{\ksp\in\hat{\T}^3}-\int\frac{\diff^3\ksp}{(2\pi)^3}\right)
    g_r (k_{\lambda};\{\psp\})
    \,,
    \label{eq:fv}
  \end{equation}
  where $\hat{\T}^3$~is the set of all vectors
taking the form $\ksp=\frac{2\pi}{L}\nsp$ where $\nsp$ has integer components,
and the ``primed'' sum means that the null vector $\mathbf{0}$ is
excluded, implementing the QED$_{\mathrm{L}}$~\citep{Hayakawa:2008an,Davoudi:2018qpl} prescription. For later use we introduce the short-hand notation
  \begin{align}
  \Delta '_{\ksp}= 
    \frac{1}{L^3}\sump_{\ksp\in\hat{\T}^3}-\int\frac{\diff^3\ksp}{(2\pi)^3}
     \,.
    \label{eq:kdeltaprime}
  \end{align}
  We next define the limit
  \begin{equation}
    f(\ksp;\{\psp\})=\lim_{\lambda\to 0}f(k_{\lambda};\{\psp\})\,,
  \end{equation}
  which is uniformly convergent in the vicinity of $|\ksp|=0$. We also write the radial expansions
  \begin{equation}
    f(k_{\lambda};\{\psp\})=\sum_{i=0}^{+\infty}f_i(\hat{k}_{\lambda};\{\psp\})\omg(\ksp)^i
    \qquad\text{and}\qquad
    f(\ksp;\{\psp\})=\sum_{i=0}^{+\infty}f_i(\kspu;\{\psp\})|\ksp|^i\,,
    \label{eq:fradexp}
  \end{equation}
  in the vicinity of $\omg(\lambda)=0$ and $|\ksp|=0$, respectively. Because of the
  analyticity assumption made above, one has
  \begin{equation}
    f_i(\kspu;\{\psp\})=\lim_{\lambda\to 0}f_i(\hat{k}_{\lambda};\{\psp\})\,,
  \end{equation}
  uniformly. Substituting the expansion in~\cref{eq:fradexp} into~\cref{eq:fv} and 
  using the substitution $\ksp=\frac{2\pi}{L}\nsp$ leads to
  \begin{equation}
    F_r[f(k_{\lambda};\{\psp\})]=\sum_i\frac{\gamma_{r-i,i}(\{\psp\};\xi)}{(2\pi)^{r-i}}
    \frac{1}{L^{3-r+i}}
    \,.\label{eq:fvexp}
  \end{equation}
  Here
  \begin{equation}
   \gamma_{j,k}(\{\psp\};\xi)=\Delta_{\nsp}'\left[\frac{f_k(\hat{n}_{\xi},\{\psp\})}{\omega_{\xi}(\nsp)^{j}}\right]
   \,,\label{eq:gammajk}
  \end{equation}
  \begin{equation}
      \xi=\frac{L\lambda}{2\pi}\,,\qquad n_\xi=\frac{L}{2\pi}k_\lambda=(\xi,\nsp)\,,
      \qquad \omega_{\xi}(\nsp)=|n_\xi|=\sqrt{\nsp^2+\xi^2}\,,
      \label{eq:defnxi}
  \end{equation}
  and $\Delta_{\nsp}'$ is the sum-integral difference operator over $\nsp$
  \begin{equation}
    \Delta_{\nsp}'=\sump_{\nsp\in\Z^3}-\int\diff^3\mathbf{n}\,.
  \end{equation}
  In~\cref{eq:gammajk}, the sum over $\nsp$ is regulated in the IR by removing the zero mode. It does not require additional infrared regularization and can be considered directly at $\xi=0$. The integral is infrared divergent for $r-i\geq 3$ and $\xi\to 0$. Both the sum and the integral are ultraviolet divergent for $r-i\leq 3$. Below, we derive formulae for arbitrary values of $i$.

  \subsection{Infrared-finite terms}
  We begin with terms in~\cref{eq:fvexp} with $r-i<3$. These terms can be computed
  directly at $\xi=0$
  \begin{equation}
    \gamma_{r-i,i}(\{\psp\})=\Delta_{\nsp}'\left[\frac{f_i(\hat{\nsp},\{\psp\})}{|\nsp|^{r-i}}
    \right]\,.
  \end{equation}
  One notices that for $\xi=0$ the integrand/summand is factorisable in the spherical
  coordinates $(|\nsp|,\hat{\nsp})$ of the 3-vector $\nsp$, simplifying the
  evaluation of $\gamma_{r-i,i}(\{\psp\})$ for a given explicit numerator
  $f_i(\hat{\nsp},{\psp})$.
  \subsection{Infrared power divergences}
  We next consider terms in~\cref{eq:fvexp} with $r-i>3$. In that case both the sum and
  the integral are ultraviolet finite and can be evaluated separately.
  The sum can be evaluated directly at $\xi=0$
  \begin{equation}
    \bar{\gamma}_{r-i,i}(\{\psp\})=\sump_{\nsp\in\Z^3}
    \frac{f_i(\hat{\nsp},\{\psp\})}{|\nsp|^{r-i}}\label{eq:defgammabar}
  \end{equation}
  Regarding the integral, we first express it in spherical coordinates ($n= |\nsp|$)
  \begin{equation}
    \int\diff^3\mathbf{n}\,\frac{f_i(\hat{n}_{\xi},\{\psp\})}{\omega_{\xi}(\nsp)^{r-i}}
    =\int_0^{+\infty}\diff n\,\int_{S^2}\diff^2\hat{\nsp}\, 
    \frac{n^2f_i(\hat{n}_{\xi},\{\psp\})}{\omega_{\xi}(\nsp)^{r-i}}\,,
  \end{equation}
  then we change the radial integration variable to $\xi n$ obtaining
  \begin{equation}
    \int\diff^3\mathbf{n}\,\frac{f_i(\hat{n}_{\xi},\{\psp\})}{\omega_{\xi}(\nsp)^{r-i}}=
    \frac{\phi_{r-i,i}(\{\psp\})}{\xi^{r-i-3}}\,,
  \end{equation}
  with
  \begin{equation}
    \phi_{r-i,i}(\{\psp\})=\int_0^{+\infty}\diff n\,\int_{S^2}\diff^2\hat{\nsp}\, 
    \frac{n^2f_i\left[\frac{1}{\sqrt{1+n^2}}(1,n\hat{\nsp}),\{\psp\}\right]}{(1+n^2)^{\frac{r-i}{2}}}\,.
    \label{eq:defphijk}
  \end{equation}
  Finally we obtain
  \begin{equation}
    \gamma_{r-i,i}(\{\psp\};\xi)=\bar{\gamma}_{r-i,i}(\{\psp\})-\frac{\phi_{r-i,i}(\{\psp\})}{\xi^{r-i-3}}\,.
    \label{eq:gammapowerir}
  \end{equation}
  Both the sum $\bar{\gamma}_{r-i,i}(\{\psp\})$ and the integral $\phi_{r-i,i}(\{\psp\})$
  have to be evaluated explicitly for a given Feyman integrand. However they are independent
  of $\xi$, whose contribution appears explicitly in~\cref{eq:gammapowerir} as a power infrared
  divergence $\xi^{3+i-r}$, as expected from power-counting.
  \subsection{Logarithmic infrared divergences}
  Finally we turn to the special case of~\cref{eq:fvexp} with $r-i=3$. This case is the most challenging
  as the integral is both infrared and ultraviolet divergent. However,
  from counting the superficial degree of divergence we know that the sum and integral each diverge at most logarithmically, so
  capturing only leading divergences is enough. This allows us to regulate both the sum and
  the integral in the ultraviolet by imposing a hard cut-off $R$ on the norm $|\nsp|$.
  Like in the previous case, the sum can be safely evaluated at $\xi=0$, and is expected
  to have the asymptotic behavior
  \begin{equation}
    \sump_{|\nsp|<R}\frac{f_i(\hat{\nsp},\{\psp\})}{|\nsp|^3}\underset{R\to+\infty}{=}
    \ell_i (\{\psp\})\log(R)+C_i^{(S)}(\{\psp\}) + \bigo\left(\frac{1}{R}\right)\,.
  \end{equation}
  Regarding the integral, as in the previous case we can change the radial integration variable to $\xi n$
  \begin{equation}
    \int_{|\nsp|<R}\diff^3\mathbf{n}\,\frac{f_i(\hat{n}_{\xi},\{\psp\})}{\omega_{\xi}(\nsp)^3}
    =\int_0^{\frac{R}{\xi}}\diff n\,\int_{S^2}\diff^2\hat{\nsp}\, 
    \frac{n^2f_i\left[\frac{1}{\sqrt{1+n^2}}(1,n\hat{\nsp}),\{\psp\}\right]}{(1+n^2)^{\frac{3}{2}}}\,.
    \label{eq:logirrescaledint}
  \end{equation}
  So the integral is only a function of $R/\xi$, which enters as the upper bound of the radial
  integral. The $R\to+\infty$ leading behavior of the sum, i.e.~the coefficient $\ell_i(\{\psp\})$, has to be identical in the case of the integral. In summary the $R/\xi\to+\infty$ asymptotic behavior of the integral has the form
  \begin{equation}
    \int_{|\nsp|<R}\diff^3\mathbf{n}\,\frac{f_i(\hat{n}_{\xi},\{\psp\})}{\omega_{\xi}(\nsp)^3}
    \underset{R\to+\infty}{=}\ell_i(\{\psp\})\log\left(\frac{R}{\xi}\right)+C_i^{(I)}(\{\psp\})\,.
  \end{equation}
  Now, defining $r=R/\xi$, the coefficient $\ell_i(\{\psp\})$ can be obtained 
  as the logarithmic derivative in $r$ of~\cref{eq:logirrescaledint} in the
  $r\to+\infty$ limit. Let us start by computing the logarithmic derivative
  \begin{equation}
    r\frac{\partial}{\partial r}\int_{|\nsp|<R}\diff^3\mathbf{n}\,\frac{f_i(\hat{n}_{\xi},\{\psp\})}{\omega_{\xi}(\nsp)^3}=\frac{r^3}{(1+r^2)^{\frac{3}{2}}}\int_{S^2}\diff^2\hat{\nsp}\, 
    f_i\left[\frac{1}{\sqrt{1+r^2}}(1,r\hat{\nsp}),\{\psp\}\right]\,.
    \label{eq:logirlogder}
  \end{equation}
  Then we have the limit
  \begin{equation}
    \lim_{r\to+\infty}\frac{r^3}{(1+r^2)^{\frac{3}{2}}}f_i\left[\frac{1}{\sqrt{1+r^2}}(1,r\hat{\nsp}),\{\psp\}\right]
    =f_i(\hat{\nsp},\{\psp\})\,.
  \end{equation}
  Since the remaining integral in~\cref{eq:logirlogder} is operating on a continuous
  function over a compact manifold, the limit above can be interchanged with the integral
  (bounded convergence theorem) to give $\ell_i(\{\psp\})$
  \begin{equation}
    \ell_i(\{\psp\})=\lim_{r\to+\infty}r\frac{\partial}{\partial r}\int_{|\nsp|<R}\diff^3\mathbf{n}\,\frac{f_i(\hat{n}_{\xi},\{\psp\})}{\omega_{\xi}(\nsp)^3}=\int_{S^2}\diff^2\hat{\nsp}\, f_i(\hat{\nsp},\{\psp\})\,.
    \label{eq:bj}
  \end{equation}
  Then the coefficients $C_i^{(S)}(\{\psp\})$ and $C_i^{(I)}(\{\psp\})$ are given
  as the finite limits
  \begin{align}
    C_i^{(S)}(\{\psp\})&=
    \lim_{R\to+\infty}\left[\sump_{|\nsp|<R}\frac{f_i(\hat{\nsp},\{\psp\})}{|\nsp|^3}-\ell_i(\{\psp\})\log(R)\right]\,,\\
    C_i^{(I)}(\{\psp\})&=
    \lim_{r\to+\infty}\left\{\int_0^{r}\diff n\,\int_{S^2}\diff^2\hat{\nsp}\, 
    \frac{n^2f_i\left[\frac{1}{\sqrt{1+n^2}}(1,n\hat{\nsp}),\{\psp\}\right]}{(1+n^2)^{\frac{3}{2}}}-\ell_i(\{\psp\})\log(r)\right\}\,,\label{eq:bji}
  \end{align}
  finally yielding
  \begin{equation}
    \gamma_{3,i}(\{\psp\};\xi)=C_i^{(S)}(\{\psp\})-C_i^{(I)}(\{\psp\})+\ell_i(\{\psp\})\log(\xi)\,.
    \label{eq:gamma3jform}
  \end{equation}
  The integral limit~\cref{eq:bji} can be put in a somewhat more convenient form
  for explicit evaluations. Combining the identity
  \begin{equation}
    \int_0^r\diff n\,\frac{n^2}{(1+n^2)^{\frac{3}{2}}}\underset{r\to+\infty}{=}
    \log(r)+\log(2)-1\,,
  \end{equation}
  with~\cref{eq:bji}, one can show
  \begin{equation}
    C_i^{(I)}(\{\psp\})=
    \int_0^{\infty}\diff n\,\int_{S^2}\diff^2\hat{\nsp}\, 
    \frac{n^2\left\{f_i\left[\frac{1}{\sqrt{1+n^2}}(1,n\hat{\nsp}),\{\psp\}\right]-f_i(\hat{\nsp},\{\psp\})\right\}}{(1+n^2)^{\frac{3}{2}}}-[1-\log(2)]\ell_i(\{\psp\})\,.
  \end{equation}
These identities will be used in Appendix~\ref{sec:cj} for deriving finite-volume coefficients appearing in the physical calculations presented below. 

  \section{Self-energy of pseudoscalar mesons}
  \label{sec:se}
In this section we consider the FVEs in the pseudoscalar mass at leading order in QED. Most of the results presented here have been derived previously and already used in lattice calculations, where they play a crucial role in the determination of physical quark masses in lattice QCD+QED calculations. In particular, we determine the leading structure-dependent corrections, which starts at order $1/L^3$, and establish some of the concepts needed to handle the leptonic decays in the next section. One key result described in details in this work is the contribution of a term at $1/L^3$ dictated by an integral along the branch cut of the forward Compton amplitude.

\subsection{The electromagnetic self-energy and its finite-size effects}
We consider an interpolating operator $\phi$ which couples to a charged, spin-$0$, stable hadronic state $P$ (e.g.~a pion or kaon) with mass $m_P$ in the full QCD+QED theory. We define the infinite-volume (IV) and finite-volume (FV) Euclidean momentum-space 2-point functions of $\phi$ as
\begin{align}
C^\infty_2(p) & =\int\diff^4 x\,\bra{0}\mathrm{T}[\phi(x)\phi^\dagger(0)]\ket{0}
e^{-ipx}\,,\label{eq:c2def}  \\
C^L_2(p) & = \int \diff x_0 \int_{\T^3}\diff^3 \textbf x\,\bra{0}\mathrm{T}[\phi(x)\phi^\dagger(0)]\ket{0}_L
e^{-ipx}\,,\label{eq:c2defFV} 
\end{align}
where the expectation value is understood to be in QCD+QED and in the FV case this is implemented via QED$_{\mathrm{L}}$. As indicated, the FV quantity is defined with periodic boundary conditions on the three-torus $\T^3$ and the spatial integral runs over this domain, whereas the spatial integral defining $C^\infty_2(p)$ runs over $\R^3$. We work throughout in a continuum theory and also take the temporal extent to be infinite.

While $C^\infty_2$ only depends on $p^2 = p_0^2 + \psp^2$, for $C^L_2$ separate dependence on $p_0$ and $\psp$ is induced by the reduced symmetry. 
In this work we consider both the IV and FV two-point functions in the complex $p_0$ plane, but only in the neighborhood of the on-shell point, $p_0 = i \sqrt{m_P^2 + \psp^2}$. In this region, $C^\infty_2$ contains a pole corresponding to $P$ together with a branch cut, running from this pole up the imaginary axis and corresponding to multi-particle states involving any number of photons together with $P$. Here we are interested in the pole position, from which one can define the physical mass $m_P$ and the operator-state overlap via
\begin{equation}
\label{eq:C2fulloverlap}
\lim_{p^2 \to - m_P^2} (p^2 + m_P^2) \, C^\infty_2(p) =  Z_P^2  \,, \qquad \qquad Z_P = \bra{0}\phi(0)\ket{P,\psp}\,.
\end{equation}
Without loss of generality we choose the phase in $\phi(0)$ and the states such that $Z_{P}$ is real and positive. Similarly, the finite-volume 2-point function, $C^L_2$, contains a tower of poles along imaginary $p_0$. For $\psp = \textbf 0$, the lowest lying of these is denoted by $p_0 = i m_P(L)$ where $m_P(L)$ is referred to as the FV mass since it satisfies
\begin{equation}
\lim_{L \to \infty} m_P(L) = m_P \,.
\end{equation}
The difference $\Delta m_P^2(L) = m_P(L)^2 - m_P^2$ is known to satisfy a series expansion in $1/L$ to which all positive integer powers contribute~\citep{Borsanyi:2014jba}. The main aim of this section is to review the determination of the $1/L$, $1/L^2$ and $1/L^3$ terms in QED$_{\mathrm{L}}$, while setting up the formalism for the next section. In the following paragraphs we first focus on QED corrections in the IV theory before returning to $C^L_2(p)$ at the end of the subsection.

\subsubsection{Electromagnetic self-energy}
Because of the perturbative nature of QED, we will work at the leading order $\mathcal{O}(e^2)$ in the elementary electric charge $e$. To define this expansion it is necessary to make reference to QCD-only quantities, which is inherently ambiguous. We assume a suitable separation scheme has been used to set the quark mass values $m_u \neq m_d$ in the theory without photons, and a review of the schemes used in lattice QCD+QED calculations can be found in~\citep{Aoki:2019cca}. The 2-point function in the QCD-only set-up has a shifted pole position with location denoted by $m_{P,0}$, i.e.
\begin{equation}
\label{eq:Z0def}
\lim_{p^2 \to - m_{P,0}^2} (p^2 + m_{P,0}^2) \, C^\infty_2(p)_{e=0} =  Z_{P,0}^2 \,, \qquad \qquad Z_{P,0} = \bra{0}\phi(0)\ket{P,\psp}_{e=0} \,,
\end{equation}
where again we take $Z_{P,0}$ to be real and positive.

More generally, the QCD-only correlator can be written as
\begin{equation}
\label{eq:QCDse}
C^\infty_2(p)_{e=0} = \frac{1}{p^2 + m_{P,0}^2 - \Sigma_{\text{QCD}}(p^2)} \,,
\end{equation}
where $\Sigma_{\text{QCD}}(p^2)$ is the renormalized self-energy of $P$ from the strong interaction. Here we have chosen the convention of setting $\Sigma_{\text{QCD}}(- m_{P,0}^2) = 0$ such that there is no distinction between the renormalized mass and the pole mass. Matching Eqs.~\eqref{eq:Z0def} and \eqref{eq:QCDse} then further gives
\begin{align}
Z_{P,0}^{-2} &= 1 - \frac{\partial\Sigma_{\text{QCD}}}{\partial p^2}(-m_{P,0}^2)\,,
\end{align}
which encodes a second conventional freedom in the theory, e.g.~one can set the derivative to vanish such that $ Z_{P,0} = 1$. We choose to keep $Z_{P,0}$ general to show that it has no effect on physical quantities. We further find it convenient to define
\begin{equation}
- \Sigma_{0}(p^2) = - \Sigma_{\text{QCD}}(p^2) - (    Z_{P,0}^{-2} - 1) (p^2 + m_{P,0}^2) \,,
\end{equation}
which simply amounts to removing the $\mathcal O[(p^2 + m_{P,0}^2)]$ term so that $\Sigma_{0}(p^2) = \mathcal O[(p^2 + m_{P,0}^2)^2]$ near the pole. Substituting this into Eq.~\eqref{eq:QCDse} gives
\begin{equation}
\label{eq:C2QCDsep}
C^\infty_2(p)_{e=0} = Z_{P,0} \cdot D_0(p) \cdot Z_{P,0} \,,
\end{equation}
with
\begin{equation}
D_0(p) = \frac{1}{p^2 + m_{P,0}^2 - \Sigma_{0}(p^2) } \,.
\end{equation}
This simple factorization into overlaps and the fully dressed QCD-propagator with unit residue can be represented diagrammatically as
\begin{equation}
\raisebox{-1.3ex}{\includegraphics{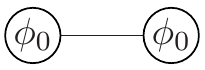}} = C^\infty_2(p)_{e=0} \,,
\end{equation}
where the line is $D_0(p)$ and the circles are the two overlap factors. One can also introduce $Z_0(p^2)$ as a (trivial) alternative to $\Sigma_0(p^2)$
\begin{equation}
Z_0(p^2) = \bigg [ 1 - \frac{ \Sigma_{0}(p^2)}{p^2 + m_{P,0}^2} \bigg ]^{-1} \,,
\end{equation}
such that
\begin{equation}
\label{eq:DeltaQCD}
D_0(p) = \frac{Z_0(p^2)}{p^2 + m_{P,0}^2} \,.
\end{equation}

Returning to the full QCD+QED theory, the 2-point function can be represented diagramatically as
\begin{equation}
C^\infty_2(p)=\raisebox{-1.3ex}{\includegraphics{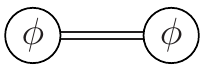}} \,,
\end{equation}
and expanding to leading order in the elementary charge squared gives
\begin{equation}
\raisebox{-1.3ex}{\includegraphics{axo_full2pt.pdf}}=
\raisebox{-1.3ex}{\includegraphics{axo_se0.pdf}}+
\raisebox{-2.1ex}{\includegraphics{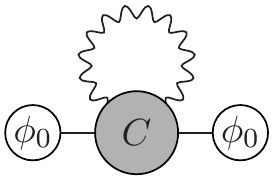}}+\bigo(e^4)\,.
\label{eq:2ptexp}
\end{equation}
We will implicitly neglect relative $\bigo(e^4)$ corrections to observables throughout this work. In~\cref{eq:2ptexp}, the grey blob labelled $C$ represents the Compton scattering kernel. In the limit that the external pseudo-scalar legs are on-shell, this becomes the forward Compton scattering amplitude
\begin{gather}\label{eq:comptonkernelcorrfcn}
\raisebox{-2.1ex}{\includegraphics{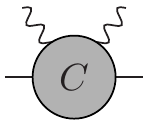}} = C_{\mu\nu}(p,k, q)
\,, \\[5pt]
\lim_{p^2 \to - m_{P}^2} C_{\mu\nu}(p,k, -k) =
 \int \diff^{4}x \, e^{-ikx} \, \bra{P,\mathbf{p}} T\left\{ J_{\mu}(x) J_{\nu}(0) \right\} \ket{P,\mathbf{p}}
\,,
\end{gather}
where $J_{\mu}$ is the Euclidean quark electromagnetic current. We absorbed the electric charge factor $e$ within the current. Here we have chosen the on-shell point in the full theory, i.e.~$m_{P}^2$ rather than $m_{P,0}^2$. The difference between these two choices within the Compton amplitude leads to a $\mathcal O(e^4)$ effect that is beyond the order we control. In general, the off-shell continuation of $C_{\mu \nu}$ is ambiguous and depends on the arbitrary choice of $\phi$:
 \begin{multline}
\label{eq:Cmunuoff}
 C_{\mu\nu}(p,k, q) =    Z^{-2}_{P,0} \, D_0(p)^{-1}  D_0(p + k + q)^{-1}  \int\diff^4 x \, \diff^4 y \,\diff^4 z \, e^{i p z + i k x + i q y} \\[-8pt] \times  \bra{0}\mathrm{T}[\phi(0) J_\mu(x)  J_\nu(y) \phi^\dagger(z)]\ket{0}
\,.
 \end{multline}
However, any such operator dependence must cancel in any spectral quantity, including $m_P$, $m_P(L)$ and, in particular, any coefficient multiplying a power of $1/L$ in the latter.

In order to relate the electromagnetic corrections in the 2-point function to those in the mass of $P$, one must sum the usual infinite subset of diagrams
\begin{align}
\raisebox{-1.3ex}{\includegraphics{axo_full2pt.pdf}}&=
\raisebox{-1.3ex}{\includegraphics{axo_se0.pdf}}+
\raisebox{-2.1ex}{\includegraphics{axo_se1.pdf}}+
\raisebox{-2.1ex}{\includegraphics{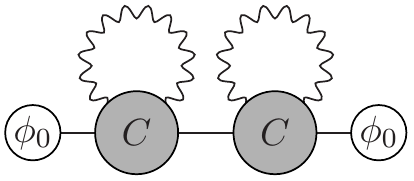}}\notag\\
&\qquad+\cdots+\raisebox{-2.1ex}{\includegraphics{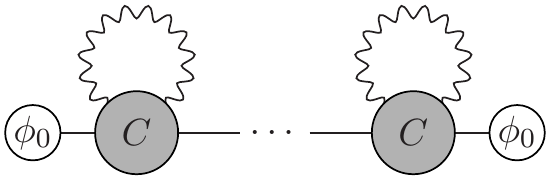}}+\cdots \,,
\label{eq:geom}
\end{align}
where the self-contracted kernel defines the $\mathcal O(e^2)$ self-energy function
\begin{align}\label{eq:defC}
\Sigma(p^2)=\raisebox{-2.1ex}{\includegraphics{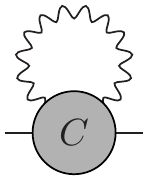}}\,.
\end{align}
Performing the summation in~\cref{eq:geom}, one obtains
\begin{equation}
\label{eq:fullSE}
C_2^\infty(p)  = \frac{Z_{P,0}^2}{p^2+m_{P,0}^2 - \Sigma_{0}(p^2) - \Sigma(p^2)}  \,.
\end{equation}
The value of $\Sigma(- m_{P}^2)$ and its derivative is specified by the chosen scheme for defining the $e\to 0$ limit of QCD+QED. The full QCD+QED mass is given by solving
\begin{equation}
    p^2 + m_{P,0}^2  - \Sigma_{0}(p^2) - \Sigma(p^2) \bigg \vert_{p^2 = -m_P^2} = 0 \,,
\end{equation}
which reduces to
\begin{equation}
    \Delta m_P^2 = m_P^2 - m_{P,0}^2 = - \Sigma(- m_P^2) + \mathcal O(e^4) \,.
    \label{eq:mren}
\end{equation}
Here we have used that $\Sigma_0(p^2) = \mathcal O[(p^2 + m^2_{P,0})^2]$ (by construction) and thus only contributes at $\mathcal O(e^4)$. 

Following Eq.~\eqref{eq:C2QCDsep} above, we also define
\begin{equation}
 C_2^\infty(p)  = Z_P \cdot D(p) \cdot Z_P \,,
\end{equation}
where $Z_P$ is already defined in Eq.~\eqref{eq:C2fulloverlap} and
\begin{equation}
\label{eq:DeltaFull}
    D(p) = \frac{Z(p^2)}{p^2 + m_P^2} \,,
\end{equation}
with $Z(p^2) = 1 + \mathcal O[(p^2 + m_P^2)]$. A particularly important quantity in the following section will be the ratio between operator overlaps in the QCD-only and full QCD+QED theories. We parametrize this via 
\begin{align}
Z_P &=   Z_{P,0} (1 + \delta_{Z_P})   \,.
\label{eq:wfren}
\end{align}
One can readily show
\begin{equation}
\label{eq:dZPdef1}
      \delta_{Z_P}  = \frac12 \big [ \Sigma_0'(- m_P^2) + \Sigma'(- m_P^2) \big ] \,. 
\end{equation}
In contrast to the pole shift, both $\Sigma_0(p^2)$ and $\Sigma(p^2)$ contribute to the overlap at the order we work.

\bigskip

Returning to the finite-volume system, an identical argument can be applied to reach a finite-volume version of \cref{eq:fullSE} in which $C_2^\infty \to C_2^L$ and the two self energies on the right-hand side both receive $L$ dependence. As was shown in Ref.~\citep{Luscher:1985dn}, the finite-volume QCD-only self-energy, call it $\Sigma^L_{0}(p_0^2, \psp = \textbf 0)$, vanishes as $e^{- m_{P,0} L}$ when evaluated at $p_0^2 = - m_{P,0}^2$. Therefore, the leading finite-volume effects are given by the difference between the FV and IV QED contributions:
\begin{equation}
\Delta m^2_P(L) = m_P(L)^2 - m_P^2 = - \big [ \Sigma^L(-m_{P}^2, \textbf 0) - \Sigma(-m_{P}^2) \big ] \,,
\end{equation}
where the second argument of $ \Sigma^L$ indicates that we focus on $P$ at rest in the FV frame. The rest of this section could be derived in an arbitrary FV frame as done in Ref.~\citep{Davoudi:2018qpl}, however for the sake of simplicity we will only consider the rest frame.

The power-like $1/L$ scaling within $\Sigma^L(-m_{P}^2 , \textbf 0)$ is due only to the fact that the spatial part of the photon momentum $k$ is summed over the discrete modes satisfying the periodic boundary conditions, with $\ksp = \textbf 0$ removed. In particular, one can take the IV definition of $C_{\mu \nu}$ within $\Sigma^L$ as the difference to the FV quantity is again exponentially suppressed. One finds
\begin{equation}\label{eq:DmPL}
\Delta m^2_P(L) = - \frac{e^2}{2} \lim_{p_0^2 \to - m_{P}^2} \Delta_{\ksp}'
\int\frac{\diff k_0}{ 2\pi }
\frac{C_{\mu\mu}(p,k,-k)}{k^2} \bigg \vert_{\psp = \boldsymbol 0}
\,,
\end{equation}
where $\Delta_{\ksp}'$ is defined in \cref{eq:kdeltaprime} above.
This implies that the FV effects on the mass, including structure-dependent contributions, can be related to the physical properties of the Compton scattering amplitude. In particular, it is clear that the finite-size effects on the physical mass cannot depend on the arbitrary choice of the interpolating operator $\phi$, and we expect any term depending on $\phi$ to cancel in the final result. To obtain the large-volume expansion of~\cref{eq:DmPL}, one can use the summation formulae derived in the previous section. This requires to discuss the reduction of the Compton kernel which is the purpose of the next section.

\subsubsection{Irreducible electromagnetic vertex functions}
\label{sec:irredv}
It is now useful to decompose the Compton kernel in irreducible diagrams as
\begin{equation}\label{eq:skelexpcompton}
\raisebox{-2.1ex}{\includegraphics{axo_sigker.pdf}}=
\raisebox{-2.1ex}{\includegraphics{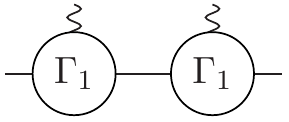}}+\raisebox{-2.1ex}{\includegraphics{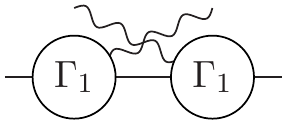}}+
\raisebox{-2.1ex}{\includegraphics{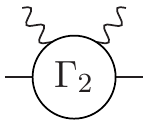}}\,,
\end{equation}
where the white blobs, labelled $\Gamma_1$ and $\Gamma_2$, correspond to $\Gamma_\mu$ and $\Gamma_{\mu\nu}$ the $P\to P\gamma^* $ and $P   \to P\gamma^* \gamma^*$
irreducible electromagnetic vertex functions, respectively,
\begin{multline}
\label{eq:CmunuDecompose}
C_{\mu\nu}(p,k,q) =  \Gamma_\mu(p,k)  D_0(p+k)  \Gamma_\nu(p+k,q) +  \Gamma_\nu(p, q)  D_0(p+q) \Gamma_\mu(p+q,k)
+  \Gamma_{\mu\nu}(p,k,q) \,.
\end{multline}
The subscripts in the diagramatic notation indicate the number of photon currents and thus also the number of Lorentz indices.
As the precise definition of the off-shell $C_{\mu \nu}(p,k,q)$ is given in Eq.~\eqref{eq:Cmunuoff}, we only require a definition of $\Gamma_{\mu}(p,k)$ to give a complete specification. The latter is defined as
\begin{align}
\label{eq:GammamuOffshellDef}
 \Gamma_\mu(p,k)= Z^{-2}_{P,0} \, D_0(p+k)^{-1} D_0(p)^{-1} \int\diff^4 x \, \diff^4 y \, e^{ i p x + i k y} \,\bra{0}\mathrm{T}[\phi(0) J_\mu(y) \phi^\dagger(x)]\ket{0}
\,.
\end{align}
Following the conventions of Refs.~\cite{Rudy:1994qb,Fearing:1996gs}, this off-shell vertex function can be decomposed into form factors
\begin{equation}
\label{eq:Gammamu}
    \Gamma_\mu(p,k) = (2p+k)_\mu \,  F(k^2,(p+k)^2,p^2) + k_\mu \,  G(k^2,(p+k)^2,p^2)\,.
\end{equation}
Through gauge-invariance and Eq.~\eqref{eq:GammamuOffshellDef}, one can show that $\Gamma_\mu(p,k)$ must satisfy the Ward-Takahashi identity (WTI)
\begin{eqnarray}
    k_\mu \Gamma^\mu(p,k) = D_0(p+k)^{-1} - D_0(p)^{-1}\,.
\end{eqnarray}
This implies relations for the off-shell form factors:
\begin{align}
\label{eq:FtoZrelation}
    F(0,p^2,-m_{P,0}^2) & = F(0,-m_{P,0}^2,p^2) = Z_0(p^2)^{-1} \,, \\
     G(k^2,(p+k)^2,p^2) &= \frac{D_0(p+k)^{-1} - D_0(p)^{-1}}{k^2} - \left(1+\frac{2\,p\cdot k}{k^2}\right) F(k^2,(p+k)^2,p^2)\,,
\end{align}
and these results combine to give a particularly useful form in the case that $p^2$, within $G$, is set to its on-shell value 
\begin{equation}
 G(k^2,(p+k)^2, -m_{P,0}^2 ) = 
  \frac{(p+k)^2  + m_{P,0}^2}{k^2}   \Big [  F(0, (p+k)^2, - m_{P,0}^2) - F(k^2,(p+k)^2, - m_{P,0}^2) \Big ] \,.
 \end{equation}

Analogous identities can be derived for the off-shell two-photon vertex, $\Gamma_{\mu\nu}$, which
is defined through Eq.~\eqref{eq:CmunuDecompose}.
This satisfies its own WTI
\begin{equation}
\label{eq:Gammamunu_WI}
    k^\mu \Gamma_{\mu\nu}(p,k,q) = \Gamma_{\nu}(p,q)-\Gamma_\nu(p+k,q)\,,
\end{equation}
 and, setting $q = - k$, one can expand the vertex in powers of $k_\mu$ to show
  \begin{equation}
 \label{eq:Gammamunu2}
     \Gamma_{\mu\nu}(p,k,-k) = -2\, \delta_{\mu\nu} \,F(0, p^2, p^2) - 8 \,p_\mu p_\nu\, F^{(0,0,1)}(0,p^2,p^2) + \bigo(k)\,.
 \end{equation}
 To reach this expression one makes use of the fact that transverse terms, i.e.~those not constrained by the WTI, do not appear at leading order.
 
 \subsection{Large-volume expansion and cancellation of off-shell contributions}
 \label{sec:offscanc}
We now have all required expressions to start reducing the sum-integral difference in Eq.~\eqref{eq:DmPL}. To do so, one first substitutes Eq.~\eqref{eq:CmunuDecompose} into Eq.~\eqref{eq:DmPL}, to express the finite-volume shift to the pseudo-scalar mass in terms of the irreducible vertex functions
\begin{equation}
    \Delta m_P^2(L) =  - e^2 \lim_{p^2 \to - m_{P,0}^2}  \Delta_{\ksp}'
\int\frac{\diff k_0}{ 2\pi } \frac{1}{k^2} \bigg [ \frac{1}{2} \Gamma_{\mu \mu}(p,k,-k) +  \Gamma_\mu(p, -k)  D_0(p-k) \Gamma_\nu(p-k,k) \bigg ] \,,
\end{equation}
where $\textbf p = \textbf 0$ is understood. Here we have used the $k \to - k$ invariance of the integrand to combine the two one-particle reducible terms into one.
The next step is to substitute the decompositions of the single-photon functions in terms of the form factors $F$ and $G$ to reach
\begin{align}
\Delta m^2_P(L)  = -  \lim_{p^2 \to - m_{P,0}^2}   \Delta_{\ksp}'
\int\frac{\diff k_0}{ 2\pi } & \frac{1}{k^2}     \mathcal I(k,p) \,,\label{eq:inti}
\end{align}
where we have introduced
\begin{align}
    \mathcal I(k,p) & =  \frac12 \Gamma_{\mu \mu}(p,k,-k)  +    (2p-k)^2 \, D_0(p - k) \, F(k^2, p^2, (p-k)^2)^2
    \notag \\[8pt]
&  + 2 k \cdot (2p - k) \,  D_0(p - k) \, G(k^2, p^2, (p-k)^2) \, F(k^2,p^2, (p-k)^2 ) \,, \notag \\[8pt]
&    + k^2 \, D_0(p-k) \, G(k^2, p^2, (p-k)^2)^2  \,,
\end{align}
and have used that $F$ is symmetric and $G$ is anti-symmetric with respect to interchange of the last two arguments. As known from the summation formulae discussed in~\cref{sec:sum}, the leading behavior in the $1/L$ expansion will be driven by the singularities of the integrand in \cref{eq:inti} for $\ksp\to\mathbf{0}$. These can be captured by studying the $k\to 0$ behavior of $\mathcal I(k,p)$.
Sending $p^2 \to -m_{P,0}^2$ within $\mathcal I(k,p)$ and expanding about $k=0$, one obtains
\begin{align}
 \lim_{p^2 \to - m_{P,0}^2}  \mathcal I(k,p)  & =    -  \, 4 + 4 \, m_{P,0}^2 \, F^{(0,0,1)}(0,- m_{P,0}^2, - m_{P,0}^2)\notag  \\[8pt]
 & \hspace{-20pt} +   \frac{ Z_0((p-k)^2)   }{(p-k)^2  + m_{P,0}^2 }   (2p-k)^2 
 \notag \\[8pt]
 & \hspace{-20pt} + 2   \frac{ Z_0((p-k)^2)   }{(p-k)^2  + m_{P,0}^2 }   (2p-k)^2    \big [ (p-k)^2 + m_{P,0}^2 \big ] F^{(0,0,1)}(0,- m_{P,0}^2,- m_{P,0}^2)  
 \notag \\[8pt]
& \hspace{-20pt} + \mathcal O(k) \label{eq:ikexp}
\,,
\end{align}
where we have set everywhere $F(0,- m_{P,0}^2, - m_{P,0}^2) = 1$, which is just the electric charge of $P$ in units of $e$. The first line here arises from the expansion of $\Gamma_{\mu \mu}$ while the second and third follow from the one-particle reducible term proportional to $F^2$. Though we have set $p^2 \to - m_{P,0}^2$ in all terms, the off-shell form factors still contribute.
 
 To see that all unphysical contributions explicitly cancel, note that Eq.~\eqref{eq:FtoZrelation} implies
\begin{equation}
   F^{(0,0,1)}(0,- m_{P,0}^2,- m_{P,0}^2) =    \frac{\partial Z_0(p^2)^{-1}}{\partial p^2} \bigg \vert_{p^2 = - m_{P,0}^2} = z_1  \,.
\end{equation}
The last line here is a definition that will be extended to higher orders in~\cref{sec:emdecay}. Using this last identity and continuing the expansion in $k$ of~\cref{eq:ikexp}, one finally reaches
\begin{multline}
 \lim_{p^2 \to - m_{P,0}^2}  \mathcal I(k,p)  =    -  \, 4 + 4 \, m_{P,0}^2 \, z_1 
  +   \frac{ 1 - z_1 [(p-k)^2  + m_{P,0}^2]   }{(p-k)^2  + m_{P,0}^2 }   (2p )^2 
  + 2     (2p )^2    z_1
  + \mathcal O(k) 
\,,\label{eq:iz1canc}
\end{multline}
where the first two terms arise from $\Gamma_{\mu \mu}$ and the third and fourth give the leading self-energy and off-shell form factor corrections, respectively, from the contribution proportional to $F^2$. The key point is that the $z_1$ factors cancel, since $4 m_{P,0}^2 z_1 - z_1 (2p)^2 + 2 (2p)^2 z_1 = 0$ for $p^2= - m_{P,0}^2$, and the result is therefore independent from the choice for $\phi$.

Now using~\cref{eq:iz1canc} and performing the $k_0$ integral in~\cref{eq:inti} gives
\begin{equation}
\Delta m_P^2(L)  =   e^2  \, \Delta_{\ksp}' \,\left[ \frac{ m_P  }{ \vert \ksp \vert^2    }
   +\frac{1}{\vert \ksp \vert} + \mathcal O(1)\right] 
\,.
\end{equation}
Using the summation formula~\cref{eq:fvexp} then directly leads to the well-known~\citep{Davoudi:2014qua,Borsanyi:2014jba,Lubicz:2016xro,Davoudi:2018qpl} universal FVEs to the EM self-energy
\begin{equation}
    \Delta m_P^2(L)  =e^2\left[\frac{m_Pc_2}{4\pi^2L}+\frac{c_1}{2\pi L^2}+\mathcal{O}\left(\frac{1}{L^3}\right)\right] \,,
\end{equation}
where we have used the zero velocity IR-finite finite-size coefficients for $j<3$
\begin{align}\label{eq:cjzerovel}
c_{j} = \Delta _{\mathbf{n}}' \frac{1}{\left| \mathbf{n}\right| ^{j}} \, .
\end{align}
As is shown in~\cref{sec:cj}, the numerical values of the two appearing above are $c_{1}\simeq-2.83730$ and $c_{2}=\pi c_1\simeq-8.91363$.

Demonstrating the explicit cancellation of unphysical contributions in $\Delta m_P^2(L)$ for the leading universal FVEs was the main aim of this subsection. These contributions are expected to cancel at all order since the EM self-energy cannot possibly depend on the choice of interpolating operator. In fact, one also expect the FV and IV self-energies to be individually independent from it. Therefore in order to work to higher orders in $1/L$ more easily, we turn now to an alternative approach where the decomposition of the Compton amplitude from the start does not depend on the choice of the pseudoscalar interpolating operator.

\subsection{Manifestly on-shell derivation and structure-dependent finite-size effects}
\label{sec:seonshell}
We now demonstrate how one can use freedom in the decomposition of $C_{\mu \nu}$, in order to remove all off-shell dependence at the beginning of the calculation. To achieve this we first define the forward, on-shell Compton amplitude, with vector indices contracted
\begin{equation}
    T(k^2 , k \cdot p) =  \lim_{p^2 \to - m_{P}^2} C_{\mu\mu}(p,k,-k) \,.
\end{equation}
The key idea then, is to define an alternative decomposition to Eq.~\eqref{eq:CmunuDecompose} in which $D_0(p)$ is replaced with the simple $1/(p^2 + m_{P,0}^2)$
factor and all off-shellness is absorbed into redefinitions of the $\Gamma$ functions. We write
\begin{equation}
\label{eq:Tdecom}
  T(k^2 , k \cdot p)
  =  \frac{\Gamma^{\sf on}_{\mu}(p,k) {\Gamma^{\sf on}_\mu}(p+k,-k) }{(p+k)^2  + m_{P}^2 } + \frac{\Gamma^{\sf on}_{\mu}(p,-k) {\Gamma^{\sf on}_\mu}(p-k,k) }{(p-k)^2   + m_{P}^2 }  + \Gamma^{\sf on}_{\mu \mu}(p,k,-k) \,,
\end{equation}
where the four-vector $p_\mu$ is understood to be on-shell but, for now, at generic spatial momentum: $p_\mu = (i\omega_{P}(\psp) , \psp)$ with $\omega_{P}(\psp) =\sqrt{\psp^2 + m_{P}^2} $. Here we have also replaced $m_{P,0}$ with $m_P$ as the difference enters the mass at $O(e^4)$, i.e. beyond the order we control.

Next decompose  $\Gamma^{\sf on}_\mu$ in direct analog to Eq.~\eqref{eq:Gammamu} above, as
\begin{align}\label{eq:g1g2formfactdec}
 \Gamma^{\sf on}_{\mu}(p,k) & = (2p+k)_{\mu} F(k^2) + k_{\mu} \, G^{\sf on}(k^2,(p+k)^2,p^2) \,,
\end{align}
where here $F(k^2)$ is the physical, on shell electromagnetic form factor of $P$
\begin{equation}
\bra{P,\psp+\ksp}J_{\mu}(0)\ket{P,\psp}=(2p_{\mu}+k_{\mu})F(k^2)\,.
\end{equation}
The second form factor, $G^{\sf on}$, is defined by
\begin{align}
G^{\sf on}(k^2,(p+k)^2,p^2) = \frac{ (p+k)^2 - p^2}{k^2} \big [ 1 - F(k^2) \big ]
\,,
\end{align}
where we used $F(0)=1$, which completes the specification of $ \Gamma^{\sf on}_{\mu}$ and, through Eq.~\eqref{eq:Tdecom}, the definition of $\Gamma_{\mu \nu}^{\sf on}$ as well.

This decomposition leads to a new, manifestly on-shell expression for the finite-volume mass shift $\Delta m^2_P(L)$:
\begin{align}
\begin{split}
\Delta m^2_P(L)  = - e^2  \, \Delta_{\ksp}'
\int\frac{\diff k_0}{ 2\pi } & \frac{1}{k^2}  \bigg [   \frac12 \Gamma^{\sf on}_{\mu \mu}(p,k,-k)   +  \frac{  (2p-k)^2   }{(p-k)^2  + m_P^2 } \, F(k^2)^2
\\[8pt]
& -   \frac{2  k \cdot (2p - k)  }{(p-k)^2 + m_P^2 }  \, \frac{(p-k)^2 - p^2}{k^2} \, F(k^2) \, [1 - F(k^2)]
\\[8pt]
& +  \frac{ k^2 }{(p-k)^2 + m_P^2 }   \frac{[(p-k)^2 - p^2]^2}{k^4} [1 - F(k^2)]^2  
\bigg ]
\,,
\end{split}
\end{align}
now with $p_\mu = (i m_P, \textbf 0)$ and $k_\mu = (k_0, \ksp)$. Here the origin of the terms can be easily read off from the dependence on $F(k^2)$. We stress that this result holds to all orders in $1/L$. See also Ref.~\cite{Lucini:2015hfa} for similar expressions in the context of finite-volume QED with C$^\star$ boundary conditions.

\begin{figure}
    \centering
    \includegraphics[width=0.7\textwidth]{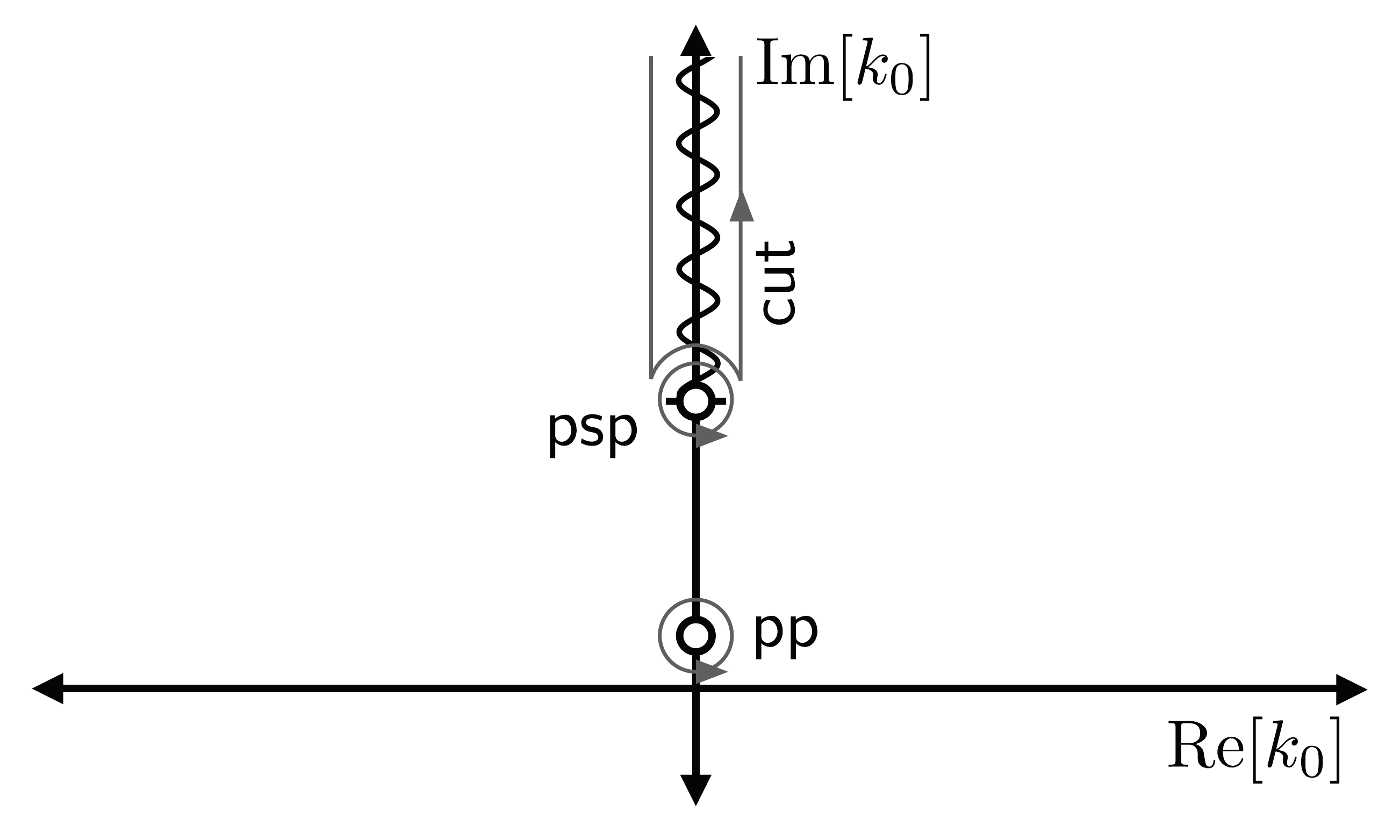}
    \caption{Analytic structure of the integrand defining $\Delta m_{P}^2(L)$. Closing the $k_0$ integral in the upper-half plane leads to three contributions: the photon pole (${\sf pp}$), the pseudo-scalar pole (${\sf psp}$) and the remaining contribution including both the branch-cut and the arc at infinity (${\sf cut}$).}
    \label{fig:contour}
\end{figure}

To complete the derivation one evaluates the $k_0$ integral by closing in the upper half of the complex plane. Doing so leads to three terms, as illustrated in Fig.~\ref{fig:contour}. This first arises from encircling the pole at $k_0 = i \vert \ksp \vert$, call it $\Delta m_{{\sf pp}}^2(L)$, where {\sf pp} stands for photon pole. The second term arises from encircling the pseudo-scalar pole at $(p-k)^2 + m_P^2 = 0$, equivalently at $k_0 =  i m_P + i \omega_P(\ksp)$ where $\omega_P(\ksp) = \sqrt{\ksp^2 + m_P^2}$. 
Refer to this contribution as $\Delta m_{\sf psp}^2(L)$, where ${\sf psp}$ stands for pseudo-scalar pole. The final term then arises from the remaining analytic structure, in the upper half of the complex $k_0$ plane, and is denoted by $\Delta m^2_{\sf cut}(L)$. In short:
\begin{equation}
    \Delta m_P^2(L) =  \Delta m^2_{\sf pp}(L) + \Delta m^2_{\sf psp}(L) + \Delta m^2_{\sf cut}(L) \,.
\end{equation}

Beginning with the photon pole, the integral readily evaluates to
\begin{equation}
\Delta m_{{\sf pp}}^2(L)  =   e^2  \, \Delta_{\ksp}' \,
   \frac{1}{2 \vert \ksp \vert}  \bigg [   4   +  \frac{  4 m_P^2  - 4 m_P \vert \ksp \vert   }{  2 m_P \vert \ksp \vert    }
+ 4 \vert  \ksp \vert  m_PF'(0) + \mathcal O(\ksp^2) \bigg ]
\,.
\end{equation}
 Here the zero momentum derivative of the EM form factor $F'(0)$ appears due to the term proportional to $F(0) - F(k^2)$, ultimately arising from the contribution linear in $G^{\sf on}$. It is through this term that the finite-volume mass shift inherits its first structure-dependent piece, proportional to the squared charge radius $\braket{r_P^2}$ of the meson $P$, via the standard definition
\begin{equation}
    F'(0) = - \frac{\braket{r_P^2}}{6} \,.
\end{equation}
The result simplifies to
\begin{equation} 
\Delta m_{\sf pp}^2(L) = e^2 \bigg [  \frac{  m_P}{4\pi^2 L}  c_{2} +  \frac{1 }{2\pi  L^2} c_{1} -   \frac{m_P \braket{r_P^2}}{3L^3} c_{0} +\bigo\left(\frac{1}{L^4}\right)   \bigg ] \, ,
\end{equation}
 where $c_0=\Delta'_{\nsp}(1)=-1$. Note that the structure-dependent $1/L^3$ term derived here within a relativistic approach is the same as in non-relativistic scalar QED~\cite{Davoudi:2014qua}. It is satisfying to see the the same quantity arises in the general model-independent context of this work, via the constraints of the Ward-Takahashi identity.  As expected, the point-like and universal contributions agree with the previous section. The numerical effect of the structure-dependent term in~\cref{eq:fveffmass} is investigated in~\cref{sec:num}.

Continuing the exercise, one can show that for the $\psp = \textbf 0$ case considered here, the pseudoscalar-pole term contributes beyond the order we keep: $\Delta m^2_{\sf psp}(L) = \bigo ( 1/L^4 )$, 
although it is known to contribute at $O(1/L^3)$ when $P$ has nonzero spatial momentum in the FV frame~\citep{Davoudi:2018qpl}. It therefore remains only to consider the contribution from all additional analytic structure within the integrand. This, in fact, leads to an additional $1/L^3$ term, expressed as an integral of the discontinuity across the branch cut within $T(k^2, k \cdot p)$, as we describe in the next subsection. 

The branch-cut term will prove challenging to predict in practice, meaning that the primarily useful terms in the mass shift have already been identified. We therefore summarize the full result here, before moving to the details of the final contribution
\begin{align}\label{eq:fveffmass}
\Delta m_P^2(L) = e^2m_P^2\left\{\frac{c_{2}}{4\pi^2m_PL} +\frac{c_{1}}{2\pi(m_PL)^2} + \frac{\braket{r_P^2}}{3m_PL^3} + \frac{\mathcal C}{(m_P L)^3} +\bigo\left[\frac{1}{(m_PL)^4}\right]\right\}
\,,
\end{align}
where
\begin{equation}
\label{eq:calCdef}
    \mathcal C = \lim_{L \to \infty} L^3 \, \frac{ m_P}{e^2} \, \Delta m^2_{\sf cut}(L) \,,
\end{equation}
is the $\bigo(1/L^3)$ contribution from the branch cut. This structure-dependent term is only present in the $1/L$ expansion because of the spatial non-locality of the $\qedl$ theory. It is contained in the residual~$\bigo(1/L^3)$ FV effect Eq.~(S35) of Ref.~\citep{Borsanyi:2014jba}, and also described in Eq.~(2.13) of Ref.~\citep{Tantalo:2016vxk}. The focus on the next sub-section is to describe more precisely this contribution from the physical properties of the Compton amplitude.

\subsection{Branch-cut contribution to the finite-size effects}

To describe the contribution to $\Delta m_P^2(L)$ arising from the branch cut, it is most straightforward to revert to an expression similar to Eq.~\eqref{eq:DmPL}, in which the decomposition into vertex functions has not been performed
\begin{equation}\label{eq:DmPLcut}
\Delta m^2_{\sf cut}(L) = - \frac{e^2}{2}   \Delta_{\ksp}'
\int_{\sf cut} \frac{\diff k_0}{ 2\pi }
\frac{T(k^2, k \cdot p) }{k^2} \bigg \vert_{\psp = \boldsymbol 0}
\,,
\end{equation}
where the label ${\sf cut}$ can be understood, for now, as the original $k_0$ integral with the contours around the photon and pseudoscalar poles removed. After $k_0$ integration, this contribution contains only non-negative powers of $\vert \ksp \vert$ and thus the only contribution at $1/L^3$ arises from the subtracted zero-mode. Applying the definition of $\mathcal C$ in Eq.~\eqref{eq:calCdef}, one finds
\begin{equation}
    \mathcal C = \frac{m_{P}}{2} \int_{\sf cut} \frac{\diff k_0}{ 2\pi } \frac{T(k_0^2, i m_{P} k_0) }{k_0^2} \,.
\end{equation}
In Fig.~\ref{fig:contour}, we illustrate the analytic structure of ${T(k_0^2, i m_{P} k_0) }/{k_0^2}$ in the complex plane and highlight the integration contour leading to this contribution.

Physically, the cut corresponds to all multi-hadron states formed when the pseudoscalar at rest collides with an off-shell photon, with zero spatial momentum and energy $E_{\gamma}$ where $k_0 = i E_{\gamma}$. The contribution from a multi-particle state with energy $\sqrt{s}$ starts at $E_{\gamma}  = \sqrt{s} -  m_{P}$ and, since the lowest lying coupled state has energy $\sqrt{s} = m_{P} + 2 m_\pi$, the cut runs from $k_0 = 2 i m_{\pi}$. As indicated in the figure, the pole at $k_0 = 2 i m_{P}$, has already been considered above as $\Delta m_{{\sf psp}}^{2}(L)$ and is therefore not included here.

We now prove that $\mathcal C \geq 0$. This is significant as it implies that the radius term in Eq.~\eqref{eq:fveffmass} contributes with the same sign as the branch-cut term and thus that (a) they cannot cancel and (b) subtracting the former will reduce the volume effects. First substitute the definition of the forward Compton amplitude, $T(k_0^2, i m_{P} k_0)$, to write
\begin{equation}
    \mathcal C = \frac{m_P}{2} \int_{\sf cut} \frac{\diff k_0}{2 \pi} \frac{1}{k_0^2}    \int \diff x_0 \, e^{-ik_0 x_0} \, \bra{P,\mathbf{0}} T\left\{ \tilde J_{\mu}(x_0, \mathbf{0}) J_{\mu}(0) \right\} \ket{P,\mathbf{0}}
\,,
\end{equation}
where $\tilde J_\mu(x_0,\mathbf 0) = \int \diff^3 \mathbf x \, J_\mu(x)$. 

Separating the two time orderings and inserting a complete set of states between the currents, one finds
\begin{multline}
\label{eq:Cintform}
    \mathcal C = \frac{m_P}{2} \int_{\sf cut} \frac{\diff k_0}{2 \pi} \frac{1}{k_0^2}  
    \int_{2 m_\pi + m_P}^\infty \diff \omega \, \frac{ \rho(\omega)}{2 \omega}
  \bigg [  \int_0^\infty   \diff x_0    \, e^{-ik_0 x_0  - M_\alpha x_0 + m_P x_0}   \\ +   \int_{- \infty}^0   \diff x_0    \, e^{-ik_0 x_0  + M_\alpha x_0 - m_P x_0}  \bigg ] 
\,,
\end{multline}
where we have introduced
\begin{align}
    \rho(\omega) & = \sum_\mu \int \diff \alpha \,   \delta(\omega - M_\alpha)  \bra{P,\mathbf{0}}    J_{\mu}(0) \ket{ \alpha,  \mathbf 0 } \bra { \alpha, \mathbf 0 } J_{\mu}(0)  \ket{P,\mathbf{0}} \,, \\
    & =  \int \diff \alpha \,   \delta(\omega - M_\alpha) \, \Big \{ \big \vert \bra { \alpha, \mathbf 0 } J_{0}(0)  \ket{P,\mathbf{0}} \big \vert^2 - \sum_k \big \vert \bra { \alpha, \mathbf 0 } J_{k}(0)  \ket{P,\mathbf{0}} \big \vert^2 \Big \} \,.
\end{align}
Here the integral over $\alpha$ runs over all internal degrees of freedom.\footnote{This can be made explicit as follows (though these details are not required for the derivation):
\begin{equation}
    \int d \alpha = \sum_{i} \frac{1}{S_i}  \int \! \frac{\diff^3 \ksp_1^i}{(2 \pi)^3 2 \omega_{i1}(\ksp_{1}^i)} \cdots \frac{\diff^3 \ksp_{N_i}^i}{(2 \pi)^3 2 \omega_{iN_{i}}(\ksp_{N_i}^i)}  (2 \pi)^3 \delta^3(\ksp_1^i + \cdots + \ksp_{N_i}^i)\,,
\end{equation}
where the sum over $i$ runs over all multi-particle channels with the relevant quantum numbers. Here $S_i$ is the channel's symmetry factor, $N_i$ is the number of particles and $\omega_{in}(\ksp) = \sqrt{m_{in}^2 + \ksp^2}$ is the relativistic energy for the $n$th particle in channel $i$.}
In the first line we have used the momentum projection on the current to project to zero momentum on the inserted states and have introduced $M_\alpha$ as the the center-of-mass energy of the state $\alpha$. In the second line we use that $J_0(0) = J_0(0)^\dagger$ and $J_k(0) = - J_k(0)^\dagger$. Next note that $J_0(0)$ is proportional to the charge operator when sandwiched between zero-momentum states and, since we only require $\rho(\omega)$ for $\omega > m_P + 2 m_\pi$, this leads to $\bra { \alpha, \mathbf 0 } J_{0}(0)  \ket{P,\mathbf{0}} \propto \langle \alpha, \mathbf 0  \vert P,\mathbf{0}\rangle  = 0$ and thus
\begin{equation}
 \rho(\omega)  =  - \sum_k\int \diff \alpha \,   \delta(\omega - M_\alpha) \,    \big \vert \bra { \alpha, \mathbf 0 } J_{k}(0)  \ket{P,\mathbf{0}} \big \vert^2   \,.
 \end{equation}
 The key point is that $\rho(\omega)$ is non-positive.
Finally, evaluating the $x_0$ and $k_0$ integrals in Eq.~\eqref{eq:Cintform}, we reach
\begin{align}
    \mathcal C 
      & =  -  \frac{m_P}{2}
   \int_{2 m_\pi + m_P}^\infty \diff \omega  \,   \frac{1}{  ( \omega  - m_P)^2}   \, \frac{ \rho(\omega)}{2 \omega} \geq 0 \,.
\end{align}
This concludes the demonstration that $\mathcal C \geq 0$.

Giving detailed predictions of this term is challenging considering it depends on all possible hadronic scales coupling to $P$ via scattering with a virtual photon. This fact in principle appears as a limitation of $\qedl$ for quantitative predictions compared to local approaches, since $\bigo(1/L^3)$ FVEs will in general be very challenging to predict because of the systematic presence of such non-local effects. However it is not clear if this is an issue in practice. Indeed, current lattice simulations are generally performed with $m_PL\gtrsim 4$. On the one hand for small volumes in that range $\mathcal{O}[1/(m_PL)^3]$ corrections may well be of a comparable size to to unknown exponentially suppressed finite-size effects, and on the other hand for larger volumes $\mathcal{O}[1/(m_PL)^3]$ corrections are expected to be at the percent level where higher-order QED contributions becomes relevant. Finally, this term can also be determined by directly fitting lattice data across several volumes. 

This completes our discussion of the finite-volume mass shift in $\qedl$ and we turn now to the main focus of this work, the $1/L$ expansion of the finite-volume matrix elements defining the leptonic decay rate.

  \section{Pseudoscalar mesons leptonic decay rate}
  \label{sec:kl2}
In this section we compute the EM finite-volume effects on radiative corrections to meson leptonic decay rates. These decays are of the form $P^{-}\rightarrow
\ell ^{-}\bar{\nu} _{\ell}$ (as well as the conjugated decay) for a given pseudoscalar meson~$P$, lepton $\ell$ and corresponding neutrino $\nu_{\ell}$. Theoretical knowledge of these amplitudes allows to extract CKM matrix elements by comparing to the experimentally measured decay rates. In the isospin symmetric limit, the lepton-neutrino pair contribution factorises and leptonic decay rates can be simply expressed in terms of the meson decay constant $f_P$. In the case of light mesons, decay constants are now predicted from lattice QCD to sub-percent accuracy~\cite{Aoki:2019cca}, and the inclusion of isospin breaking effects is necessary. Once electromagnetic interactions are present, the lepton can interact with the meson and the factorisation of the amplitude is not possible anymore. A method to overcome this issue was developed and successfully applied in a lattice calculation in Refs.~\cite{Carrasco:2015xwa,Lubicz:2016xro,DiCarlo:2019thl,Desiderio:2020oej,Frezzotti:2020bfa}.

Beyond precision considerations, EM finite-size corrections on radiative corrections to leptonic decays are particularly important as the volume acts as an IR regulator for the virtual amplitude. In this section we focus on predicting higher-order IR finite and structure-dependent contributions which will allow one to reduce the systematic uncertainty associated with finite-volume effects for a given set of numerical data.

We restrict attention to the case that $P^{-}$ has zero spatial momentum in the finite-volume frame and denote by $\ml$ and $\pspl$ the mass and the momentum of the lepton $\ell^-$, respectively. With $p$ and $p _\ell$ the 4-momenta of $P^-$ and $\ell ^{-}$, respectively, the neutrino has momentum $p_{\nu _{\ell}} = p-p_{\ell}$. We also define the lepton energy $\oml=\sqrt{m_{\ell}^2+\pspl^2}$, the lepton velocity $\vell=\pspl/\oml$, and the ratio $\rl=\ml/m_{P}$.
Using momentum and energy conservation, one then obtains the useful kinematical relations
\begin{align}
  |\pspl|&=\frac{m_{P}}{2}(1-\rl^2)\,,\\
  \oml&=\frac{m_{P}}{2}(1+\rl^2)\,,\\
  |\vell|&=\frac{1-\rl^2}{1+\rl^2}\,.\label{eq:vlrl}
\end{align}

\subsection{General strategy}
The purely virtual $\mathcal{O}(\alpha )$-corrected leptonic decay rate $\Gamma _{0}  = \Gamma \left( P^{-}\rightarrow \ell ^{-}\nu _{\ell}\right) $ is IR-divergent. However, in a standard fashion these divergences can be cancelled by studying instead the inclusive decay rate 
\begin{align}\label{eq:inclusiverate}
\Gamma \left(  P^{-}\rightarrow \ell ^{-}\nu _{\ell}[\gamma ]\right) =  \Gamma _{0}+\Gamma _{1}(\Delta E_{\gamma})\, ,
\end{align}
where $\Delta E_{\gamma}$ is an upper limit on the photon energy in the real radiative decay rate $\Gamma _{1}(\Delta E_{\gamma}) = \Gamma \left(  P^{-}\rightarrow \ell ^{-}\nu _{\ell}\gamma \right) $. The subscripts here refer to the number of photons in the final state and the quantity in~\cref{eq:inclusiverate} is IR-finite. 

In a finite volume, both terms in~\cref{eq:inclusiverate} acquire dependence on $L$. A strategy to calculate these EM-corrected quantities on the lattice was first laid out in Ref.~\cite{Carrasco:2015xwa}, which eventually lead to the calculation of $\Gamma _{0} (L)$ in Ref.~\cite{DiCarlo:2019thl} and $\Gamma _{1}(L, \Delta E_{\gamma})$ in Ref.~\cite{Desiderio:2020oej}. As the cancellation of IR divergences has to occur numerically, it was realized in Ref.~\cite{Carrasco:2015xwa} that one may add and subtract the universal FV decay rate $\Gamma _{0}^{\mathrm{uni}} (L) $, which can be calculated in perturbation theory in the point-like approximation and has the same IR-divergences as $\Gamma _{0} (L)$ and $\Gamma _{1}(L, \Delta E_{\gamma})$. As a consequence, one can split the right-hand side of~\cref{eq:inclusiverate} into
\begin{align}
& 
\Gamma _{0}+\Gamma _{1}(\Delta E_{\gamma})   =  \lim _{L\to \infty} [ \Gamma _{0} (L) - \Gamma _{0}^{\mathrm{uni}} (L) ] +\lim _{L\to \infty} [\Gamma _{0}^{\mathrm{uni}} (L) +\Gamma _{1}(L, \Delta E_{\gamma}) ]\label{eq:gammadec}
 \, ,
\end{align}
where now each of the two bracketed terms is separately IR-finite. At $\bigo(e^2)$ in QED, the photon only appears in the real radiative decay as an external state and therefore $\Gamma _{1}(\Delta E_{\gamma})$ is purely a QCD matrix element without photon loops. It is simpler to just choose a different IR-regulator in the second term of~\cref{eq:gammadec}. Using a photon mass $\lambda$ the equation takes the form
\begin{align}\label{eq:rm123strategy}
& 
\Gamma _{0}+\Gamma _{1}(\Delta E_{\gamma})   =  \lim _{L\to \infty} [ \Gamma _{0} (L) - \Gamma _{0}^{\mathrm{uni}} (L) ] +\lim _{\lambda \to 0} [\Gamma _{0}^{\mathrm{uni}} (\lambda ) +\Gamma _{1}(\lambda , \Delta E_{\gamma}) ]
 \, .
\end{align}
Note that all the volume-dependence now sits in the first term in brackets on the right-hand side, which is the one of interest to us.

As was shown in Ref.~\cite{Lubicz:2016xro}, the universal decay rate $\Gamma _{0}^{\mathrm{uni}} (L)$ only includes FV corrections up to $\bigo(1/L)$. The difference $\Gamma _{0} (L) - \Gamma _{0}^{\mathrm{uni}} (L)$ scales then as $1/L^2$, but at this level the point-like approximation is no longer valid and the structure of the decaying meson starts playing a role. Our goal is to extend the formalism from the previous section to systematically compute the finite-size scaling in $\Gamma _{0} (L) - \Gamma _{0}^{\mathrm{uni}} (L)$ order by order in $1/L$, including structure-dependent corrections. We therefore generalize~\cref{eq:rm123strategy} by subtracting $\Gamma _{0}^{(n)} (L)$ defined through
\begin{align}\label{eq:decratefve}
\Gamma _{0}^{(n)} (L) = \Gamma _{0}^{\mathrm{uni}} (L) + \sum _{j=2}^{n}\Delta \Gamma _{0}^{(j)}(L) \, .
\end{align}
Here $\Delta \Gamma _{0}^{(j)}(L)$ are the effects of order $1/L^j$ with $j\geq 2$. As the latter extra terms vanish in the infinite-volume limit,~\cref{eq:rm123strategy} can be rewritten as
\begin{align}
	\label{eq:nhtapstrategy}
& 
\Gamma _{0}+\Gamma _{1}(\Delta E_{\gamma})   =  \lim _{L\to \infty} [ \Gamma _{0} (L) - \Gamma _{0}^{(n)} (L) ] +\lim _{\lambda \to 0} [\Gamma _{0}^{\mathrm{uni}} (\lambda ) +\Gamma _{1}(\lambda , \Delta E_{\gamma}) ]
 \, ,
\end{align}
now with residual higher-order FVEs starting from
\begin{align}
\Gamma _{0} (L) - \Gamma _{0}^{(n)} (L)  \sim \mathcal{O}\left( \frac{1}{L^{n+1}} \right) \, . 
\end{align}
By next writing the tree-level decay rate as
\begin{align}
\label{eq:Gamma0tree}
\Gamma_{0}^{\mathrm{tree}} = \frac{G_{F}^2}{8\pi}\left| V_{ij}\right| ^2 f_{P}^{2} \, m_{P}m_{\ell}^2(1-r_\ell^2) ^2 \, ,
\end{align}
where $ V_{ij}$ is the CKM matrix element relevant for $P^{-}$, $m_{P}$ and $m_\ell$ the physical masses of meson and lepton, respectively, and $f_{P}$ the QCD decay constant, we may write $\Gamma _{0}^{(n)} (L) $ as 
\begin{align}\label{eq:y2def}
	\Gamma _{0}^{(n)}(L)  = \Gamma _{0}^{\textrm{tree}} \left[ 1+2 \frac{\alpha }{4\pi } \, Y^{(n)}(L) \right]+\mathcal{O} \left( \frac{1}{L^{n+1}} \right)
	\, .
\end{align}
The above equation, together with~\cref{eq:decratefve}, defines $Y^{(n)}(L)$. In the following our aim is to derive $Y^{(2)}(L)$ in~\cref{eq:y2def}, but our method in principle allows to determine $Y^{(n)}(L)$ to an arbitrarily high order in $1/L$. An important check will be to reproduce the point-like results from Refs.~\cite{Lubicz:2016xro,Tantalo:2016vxk}. Note that in \cref{eq:Gamma0tree} we choose $f_P$ to be the decay constant of the meson in QCD, \ie including the $SU(2)$-breaking corrections and assuming a suitable separation scheme has been chosen to separate such effects from the electromagnetic corrections. In principle one could similarly choose $\Gamma_0^\mathrm{tree}$ to be defined in terms of the decay constant $f_P^\mathrm{(0)}$ computed in the isospin-symmetric theory, however this does not have any impact on the final result for the FV effects when working at first order in the isospin-breaking corrections.

The inclusion of QED corrections at $\bigo(\alpha)$ also generates new UV divergences. These are removed in the infinite volume by using the $W$-regularization scheme to define the Fermi constant $G_F$~\cite{Sirlin:1980nh}. In this paper we are interested in computing FV corrections to the decay rate that, as explained in Sec.~\ref{sec:sum}, appear from the $1/L$ expansion of sum-integral differences which are UV finite. The $W$-regularization of the IV integrals will not be discussed here and we refer to Refs.~\cite{Carrasco:2015xwa,Lubicz:2016xro} for further details.

In the following we assume that the lepton mass $m_\ell$ has been renormalized perturbatively following a usual on-shell scheme introducing an appropriate counterterm in the infinite-volume QCD+QED Lagrangian. Moreover, since the lepton self-energy contribution can be factorized and treated analytically in infinite-volume we will not consider it in the FV calculation. We will assume in the rest of the section that the Euclidean lepton propagator is given
\begin{equation}
	S_{\ell}(p_{\ell}) =\frac{m_{\ell} - i \slashed{p}_{\ell }}{p_{\ell}^2+m_{\ell}^{2}} \,.
\end{equation}

\subsection{Electromagnetic corrections to the decay width}
\label{sec:emdecay}
In this section we perform the analytic calculation of the FVEs on leptonic decay rates. We start by defining the kernels of interest, and then proceed to the separation into irreducible vertices and discuss their structure, both point-like and structure-dependent. Finally, we present and discuss the result for $Y^{(2)}(L)$.
\subsubsection{Formal description of the leptonic decay amplitude}
\label{sec:FVformaldescr}
As in the previous discussion of pseudoscalar mesons EM self-energy, we will use Euclidean space-time. For the specific process of leptonic decays, Euclidean amplitudes can be trivially continued to the Minkowski ones though a multiplicative factor of $i$.
In the following we study the Euclidean correlation function
\begin{align}
\label{eq:CWfull}
C_{W}^{rs}(p,p_{\ell}) = \int\diff^{4}z\, e^{ip z}\,\bra{\ell^-,\psp_\ell,r;\nu_{\ell},\psp_{\nu _{\ell}},s}\mathrm{T}[\mathcal{O}_{W}(0)\phi^{\dagger}(z)]\ket{0}\, ,
\end{align}
where $r$ and $s$ are the polarizations of the lepton and neutrino, respectively, and $\mathcal{O}_{W}(0)$ is the four-fermion operator entering the effective weak Hamiltonian density responsible for the decay  $P\rightarrow \ell\bar{\nu}$, i.e.~
\begin{align}\label{eq:weakH}
\mathcal{H}_{W} = 
\frac{G_{F}}{\sqrt{2}}
V_{ij} \ \mathcal{O}_W=  \frac{G_{F}}{\sqrt{2}}
V_{ij}  \ [ \bar{q}_{1}\gamma _{\rho} (1-\gamma _{5}) q_{2}][ \bar{\ell}\gamma _{\rho} (1-\gamma _{5})\nu _{\ell}] \, . 
\end{align}
The correlation function $C_{W}^{rs}(p,p_{\ell})$ is understood to be in the full QCD+QED theory. 
The polarized matrix element $\mathcal{M}^{rs}$ of the $P^-\to\ell^-\bar{\nu}_\ell$ decay is then given by the reduction formula
\begin{equation}\label{eq:cwred}
	\mathcal{M}^{rs}=\lim_{p^2\to -m_P^2}Z_P^{-1}D(p)^{-1} \, C_W^{rs}(p,p_\ell)\,,
\end{equation}
with $Z_P$ and $D(p)$ defined in \cref{eq:C2fulloverlap,eq:DeltaFull}, respectively, and assuming the external lepton and neutrino propagators to be already amputated. 
The matrix element $\mathcal{M}^{rs}$ can be written in term of external state spinors $\bar{u}^r_\ell=\bar{u}^r(p_\ell)$ and $v^s_\nu=v^s(p_{\nu_\ell})$ as follows
\begin{equation}
	\mathcal{M}^{rs}=\bar{u}^r_\ell \,\tilde{\mathcal{M}}\,v^s_\nu\,,
\end{equation}
where now $\tilde{\mathcal{M}}$ is a $4\times 4$ spin matrix. 
The full correlator $C_W^{rs}(p,p_\ell)$ in~\cref{eq:cwred} can then be expressed in the diagrammatic language defined in~\cref{sec:se} as
\begin{equation}\label{eq:cwfull}
	C_W^{rs}(p,p_\ell)=\raisebox{-3.7ex}{\includegraphics{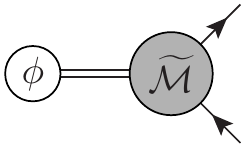}}\,.
\end{equation}
Up to order $\bigo(e^4)$ corrections, it can be expanded as
\begin{equation}\label{eq:cwexp}
	C_W^{rs}(p,p_\ell)=C_{W,0}^{rs}(p,p_\ell)+  C_{W,1}^{rs}(p,p_\ell)=\raisebox{-3.7ex}{\includegraphics{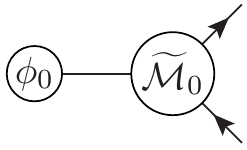}}+\raisebox{-3.7ex}{\includegraphics{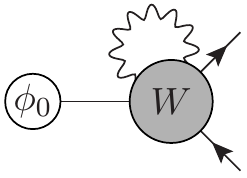}}\,,
\end{equation}
where the correlation functions $C_{W,0}^{rs}(p,p_\ell)$ and $C_{W,1}^{rs}(p,p_\ell)$ can be obtained from the one in \cref{eq:cwfull} as
\begin{equation}
	C_{W,0}^{rs}(p,p_\ell) = \left. C_W^{rs}(p,p_\ell) \right|_{e=0}~,
\end{equation}
and 
\begin{equation}
	C_{W,1}^{rs}(p,p_\ell) = \frac{e^2}{2}\left.  \frac{\partial^2}{\partial e^2}  C_W^{rs}(p,p_\ell) \right|_{e=0}\,.
\end{equation}
The QCD correlation function $C_{W,0}^{rs}(p,p_\ell)$ takes the form
\begin{equation}
	C_{W,0}^{rs}(p,p_\ell) = \raisebox{-3.7ex}{\includegraphics{axo_treecw.pdf}} =  C_{W,0}^\rho(p) \,  \mathcal{L}^{rs}_\rho(p,p_\ell)
	\, ,
\end{equation}
where 
\begin{equation}
	\label{eq:Lrho}
	\mathcal{L}_\rho^{rs}(p,p_\ell) = \bar{u}^r(p_\ell) \, \gamma_\rho (1-\gamma_5)\, v^s(p_{\nu_\ell}) \, , 
\end{equation}
and $C_{W,0}^\rho(p)$ corresponds to the correlation function
\begin{equation}
	C_W^\rho(p) = \int\diff^{4}z\, e^{ip z}\,\bra{0}\mathrm{T}[J_{W}^\rho(0)\phi^{\dagger}(z)]\ket{0}
	\, ,
\end{equation}
evaluated at $e=0$, i.e. $C_{W,0}^\rho(p)=\left[C_W^\rho(p)\right]_{e=0}$. Here $J_W^\rho = \bar{q}_1 \, \gamma^\rho(1-\gamma^5)\, q_2$ is the $V-A$ quark current entering the weak Hamiltonian of~\cref{eq:weakH}.
The QCD+QED correlator $C_W^\rho(p)$ has the following spectral decomposition in the vicinity of $p^2 = -m_{P}^2$,
\begin{equation}\label{eq:Crho}
	C_W^\rho(p) = Z_P \, D(p)\, W^\rho(p)\, ,
\end{equation}
and through Lorentz covariance we can define the weak vertex as 
\begin{equation}\label{eq:Wrho}
	W^\rho(p)=-p^\rho F_W(p^2)\,,
\end{equation}
where the generic off-shell function $F_W(p^2)$ is such that in QCD one gets $F_W(-m_{P,0}^2)=f_P$. 

The reducible $\bigo(e^2)$ kernel $W$ in \cref{eq:cwexp} can be decomposed as follows
\begin{align}
	\raisebox{-3.7ex}{\includegraphics{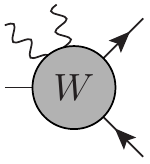}}&=\raisebox{-3.7ex}{\includegraphics{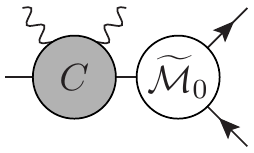}}+
	\raisebox{-3.7ex}{\includegraphics{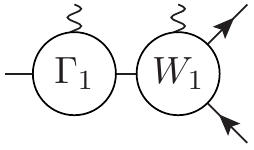}}+\raisebox{-3.7ex}{\includegraphics{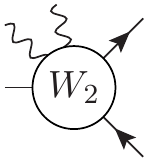}}+
	\raisebox{-3.7ex}{\includegraphics{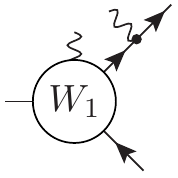}}+\raisebox{-3.7ex}{\includegraphics{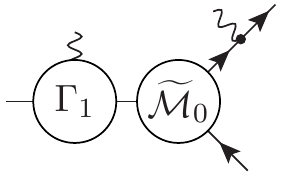}}\notag
	\\
	&\qquad+\raisebox{-3.7ex}{\includegraphics{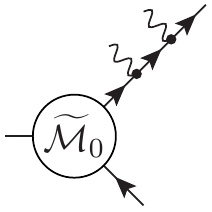}} + \mathrm{crossings}\,,
	\label{eq:m1exp}
\end{align}
where the Compton kernel $C$ and electromagnetic kernel $\Gamma_1$ have been introduced in~\cref{sec:se}, while $W_1$ and $W_2$ are two new weak irreducible kernels for $P^-\to\ell^-\bar{\nu}_\ell\gamma^*$ and $P^-\to\ell^-\bar{\nu}_\ell\gamma^*\gamma^*$ decays, respectively. 
Finally, once the photon lines are contracted in~\cref{eq:m1exp}, the last diagram in the expansion of $W$ becomes the self-energy of the charged lepton that, as discussed above, we do not consider in this calculation. 
The multiplicity from photon crossings of each diagram in \cref{eq:m1exp} is understood. From the decomposition of the kernel $W$ in \cref{eq:m1exp} it is clear that the correlation function $C_{W,1}^{rs}$  takes the form
\begin{align}
\label{eq:CW1}
    C_{W,1}^{rs}(p,p_\ell) = C_{W,P}^{rs}(p,p_\ell) + C_{W,\ell}^{rs}(p,p_\ell)\,,
\end{align}
where $C_{W,P}^{rs}(p,p_\ell)$ includes contributions where a photon is emitted and reabsorbed by the meson $P$, while $C_{W,\ell}^{rs}(p,p_\ell)$ denotes the correlation function where a photon is exchanged between the meson and the lepton. We will refer to these as factorisable and non-factorisable contributions, respectively. The correlation functions defined in \cref{eq:CW1} can be represented diagramatically as
\begin{eqnarray}
	{C}_{W,P}^{rs}(p,p_\ell) = & \ \raisebox{-3.7ex}{\includegraphics{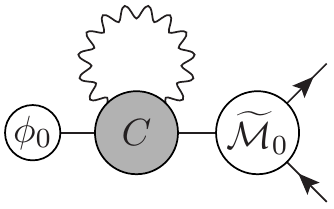}} \  + & \ \raisebox{-3.7ex}{\includegraphics{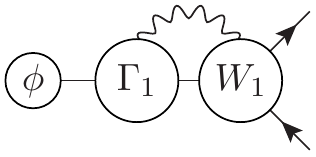}} \ \ + \ \ \raisebox{-3.7ex}{\includegraphics{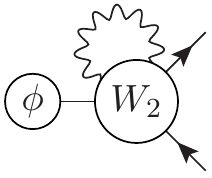}}\ \, ,\\
	C_{W,\ell}^{rs}(p,p_\ell) = & \ \raisebox{-3.7ex}{\includegraphics{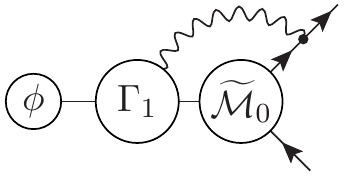}} + & \ \raisebox{-3.7ex}{\includegraphics{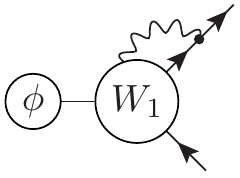}}\,.
\end{eqnarray}
Let us now turn to the $\bigo(e^2)$ contributions to the reduction formula~\cref{eq:cwred} and define
\begin{equation}
	\mathcal{M}^{rs} = \mathcal{M}_0^{rs} +  \mathcal{M}_1^{rs} + \bigo(e^4)~.
\end{equation}
We consider first the contribution to $\mathcal{M}_1^{rs}$ coming from the meson self-energy, namely $ \mathcal{M}^{rs}_{\mathrm{self}}$. This is given by picking from the correction $ C_{W,P}^{rs}(p,p_\ell)$ only the contribution $ C_{W,\mathrm{self}}^{rs}(p,p_\ell)$ given by the Compton kernel $C$.
Adding $ C_{W,\mathrm{self}}^{rs}(p,p_\ell)$ to the tree-level correlation function we get
\begin{equation}
	\raisebox{-3.7ex}{\includegraphics{axo_treecw.pdf}}+\raisebox{-3.7ex}{\includegraphics{axo_wcchain1.pdf}}
	= Z_{P,0} \,D_0(p) \, \big[1+ \Sigma(p^2)D_0(p)\big] \, W^\rho(p) \mathcal{L}^{rs}_\rho(p,p_\ell)\,,
	\label{eq:cw0cwself}
\end{equation} 
where $\Sigma(p^2)$ is defined in terms of the Compton amplitude $C_{\mu\mu}(p,k,-k)$ in \cref{eq:comptonkernelcorrfcn} as
\begin{equation}
    \Sigma(p^2) = \frac{1}{2} \int \frac{\diff^4 k}{(2\pi)^4} \, \frac{C_{\mu\mu}(p,k,-k)}{k^2}\,.
\end{equation}
Inserting \cref{eq:cw0cwself} into the reduction formula of \cref{eq:cwred} requires the evaluation about the on-shell point $p^2=-m_P^2$ of the following quantity
\begin{equation}
    Z_P^{-1}D(p)^{-1}\cdot Z_{P,0} \,D_0(p) \, \big[1+ \Sigma(p^2) D_0(p)\big]F_W(p^2)\,.
    \label{eq:red_c0cself}
\end{equation}
The leptonic tensor $-p^\rho \mathcal{L}^{rs}_\rho(p,p_\ell)$ in \cref{eq:cw0cwself} is factorized here to simplify the discussion.
By rewriting $Z_{P,0}=Z_P(1-\delta_{Z_P})$ and $m_{P,0}^2 = m_P^2-\Delta m^2$, the evaluation of \cref{eq:red_c0cself} at $\bigo(e^2)$ and at the on-shell point $p^2=-m_P^2$ gives
\begin{equation}
\label{eq:red_c0cself_expanded}
    f_P\left[ 1 +(2\,z_1-f_1)\Delta m_P^2 -\delta_{Z_P}+ \Sigma'(-m_P^2) \right]\,,
\end{equation}
where the quantities $z_1$ and $f_1$ are unphysical off-shell contributions related to the meson propagator
and to the weak vertex function, respectively, as
\begin{equation}
	\label{eq:znfn}
	z_n = \left.\frac{\partial^n Z_0(p^2)^{-1}}{\partial (p^2)^n}\right|_{p^2=-m_{P,0}^2}\,, \quad
	f_n = \frac{1}{f_P} \left.\frac{\partial^n F_W(p^2)}{\partial (p^2)^n}\right|_{p^2=-m_{P,0}^2}\,,
\end{equation}
matching the notation of Ref.~\cite{Lubicz:2016xro}.
The overlap shift $\delta_{Z_P}$ entering \cref{eq:red_c0cself_expanded} was obtained in \cref{eq:dZPdef1} and depends on $\Sigma_0'(-m_P^2)$. This quantity can be rewritten in terms of $z_{n}$ and the mass shift $\Delta m_P^2$ by using the relation
\begin{equation}
    \Sigma_0'(-m_P^2) = -\Delta m_P^2 \, \Sigma_0''(-m_{P,0}^2) + \dots \, ,
\end{equation}
and by solving
\begin{equation}
    Z_0(p^2)^{-1} = 1 - \frac{\Sigma_0(p^2)}{p^2 + m_{P,0}^2} \, 
\end{equation}
for $\Sigma_0(p^2)$. Together with the definition of the $z_n$ in~\cref{eq:znfn} one then finds the overlap shift
\begin{eqnarray}
    \delta_{Z_P}  = z_1 \, \Delta m_P^2  + \frac12 \left. \frac{\partial \Sigma(p^2)}{\partial p^2}\right|_{p^2=-m_P^2} \,.
\end{eqnarray}
Combining all previous equations we get the following correction to the matrix element
\begin{equation}
	\mathcal{M}_0^{rs}+\mathcal{M}_\mathrm{self}^{rs} =
	\left[ 1+ \overline{\delta}_{Z_P}\right] \times \raisebox{-3.7ex}{\includegraphics{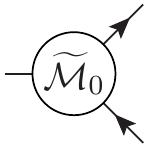}}
\end{equation}
where $\overline{\delta}_{Z_P} = \delta_{Z_P} - f_1 \, \Delta m_P^2$ and the (on-shell) tree-level matrix element is given by
\begin{equation}
	 \raisebox{-3.7ex}{\includegraphics{axo_m0ker.pdf}} = \bra{0} J_W^\rho(0) \ket{P^-,\mathbf{p}} \, \mathcal{L}_\rho^{rs}(p,p_\ell)= - p^\rho f_P \,\mathcal{L}_\rho^{rs}(p,p_\ell)\,.
\end{equation}
Note that in Ref.~\cite{Lubicz:2016xro} the proof of the universality of FVEs up to $\bigo(1/L)$ relies on the cancellation of the unphysical terms $f_1$ and $z_1$. However, similarly to what was discussed in~\cref{sec:offscanc}, the final result for physical observables cannot depend on such terms since they are related to the meson interpolating operator $\phi(x)$. Therefore, $z_n$ and $f_n$ must cancel at all orders in $1/L$. One could in principle perform the whole calculation that follows with the simplification $z_n=f_n=0$ without loss of generality. However we found that keeping those terms and expecting their cancellation is a useful way of controlling the correctness of the final result.

\begin{figure}[t]\centering
	\includegraphics{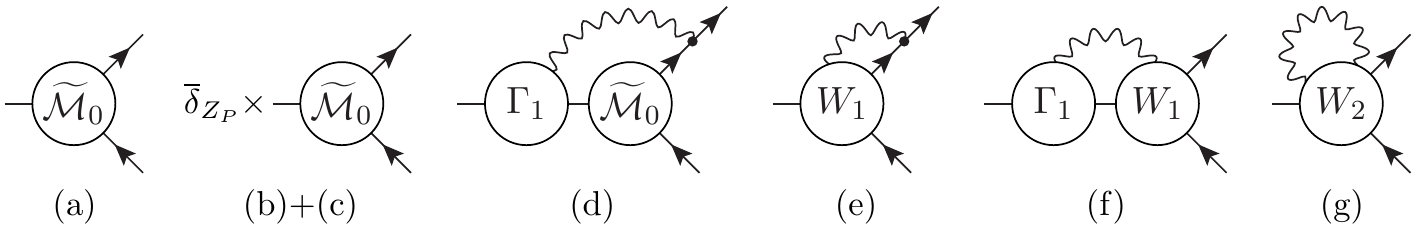}
	\caption{The various diagrams contributing to the leptonic decay width at order $\bigo(e^2)$. The labelling of the diagrams has been chosen to match the one used in Ref.~\citep{Lubicz:2016xro}.}\label{fig:kl2diagrams}
\end{figure}

All the $\bigo(e^2)$  corrections other than the self-energy
are simply obtained by amputating the $P^-$ propagator and wave function in~\cref{eq:cwexp} from the correlation function ${C_{W,1}^{rs}(p,p_\ell)-C_{W,\mathrm{self}}^{rs}(p,p_\ell)}$.
In summary, all the amplitudes to consider are listed in~\cref{fig:kl2diagrams}, using a notation matching Ref.~\citep{Lubicz:2016xro}.

 Let us conclude this part by relating all the diagrams to the FV decay width $\Gamma _{0}^{(n)}(L)$ in~\cref{eq:decratefve}. The decay rate is related to the squared matrix element
\begin{align} \label{eq:M2def}
	\left| \mathcal{M}\right| ^{2} = \sum_{r,s}\left| \mathcal{M}^{rs}\right| ^{2}&=
	\sum_{r,s}|\mathcal{M}_{0}^{rs}|^2+\sum_{r,s}\left[\mathcal{M}_{1}^{rs}(\mathcal{M}_{0}^{rs})^{\dagger}+\textrm{h.c.}\right]+\bigo(e^4)
\nonumber \\
&=|\text{(a)}|^2+
 2 \, \left[ 	 \text{(b)}+\text{(c)}
	 +\text{(d)}+\text{(e)}+\text{(f)}+\text{(g)}\right] \times\text{(a)}^{\dagger} +\bigo(e^4)
	\,,
\end{align}
and therefore the electromagnetic finite-size effects $\Delta|\mathcal{M}|^{2}$ are given by the following sum-integral differences
\begin{align}\label{eq:M2fve}
	\Delta \left| \mathcal{M}\right|^{2} =  2 \ \Delta
	\left[ 	 \text{(b)}+\text{(c)}
	+\text{(d)}+\text{(e)}+\text{(f)}+\text{(g)}\right] \times\text{(a)}^{\dagger} \,.
\end{align}
Finally, the quantity $Y^{(n)}(L)$ defined in \cref{eq:y2def} can be obtained by adding the universal IV contribution evaluated in the point-like theory to the FV corrections computed up to terms of $\bigo(1/L^n)$, namely
\begin{equation}
	Y^{(n)}(L) = \Delta Y^{(n)}(L,\lambda) + Y_\mathrm{IV}^\mathrm{uni}(\lambda)\,.
\end{equation}
The infinite volume contribution $Y_\mathrm{IV}^\mathrm{uni}(\lambda)$ computed in the $W$-regularization scheme can be found in Ref.~\cite{Lubicz:2016xro} and is reported in \cref{eq:ivyl}  below. 
Here $\lambda$ plays the role of a photon mass to regulate in the IR the IV integrals. The quantity $Y_\mathrm{IV}^\mathrm{uni}(\lambda)$ cancels the dependence on $\lambda$ in $\Delta Y^{(n)}(L,\lambda)$, thus leaving the size $L$ as the IR regulator of the FV quantity $Y^{(n)}(L)$. The FV correction $\Delta Y^{(n)}(L,\lambda)$ can then be expressed in terms of $\Delta | \mathcal{M}|^{2}$ as
\begin{equation}
	\label{eq:YnFVE}
	\Delta Y^{(n)}(L,\lambda) = \left( 2 \frac{\alpha}{4\pi} \right)^{-1} \, 
	\frac{\Delta\left| \mathcal{M}\right|^{2}}{\left|\mathcal{M}_0\right|^2}\,,
\end{equation}
with $|\mathcal{M}_0|^2 = \sum_{r,s}|\mathcal{M}_0^{rs}|^2 = 4m_{\ell}^2 m_{P}^2(1-r_{\ell}^2)f_{P}^2$\,.

\subsubsection{The irreducible weak vertex functions}
We must now discuss the various irreducible vertex functions entering into the calculation extending what was done in~\cref{sec:irredv}, which follows a procedure similar to the one outlined in the Appendix of Ref.~\cite{Lubicz:2016xro}. Here we extend the calculation by including higher order terms in the photon momentum $k$, that are relevant for the $1/L^2$ FV corrections. 

\paragraph{Electromagnetic vertices:}
Here we use the general off-shell definition for the electromagnetic vertex $\Gamma_\mu(p,k)$ introduced above in~\cref{eq:Gammamu}.
Applying simple power-counting arguments to the diagram (b)+(c), where the vertex $\Gamma^{\mu\nu}(p,k,-k)$ appears, we deduce that only terms of $\bigo(1)$ in the photon momentum contribute to the FV corrections at $\bigo(1/L^2)$. Therefore we can use directly the expression in \cref{eq:Gammamunu2} obtained  from the WTI up to $\bigo(k)$.

\paragraph{Weak vertex:}
The off-shell weak vertex $W^\rho(p)$ for a pseudoscalar of incoming momentum $p$ has been introduced in \cref{eq:Wrho} above. It is obtained from the amputation of the correlation function $C_W^\rho(p)$ in \cref{eq:Crho}, namely
\begin{equation}
	W^\rho(p) = Z_P^{-1} D(p)^{-1} \, C_W^\rho(p) = -p^\rho F_W(p^2)\,.
\end{equation}
In QCD and on-shell it reduces to $W^\rho(p)=-p^\rho f_P$, as in a point-like theory.

\paragraph{Weak vertex + one photon:}
The irreducible kernel $W_{1}$ in~\cref{eq:m1exp}, for a pseudoscalar and photon of incoming respective momenta $p$ and $k$, is defined in terms of the correlation function 
\begin{equation}
		C_W^{\rho\mu}(p,k) = i \int \diff^4 z \, \diff^4 x \, e^{ipz+ikx} \bra{0} \mathrm{T}[J_{W}^\rho(0) J^\mu(x)\phi^{\dagger}(z)] \ket{0}\,.
\end{equation}
When evaluated on-shell, this is strictly related to the amplitude 
of radiative decays ${P\rightarrow\ell \nu \gamma^*}$, that was studied in e.g.~Refs.~\cite{Bijnens:1992en,Desiderio:2020oej}. The weak vertex can be defined by amputating $C_W^{\rho\mu}(p,k)$ and removing the pole associated with the so-called ``inner bremsstrahlung''
\begin{equation}
	\label{eq:Wrhomu}
	W^{\rho\mu}(p,k) =  Z_P^{-1} D(p)^{-1}\, C_W^{\rho\mu}(p,k) - \Gamma^{\mu}(p,k) D(p+k) W^\rho(p+k)\,.
\end{equation}
This procedure to define $W_{1}$ is equivalent to how $\Gamma _{2}$ was defined as the regular part of $C$ above. 
This leads to the expression of the irreducible vertex function $W_{1}$ from the contraction with $\gamma_{\rho}(1-\gamma _{5})$
\begin{equation}
  \raisebox{-4.5ex}{\includegraphics{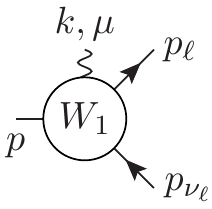}}=W^{\rho \mu}(p,k) \gamma _{\rho} (1-\gamma _{5})\,.
\end{equation} 
The vertex $W^{\rho\mu}(p,k)$ satisfies the following WI
\begin{equation}
	\label{eq:wardW1}
	k_\mu W^{\rho\mu}(p,k) = W^\rho(p) - W^\rho(p+k)\,,
\end{equation}
that can be exploited to determine the functional form of the vertex up to transverse terms. The $W_1$ kernel enters diagrams (e) and (f) in Fig.~\ref{fig:kl2diagrams} and, by applying again finite-volume power-counting arguments, one can show that in order to extract the FVEs at $\bigo(1/L^2)$ it is sufficient to know the vertex $W^{\rho\mu}(p,k)$ at $\bigo(k)$. Therefore, by expanding \cref{eq:wardW1} at $\bigo(k^2)$ we get
\begin{align}
	W^{\rho\mu}(p,k) =& \  \delta^{\rho\mu} \, F_W(p^2) + \left[2\,k^\rho p^\mu + k^\mu p^\rho + 2 p^\mu p^\rho \right] \, F_W'(p^2) + 2 \, (p\cdot k) \, p^\rho p^\mu \, F_W''(p^2) \ + \\
	& \ -  \, \frac{V_1(k^2,(p+k)^2)}{m_P} \, \varepsilon_{\mu\rho\alpha\beta} \, k^\alpha p^\beta + \frac{A_1(k^2,(p+k)^2)}{m_P} \left[ \delta^{\rho\mu} \, (p\cdot k) - k^\rho p^\mu  \right]+ \bigo(k^2)\notag\label{eq:Wrhomu2}
\end{align}
where the form factors $V_1(k^2,(p+k)^2)$ and $A_1(k^2,(p+k)^2)$ are not constrained by the WI in \cref{eq:wardW1}.  Some comments can be made here. The form factors relevant for the $\bigo(1/L^2)$ FVEs are those entering real decays with on-shell photons, \ie $A(p^2)=A_1(0,p^2)$ and $V(p^2)= V_1(0,p^2)$. When evaluated on-shell, these quantities reduce to $F_A^P = A(-m_P^2)$ and $F_V^P=V(-m_P^2)$ for $P\to \ell\nu\gamma$ decays. Additionally, the derivatives of these form factors have been estimated in chiral perturbation theory (ChPT) and measured in experiment~\cite{Cirigliano:2011ny}, but they only contribute to higher orders than $\bigo(1/L^2)$, together with additional form factors. Notice that at $\mathcal{O}(\alpha)$ the derivatives $F_W'(p^2)$ and $F_W''(p^2)$ reduce respectively to $f_1$ and $f_2$ defined in \cref{eq:znfn} when evaluated on-shell. However, we stress that it is important to define the vertex $W^{\rho\mu}(p,k)$ in terms of the off-shell form factor $F_W(p^2)$ (and its derivatives) and to take the on-shell limit only after computing the diagrams.

\paragraph{Weak vertex + two photons}
The irreducible vertex function $W_{2}$ is in an analogous fashion formally defined as the regular part of a correlation function $C_W^{\rho \mu \nu} (p,k,q)$ related to the decay $P\rightarrow \ell \nu \gamma ^{*}\gamma ^{*}$, 
\begin{align}
C_W^{\rho \mu \nu} (p,k,q) = - \int d^4z \, d^4x\, d^4y\,  e^{ipz+ikx+iqy}   \bra{0} T  [ J_W^\rho (0) J^{\mu} (x)J^{\nu} (y) \phi^\dagger(z) ] \ket{0}
\, .
\end{align}
Here the pseudoscalar and two photons are incoming, with momenta $p$, $k$ and $q$ respectively.
We may thus write
\begin{equation}
 \raisebox{-4.5ex}{\includegraphics{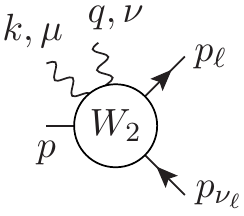}}
 =
 W^{\rho \mu \nu}(p,k,q) \gamma _{\rho}(1-\gamma _{5})
 \, .
\end{equation}
with 
\begin{align}
	W^{\rho \mu \nu}(p,k,q)  =& \  Z_P^{-1}D(p)^{-1} C_W^{\rho\mu\nu}(p,k,q)  - 
	C^{\mu\nu}(p,k,q)
	D(p+k+q) W^{\rho}(p+k+q)\notag \\ 
	& \ - \Gamma^\mu(p,k)D(p+k)W^{\nu\rho}(p+k,q) - \Gamma^\nu(p,q)D(p+q)W^{\mu\rho}(p+q,k) 
\end{align}
and $C^{\mu\nu}(p,k,q)$ defined above in \cref{eq:CmunuDecompose}.

From the Ward-Takahashi identity%
\begin{align}
	\label{eq:wardW2}
k_{\mu}W^{\rho \mu \nu} (p,k,q) = W^{\rho \nu}(p,q)-W^{\rho \nu}(p+k,q) \, ,
\end{align}
one can deduce the form of the vertex as done for $W^{\rho \mu} (p,k)$. Through power counting applied to diagram (g) of Fig.~\ref{fig:kl2diagrams} we see that only $\bigo(1)$ terms in $W^{\rho \mu\nu} (p,k,q)$ contribute at order $1/L^2$ and no $\bigo(1/L)$ corrections are produced. Therefore, by inserting \cref{eq:Wrhomu2} into \cref{eq:wardW2} and expanding at $\mathcal{O}(k)$ we get
\begin{equation}
\label{eq:Wrhomunu}
	W^{\rho \mu\nu} (p,k,q) = -2\,(\delta^{\rho\nu} p^\mu + \delta^{\rho\mu}p^\nu+\delta^{\mu\nu}p^{\rho}) F'_W(p^2) - 4 \, p^\rho p^\mu p^\nu \, F''_W(p^2) + \bigo(k,q)\,,
\end{equation}
that respects the crossing symmetry $W^{\rho \mu\nu} (p,k,q)=W^{\rho \nu\mu} (p,q,k)$, as expected. Moreover, we notice that only unphysical off-shell terms contribute to this diagram.
 
\subsubsection{Final expressions for individual Feynman diagrams} 
Having defined the irreducible vertex functions contributing at order $\alpha$, we may now write down the integrals for the diagrams in~\cref{fig:kl2diagrams}. 
The respective diagrams, evaluated choosing the Feynman gauge for the photon propagator and in units of $e^2$, are 
\begin{equation}\label{eq:diagramdefs}
\begin{split}
& \text{(a)}: \; W^{\rho}(p)  \mathcal{L}^{rs}_\rho(p,p_\ell) \, ,
 \\
&  \text{(b)+(c)}: \; \left[ (z_1-f_1) \Delta m_P^2 + \frac12 \left. \frac{\partial \Sigma(p^2)}{\partial p^2}\right|_{p^2=-m_P^2}\right] W^{\rho}(p)  \mathcal{L}^{rs}_\rho(p,p_\ell)
\, ,
 \\
&	\text{(d)}: \; \int \frac{\diff^4k}{(2\pi)^4}\,\frac{1}{k^{2}}\, \Gamma _{\mu}(p,k)  D_0 (p+k) W^{\rho}(p+k) \mathcal{L}^{rs}_{\rho\mu}(p,p_\ell,k) \, ,
 \\
& \text{(e)}: \; \int \frac{\diff^4k}{(2\pi)^4}\,	\frac{1}{k^{2}}\, W^{\rho \mu}(p,k)  \mathcal{L}^{rs}_{\rho\mu}(p,p_\ell,k)\, ,
 \\
& \text{(f)}:\; 	\int \frac{\diff^4k}{(2\pi)^4}\, \frac{1}{k^2} \, \Gamma _{\mu}(p,k) D_0(p+k) W^{\rho \mu} (p+k,-k)  \mathcal{L}_\rho^{rs}(p,p_\ell) \, ,
 \\
& \text{(g)}:\; 	\int \frac{\diff^4k}{(2\pi)^4}\, \frac{1}{k^2} \,\frac{1}{2}\, W^{\rho \mu\mu} (p,k,-k)  \mathcal{L}_\rho^{rs}(p,p_\ell) \, ,
\end{split}
\end{equation}
where the leptonic tensor $\mathcal{L}_{\rho}^{rs}(p,p_\ell)$ is defined in \cref{eq:Lrho} and  $\mathcal{L}_{\rho \mu}^{rs}(p,p_\ell,k)$ is given by
\begin{equation}
	\mathcal{L}_{\rho \mu}^{rs}(p,p_\ell,k) = i \,  \bar{u}^{r}(p_{\ell})\, \gamma_{\mu} S_{\ell}(p_{\ell}+k)\gamma_{\rho}(1-\gamma _{5}) \, v^{s}(p_{\nu_{\ell}})\,.
\end{equation}
Note that due to the appearance of three propagators in diagrams (d) and (b)+(c), arising from the derivative in the latter case, we will have here IR-divergent finite-size coefficients containing logarithms of $L$, as explained in~\cref{sec:sum} and App.~\ref{sec:cj}.
Having defined all the diagrams, we next turn to the calculation of the FVEs in $Y^{(2)}(L)$.

\subsection{Electromagnetic finite-size effects}
Here we consider the diagrams (b)+(c), (d), (e), (f) and (g) in turn to obtain the finite-volume effects to the square matrix element $\Delta |\mathcal{M} |^2$ defined in \cref{eq:M2fve}. The contraction with diagram (a)$^\dagger$ in \cref{eq:M2fve} and the sum over the final-state spins require the completeness relations for spinors in Euclidean space. These are straightforward to derive from the Euclidean Dirac equation and are given by 
\begin{align}
	&\sum_{r,r'} \, u^{r'\!}(p_\ell) \, \bar{u}^r(p_\ell) = - i \slashed{p}_\ell + m_\ell\,,\\
	&\sum_{s,s'} \, v^s(p_\nu) \, \bar{v}^{s'\!}(p_\nu) = i \slashed{p}_\nu\,,
\end{align}
and can be used to compute the following spinor traces entering~\cref{eq:diagramdefs} 
\begin{align}
	T_\rho(p,p_\ell) =& 
	\sum_{r,s,r'\!,s'} \mathcal{L}_\rho^{rs}  \left(\mathcal{M}_0^{r'\!s'}\right)^\dagger \nonumber\\
	=& - i m_\ell f_P \, \mathrm{Tr}\left[(-i\slashed{p}_\ell+m_\ell) \gamma_\rho(1-\gamma_5)(i \slashed{p}_\nu) (1+\gamma_5)\right]\,,\\
	T_{\rho\mu}(p,p_\ell,k)=& 
	\sum_{r,s,r'\!,s'} \mathcal{L}_{\rho\mu}^{rs} \left(\mathcal{M}_0^{r'\!s'}\right)^\dagger \nonumber \\
	=& 
	\, m_\ell f_P \, \mathrm{Tr}\left[ (-i\slashed{p}_\ell+m_\ell) \gamma_\mu S_\ell(p_\ell+k)\gamma_\rho(1-\gamma_5)(i \slashed{p}_\nu)(1+\gamma_5) \right]\,,
\end{align}
where we have used the equations of motion for the leptons $\bar{u}^r (p_\ell)\, \slashed{p}_\ell = i m_\ell \, \bar{u}^r(p_\ell)$ and $\slashed{p}_\nu \, v^s(p_\nu)=0$. 

Letting $(i)$ refer to any of the diagrams in Fig. 2 (i.e. $(i)\in\{\text{(a),(b),(c),(d),(e),(f),(g)}\}$), denote the integrand of the corresponding expression in \cref{eq:diagramdefs} as  $I_{(i)}^{rs}(k)$. Then the associated FV correction to $\Delta |\mathcal{M} |^2$ according to \cref{eq:M2fve} by computing the following sum-integral difference
\begin{equation}
	\label{eq:corrM2diag}
	\Delta \left[(i)\right]\times\text{(a)}^\dagger = \sum_{r,s,r'\!,s'}  \Delta_\ksp'  \int \frac{\diff k_0}{2\pi}\, I_{(i)}^{rs}\left(k_0,\ksp\right)\left(\mathcal{M}_0^{r'\!s'}\right)^\dagger\,,
\end{equation}
with $\Delta_\ksp'$ the sum-integral difference operator defined in~\cref{sec:se}.
In the following sections we use the shorthand:
\begin{equation}
	(i) :\;\;  2 \, \Delta [(i)] \times (a)^\dagger\,.
\end{equation}
The FVEs will be expressed in terms of physical quantities and finite volume coefficients, and some of them depend on the velocity $\mathbf{v}_{\ell}=\mathbf{p}_{\ell}/\omega_\ell$ of the lepton in the rest frame of the pseudoscalar meson. These will be discussed case by case below using the notation defined in App.~\ref{sec:cj}. For their calculation we make use of an accelerated numerical algorithm presented in~\cref{sec:cj}. 

\subsubsection{Diagram (b+c)}
The contribution of this diagram to the FV correction $\Delta |\mathcal{M}|^2$, with the inclusion of the off-shell terms $z_n$ and $f_n$, is obtained from 
\begin{equation}
	\label{eq:FVdiagBCdef}
\text{(b)+(c)}:\;\;  	\left[ 2(z_1-f_1) \Delta m_P^2(L) +  \left. \frac{\partial }{\partial p^2}\, \Delta \Sigma(p^2)\right|_{p^2=-m^2}\right] W^{\rho}(p) T_\rho(p,p_\ell)\,
\end{equation}
where $\Delta m_P^2(L)$ is the FV correction to the squared mass of the meson obtained in \cref{eq:fveffmass} and $\Delta \Sigma(p^2)$ is given by
\begin{equation}
	 \Delta \Sigma(p^2) = \Delta_\ksp' \int \frac{\diff k_0}{2\pi} \frac{1}{k^2}  \left[\Gamma_\mu(p,k)D_0(p+k)\Gamma_\mu(p+k,-k) + \frac{1}{2} \, \Gamma_{\mu\mu}(p,k,-k)\right]\,,
\end{equation}
with $\Gamma_\mu(p,k)$ and $\Gamma_{\mu\nu}(p,k,-k)$ defined above in \cref{eq:Gammamu,eq:Gammamunu2}, respectively.
 Note that the calculation of the derivative is simplified in the rest frame of the meson $p=(p_0,\boldsymbol{0})$, namely
\begin{equation}
	\left.\frac{\partial }{\partial p^2}\, \Delta \Sigma(p^2)\right|_{p^2=-m_P^2} = \frac{1}{2 p_0} \left.\frac{\partial }{\partial p_0}\, \Delta \Sigma(p^2)\right|_{p_0=im_P}\,.
\end{equation}
By using the procedure outlined above, together with \cref{eq:YnFVE}, we obtain the following FV correction to $Y^{(2)}(L)$
\begin{align}
	\label{eq:FVdiagBC}
	\Delta Y^{(2)}_\text{(b)+(c)}(L,\lambda) = \frac{b_3}{2 \pi} - \frac{1}{L} \, \left[ 4\, m_P  f_1 \, c_2\right] + \frac{1}{L^2} \, \left[-\frac{2\pi}{3}\,  \langle r_P^2\rangle\, c_1 - 8\pi \, f_1\,c_1\right]\,,
\end{align}
where the IR divergent coefficient $b_3$ is given by
\begin{equation}
	b_3 = c_3 + 4\pi \, \left[\log\left(\frac{L\lambda}{2\pi}\right)-\log 2 + 1\right]\,,
\end{equation}
as obtained in Appendix~\ref{sec:cj} together with the coefficients $c_j$. Here we see the logarithmic dependence on $L$ and the structure-dependence appears via the charge radius $\langle r_P^2\rangle$. Moreover, we notice that the term $2 z_1 \Delta m_P^2(L)$ in \cref{eq:FVdiagBCdef} is cancelled exactly by equal and opposite terms obtained from the derivative of $\Delta \Sigma(p^2)$, both at $\bigo(1/L)$ and $\bigo(1/L^2)$.

\subsubsection{Diagram (d)}
The contribution to $\Delta |\mathcal{M}|^2$ from this diagram is given by 
\begin{equation}
 \text{(d)}: \; \;    2\, \Delta_\ksp ' \int \frac{\diff k_0}{2\pi} \,\frac{1}{k^{2}}\, \Gamma _{\mu}(p,k)  D_0 (p+k) W^{\rho}(p+k) T_{\rho\mu}(p,p_\ell,k)
\end{equation}
which yields the following correction to $Y^{(2)}(L)$
\begin{eqnarray}
    \Delta Y^{(2)}_\text{(d)}(L,\lambda) 
    &=& -\frac{b_3(\mathbf{v}_\ell)}{\pi} + \frac{1}{L}\left[\frac{(1+r_\ell^2)(1-3r_\ell^2)\, c_2 + 4\, c_2(\mathbf{v}_\ell)}{ m_P (1-r_\ell^4)} - 4\, m_P f_1\, c_2(\mathbf{v}_\ell) \right]\\
    && + \, \frac{1}{L^2} \left\{-\frac{2\pi \, c_1}{m_P^2} + \frac{2\pi}{3}\, \langle r_P^2\rangle \, c_1 - 8\pi\, m_P^2 f_2 \, c_1(\mathbf{v}_\ell)\right. \notag \\
    && \left. \qquad \quad \ + \frac{4\pi \, f_1 \left[(1+r_\ell^2)(1-3r_\ell^2)\, c_1 + 4 \, c_1(\mathbf{v}_\ell)\right]}{1-r_\ell^4} \right\}\,.
    \notag 
\end{eqnarray}
Here the IR-divergence is encoded in the FV coefficient
\begin{eqnarray}
    b_3(\mathbf{v}_\ell) = c_3(\mathbf{v}_\ell) + 4\pi A_1(\mathbf{v}_\ell) \log\left(\frac{L\lambda}{2\pi}\right)-B_1(\mathbf{v}_\ell)
\end{eqnarray}
that depends this time on the velocity of the lepton $\mathbf{v}_\ell$. The functions $A_1(\mathbf{v}_\ell)$ and $B_1(\mathbf{v}_\ell)$ are defined in Appendix~\ref{sec:cj}, together with the finite volume coefficients $c_j(\mathbf{v}_{\ell})$. Notice that there's no contribution of $z_n$ terms in this diagram and the structure dependence, appearing at $\bigo(1/L^2)$, is completely determined by the charge radius of the meson $\langle r_P^2\rangle$.

\subsubsection{Diagram (e)}
The contribution to $\Delta |\mathcal{M}|^2$ from this diagram is obtain from  
\begin{equation}
 \text{(e)}:\; \;    2\, \Delta_\ksp ' \int \frac{\diff k_0}{2\pi} \,\frac{1}{k^{2}}\, W^{\rho\mu}(p,k) T_{\rho\mu}(p,p_\ell,k)\,.
\end{equation}
The finite-size effects contributing to $Y^{(2)}(L)$ starts at $\bigo(1/L)$ and are given by 
\begin{eqnarray}
    \Delta Y^{(2)}_\text{(e)}(L) &=& \frac{1}{L} \left[-\frac{4 \, c_2(\mathbf{v}_\ell)}{m_P(1+r_\ell^2)} + 4\, m_P f_1 \, c_2(\mathbf{v_\ell})\right] +\frac{1}{L^2} \left\{ \frac{8\pi\,  [(1+r_\ell^2)\,c_1-2\, c_1(\mathbf{v}_\ell)]}{ m_P^2(1-r_\ell^4)} + \right. \\
    &&\left. - \, \frac{F_A^P}{f_P}\,\frac{4\pi \, [(1+r_\ell^2)^2\, c_1-4 r_\ell^2 \, c_1(\mathbf{v}_\ell)]}{ m_P(1-r_\ell^4)} + 8\pi\, m_P^2 f_2 \, c_1(\mathbf{v}_\ell) \, +\right.\nonumber\\
    && - \left.\frac{4\pi \, f_1 \left[(1+r_\ell^2)(1-3r_\ell^2)\, c_1 + 4 \, c_1(\mathbf{v}_\ell)\right]}{1-r_\ell^4} \right\} \,. \notag
\end{eqnarray}
Here the structure dependence comes at $\bigo{(1/L^2)}$ from the axial form factor $F_A^P$. The vector form factor $F_V^P$, instead, does not contribute because of the anti-symmetric properties of the Levi-Civita tensor in $W^{\rho\mu}(p,k)$, see \cref{eq:Wrhomu2}. As in the case of diagram (d), here we have FV coefficients $c_j(\mathbf{v}_\ell)$ that depend on the lepton velocity and we observe that the dependence on $z_n$ is absent also in this case.

\subsubsection{Diagram (f)}
For this diagram we have to compute
\begin{equation}
 \text{(f)}:\; \;    2\, \Delta_\ksp ' \int \frac{\diff k_0}{2\pi}\, \frac{1}{k^2} \, \Gamma _{\mu}(p,k) D_0(p+k) W^{\rho \mu} (p+k,-k)  T_\rho(p,p_\ell)\,.
\end{equation}
This leads to 
\begin{eqnarray}
    \Delta Y^{(2)}_\text{(f)}(L) &=& \frac{1}{L}\left[-\frac{2\,c_2}{ m_P} + 4\,m_P f_1\,c_2\right] + \frac{1}{L^2}\left[\frac{2\pi \, c_1}{m_P^2} - 16\pi \, f_1 \, c_1 + 8\pi\, m_P^2 f_2\, c_1\right]\,.
\end{eqnarray}
As for diagram (e), here the FV corrections start at $\mathcal{O}(1/L)$, but no physical structure-dependent terms contribute in this case. The FVEs depend on the coefficients $c_j$ defined in~\cref{sec:cj}.

\subsubsection{Diagram (g)}
Finally, for diagram (g) we need to compute 
\begin{equation}
 \text{(g)}:\; \;    \Delta_\ksp ' \int \frac{\diff k_0}{2\pi}\, \frac{1}{k^2} \, W^{\rho \mu\mu}(p,k,-k) T_\rho(p,p_\ell)\,.
\end{equation}
This diagram starts contributing at $\bigo(1/L^2)$ and the FV correction to $Y^{(2)}(L)$ only depends on the off-shell quantities $f_n$, as expected from the definition of the vertex $W^{\rho\mu\nu}(p,k,q)$ in \cref{eq:Wrhomunu}. We obtain
\begin{eqnarray}
    \Delta Y^{(2)}_\text{(g)}(L) = \frac{1}{L^2} \left[24\pi\, f_1\, c_1 - 8\pi\, m_P^2 f_2 \, c_2\right]\,.
\end{eqnarray}

\subsubsection{Total finite-size effects}
Here we present our final result for the finite-size effects in $Y^{(2)}(L)$ up to and including order $1/L^2$ terms for the leptonic decay of a pseudoscalar meson $P^{-}$. This is obtained by summing the contributions from all the diagrams above. 
Rewriting the FV correction $\Delta Y^{(2)}(L,\lambda)$ as
\begin{equation}
    \Delta Y^{(2)}(L,\lambda) = Y_{\mathrm{log}} \log \frac{L \lambda }{2\pi} + Y_{0} + \frac{1}{L}\,Y_{1} +  \frac{1}{L^2}\,Y_{2}
\end{equation}
we get
\begin{eqnarray}
    Y_\mathrm{log} &=& 2\left(1-2\, A_1(\mathbf{v}_\ell)\right)\,, \\
    Y_0 &=& \frac{c_3-2\,(c_3(\mathbf{v}_\ell)-B_1(\mathbf{v}_\ell))}{2\pi} + 2 \left(1-\log 2\right)\,,\\
    Y_1 &=& - \frac{(1+r_\ell^2)^2\,c_2 - 4\, r_\ell^2 \, c_2(\mathbf{v}_\ell)}{m_P (1-r_\ell^4)}\,,\\
    Y_2 &=&  - \frac{F_A^P}{f_P} \, \frac{4\pi \,[(1+r_\ell^2)^2 \, c_1 - 4\, r_\ell^2\, c_1(\mathbf{v}_\ell)]}{m_P (1-r_\ell^4)} + \frac{8\pi \, [(1+r_\ell^2) \, c_1 - 2\,  c_1(\mathbf{v}_\ell)]}{m_P^2 (1-r_\ell^4)}\,.
\end{eqnarray}
As discussed at the end of Sec.~\ref{sec:FVformaldescr}, in order to compute $Y^{(2)}(L)$ we also need the infinite volume contribution $Y^\mathrm{uni}_\mathrm{IV}(\lambda)$ computed in the point-like approximation. This can be found in Ref.~\cite{Lubicz:2016xro} and reads
\begin{equation}
\label{eq:ivyl}
Y^\mathrm{uni}_\mathrm{IV}(\lambda) = 
-\frac54 + 2 \, \log\left(\frac{m_\ell^2}{m_W^2}\right) + \log\left(\frac{m_W^2}{\lambda^2}\right) - A_1(\mathbf{v}_\ell) \left[\log\left(\frac{m_\ell^2}{\lambda^2}\right)+\log\left(\frac{m_P^2}{\lambda^2}\right)-2\right]\,, 
\end{equation}
where we have used the relations
\begin{align}
\label{eq:A1_alt}
    |\mathbf{v}_\ell|=\frac{1-r_\ell^2}{1+r_\ell^2}\,,\quad A_1(\mathbf{v}_\ell) = \frac{\arctanh(|\vel_\ell|)}{|\vel_\ell|} = - \frac{1+r_\ell^2}{1-r_\ell^2}\frac{\log(r_\ell^2)}{2}\,.
\end{align}
It is easy to show that the coefficient of $\log(\lambda)$ in \cref{eq:ivyl} is equal and opposite to $Y_\mathrm{log}$ and therefore the FV quantity $Y^{(2)}(L)$ only depends on the IR regulator $L$. We obtain 
\begin{align} 
 \label{eq:y2finalform}
    Y^{(2)}(L) \, &= \ \frac34 + 4\, \log\left(\frac{m_\ell}{m_W}\right) + 2\,\log\left(\frac{m_W L}{4\pi}\right) \ +\frac{c_3-2\,( c_3(\mathbf{v}_\ell)-B_1(\mathbf{v}_\ell))}{2\pi}\, - \\
    & 
      - 2\,A_1(\mathbf{v}_\ell)\left[\log\left(\frac{m_P L}{2\pi}\right)+\log\left(\frac{m_\ell L}{2\pi}\right)-1\right]  - \frac{1}{m_P L} \left[ \frac{(1+r_\ell^2)^2\,c_2 - 4\, r_\ell^2 \, c_2(\mathbf{v}_\ell)}{ 1-r_\ell^4} \right] + \nonumber\\
    & + \, \frac{1}{(m_P L)^2} \left[ - \frac{F_A^P}{f_P} \, \frac{4\pi \, m_P \,[(1+r_\ell^2)^2 \, c_1 - 4\, r_\ell^2\, c_1(\mathbf{v}_\ell)]}{1-r_\ell^4} + \frac{8\pi \, [(1+r_\ell^2) \, c_1 - 2\,  c_1(\mathbf{v}_\ell)]}{ (1-r_\ell^4)} \right]\,. \nonumber
\end{align}
Several comments can be made here. We observe the expected and complete cancellation of off-shell contributions proportional to $z_n$ and $f_n$. This property must be true at all orders and the calculation could have been done assuming $z_n=f_n=0$, although conserving these terms is a practical way to detect mistakes in the construction of the final result. These terms arise from the skeleton expansion of the full QCD+QED correlator $C_W^{rs}(p,p_\ell)$ of \cref{eq:CWfull} into 1PI subdiagrams up to $\mathcal{O}(e^2)$ corrections. Such separation is although arbitrary and it is likely possible to redefine vertex functions to achieve a manifestly on-shell derivation of the FVEs similarly to what was done for the simpler self-energy case in~\cref{sec:seonshell}. It is interesting to notice that there is a perfect cancellation of off-shell terms separately in the factorizable correlation function $C_{W,P}^{rs}(p,p_\ell)$ (diagrams (b)+(c), (f), and (g)) and in the non-factorizable correlation function $C_{W,\ell}^{rs}(p,p_\ell)$ (diagrams (d) and (e)) where the photon is attached to the external charged lepton. This can also be expected on general grounds, as the factorizable correction is the $\bigo(e_q^2)$ correction to the leptonic decay amplitude where $e_q$ is the quark elementary charge, and the non-factorizable correction is the $\bigo(e_qe_\ell)$ correction where $e_\ell$ is the lepton elementary charge. Although in Nature $e_q=e_\ell=e$, in practice both charges are independent parameters of the QCD+QED Lagragian and therefore the factorizable and non-factorizable corrections are both physical amplitudes which must be independent from off-shell contributions.

It is also interesting to notice that the structure dependence in $Y^{(2)}(L)$ is only given by the axial form factor $F_A^P$, while the charge radius contribution cancels in the final result. This is related to the conservation of the electric charge in the process $P\to \ell\nu_\ell$. In fact, by keeping the meson and lepton charge factors explicit in diagrams (b)+(c) and (d), where $\langle r_P^2\rangle$ contributes, we find that $\langle r_P^2\rangle$ gets multiplied by the factor $e_P(e_P- e_\ell)$, which vanishes if the charge is conserved.

The knowledge of $Y^{(2)}(L)$ in~\cref{eq:y2finalform} allows to control the systematic FVEs in lattice calculations of $\Gamma _{0}(L)$. In~\cref{sec:num} we make a brief study of the size of the FVEs, in particular by seeing how large structure-dependent effects are. However, before that we compare our results with those obtained in the point-like approximation in Refs.~\cite{Lubicz:2016xro,Tantalo:2016vxk}.

\subsection{Comparing to known point-like results}
The finite-size effects in $Y^{(1)}(L)$, \ie up to and including order $1/L$ terms, were studied in Refs.~\cite{Lubicz:2016xro,Tantalo:2016vxk} assuming the decaying meson to be a point-like particle. The method used by the authors of Ref.~\cite{Lubicz:2016xro} to calculate the FV effects is fundamentally different from ours. In fact, in the point-like approximation, only diagrams (b), (d), (e) and (f) contribute and their evaluation reduces to the calculation of five master integrals. The master integrals give rise to finite-size coefficients $K_{ij}$ and $K_P$, defined as integrals of Jacobi theta functions. Our result in \cref{eq:y2finalform} is expressed in terms of the FV coefficients $c_j$ and $c_j(\mathbf{v}_\ell)$, and of the known functions $A_1(\mathbf{v}_\ell)$ and $B_1(\mathbf{v}_\ell)$ (see Appendix~\ref{sec:cj}). Clearly, the result for $Y^{(1)}(L)$ must coincide in the two cases, and it is therefore possible to derive useful relations between the two sets of finite-size coefficients. The FV corrections to $Y^{(1)}(L)$ in Ref.~\cite{Lubicz:2016xro} are obtained as
\begin{eqnarray}
    \Delta Y^{(1)} = 16 \pi^2 \, \left( \Delta X_1 + \Delta X_2 + \Delta X_3 + \frac{\Delta X_P}{2} \right)\,,
\end{eqnarray}
where
\begin{eqnarray}
   \label{eq:DeltaX1}
    16 \pi^2 \Delta X_1 &=& 
    \frac{4}{3} + 2 (K_{31}+K_{32})(1+r_\ell^2) - 2 A_1(\mathbf{v}_\ell)\left[ \gamma_E + \log \pi + 2 \log \left(\frac{L\lambda}{2\pi}\right)\right] +
    \\
    && + \, \frac{1}{m_P L} \left[\frac{2(K_{21}+K_{22})}{1-r_\ell^2}-\frac{4\pi(1+r_\ell+r_\ell^2)}{r_\ell(1-r_\ell^4)}+\frac{\pi(K_{11}+K_{12}-3)(1-3r_\ell^2)}{1-r_\ell^2}\right]\,, 
    \nonumber \\ \label{eq:DeltaX2}
    16 \pi^2 \Delta X_2 &=& 
    \frac{2\pi}{m_P L}\left[3-K_{11}-K_{12}\right]\,,
    \\ \label{eq:DeltaX3}
    16 \pi^2 \Delta X_3 &=& 
    -2(K_{21}+K_{22})+\frac{4\pi(1+r_\ell+r_\ell^2)}{r_\ell(1+r_\ell^2)}\,,
    \\ \label{eq:DeltaXP}
    16 \pi^2 \Delta X_P &=& -K_P+4 \log(L\lambda) \,.
\end{eqnarray}
Note the presence of the finite-size coefficients $K_{ij}$ and $K_{P}$, and $\gamma _{E}$ as the Euler-Mascheroni constant. Above we have used the relations in \cref{eq:A1_alt} to better match these expressions with our results. The correspondence between the above finite-size effects and the ones calculated in the previous section is then
\begin{eqnarray}
    \label{eq:DeltaX1Check}
     16 \pi^2 \Delta X_1 &=& \Delta Y^{(1)}_\text{(d)}\,,
     \\\label{eq:DeltaX2Check}
     16 \pi^2 \Delta X_2 &=& \Delta Y^{(1)}_\text{(f)}\,,
     \\ \label{eq:DeltaX3Check}
     16 \pi^2 \Delta X_3 &=& \Delta Y^{(1)}_\text{(e)}\,,
     \\ \label{eq:DeltaXPCheck}
     16 \pi^2 \Delta X_P &=& 2\,\Delta Y^{(1)}_\text{(b)}\,.
\end{eqnarray}
Using these matching conditions we obtain the following relations between the various FV coefficients
\begin{align}
   c_{2}  = & \  \pi ( K_{11} + K_{12}-3) \, ,
 \\
c_{3} =&\,  -\pi (4+K_{P}-4 \log 4\pi) \, ,
 \\
c_{2}(\mathbf{v}_{\ell}) = &\  \frac{1}{2}(K_{21}+K_{22})(1+r_{\ell}^2)- \frac{\pi(1+r_{\ell}+r_{\ell}^2)}{r_{\ell}} \, ,
 \\
c_{3}(\mathbf{v}_{\ell})  =&\,  
-\frac{4\pi}{3} - 2 \pi (K_{31}+K_{32})(1+r_\ell^2) + 2 \pi A_1(\mathbf{v}_\ell) (\gamma_E + \log \pi) + B_1(\mathbf{v}_\ell) 
\, .
\end{align}
By using the results in \cref{tab:cj} of Appendix~\ref{sec:cj} for the zero-velocity FV coefficients $c_j$ together with the results of $K_{11} \simeq 0.0765331$,  $K_{12} \simeq 0.0861695$ and $K_P \simeq 4.90754$ from Ref.~\cite{Lubicz:2016xro} we find a full agreement between the two calculations. Moreover, using $m_P=m_\pi=139.57018$~MeV, $m_\ell=m_\mu=105.65837$ MeV and $|\mathbf{p_\ell}|=|\mathbf{p_\mu}|=29.792$~MeV we get
\begin{eqnarray}
\begin{aligned}
    c_2(\mathbf{v_\ell^\pi}) & = -9.14489 \, , \\
    c_3(\mathbf{v_\ell^\pi}) & = 3.91764 \, ,
\end{aligned} \qquad \text{for} \quad \mathbf{\hat{v}_\ell^\pi} = (1,1,1)/\sqrt{3}
\end{eqnarray}
and 
\begin{eqnarray}
\begin{aligned}
    c_2(\mathbf{v_\ell^\pi}) & = -9.13932 \, , \\
    c_3(\mathbf{v_\ell^\pi}) & = 3.92388 \, ,
\end{aligned} \qquad \text{for} \quad \mathbf{\hat{v}_\ell^\pi} = (0,0,1)\,.
\end{eqnarray}
Taking the values of $K_{2j}$ and $K_{3j}$ evaluated at the same physical point from Ref.~\cite{Lubicz:2016xro} and using $A_1(\vell)$ and $B_{1}(\vell)$ evaluated as in Appendix~\ref{sec:cj}, we find an excellent agreement also for these velocity-dependent FV coefficients. 

In Ref.~\cite{Tantalo:2016vxk} the point-like decay rate was considered up to order $1/L^3$, but using a different representation based on generalized $\zeta$-functions for the sum-integral differences. We compare also to these results. We find the following matching conditions
\begin{equation}
\begin{split}
&    c_1 = 4 \pi \, \zeta_A\,, \\
&    c_2 = 4 \pi^2 \, \zeta_A\,,\\
&   c_3 = -4\pi (1-2\pi ^2 \zeta_C - \log 4\pi)\,,
\end{split} \quad 
\begin{split}
&    c_1(\mathbf{v_\ell}) = 4\pi \, \zeta_B^{P\ell}(\mathbf{v_\ell})\,,\\
&    c_2(\mathbf{v_\ell}) =  16 \pi^2 \, \zeta_B(\mathbf{v_\ell})\,,\\
&    c_3(\mathbf{v_\ell}) = 8 \pi^3 \, \zeta_C(\mathbf{v_\ell})  + 4 \pi \log(2 \pi)  \, A_1(\mathbf{v_\ell}) + B_1(\mathbf{v_\ell})\,,
\end{split}
\end{equation}
and observe a complete numerical agreement for both the FV coefficients as well as for the point-like FV corrections to the decay rate at $\mathcal{O}(1/L)$ and $\mathcal{O}(1/L^2)$. The numerical values for $c_1(\mathbf{v_\ell})$ at the physical point are
\begin{equation}
\begin{split}
& c_1(\mathbf{v_\ell^\pi}) = -2.91210 \qquad \text{for} \quad \mathbf{\hat{v}_\ell^\pi} = (1,1,1)/\sqrt{3}\,,\\
& c_1(\mathbf{v_\ell^\pi}) = -2.90736 \qquad \text{for} \quad \mathbf{\hat{v}_\ell^\pi} = (0,0,1)\,.
\end{split}
\end{equation}
The $\mathcal{O}(1/L^3)$ correction to the point-like decay rate takes a particularly simple form. Denoting the coefficient of the $1/(m_P L)^3$ term in the expansion by $d_{3}$, we obtain
\begin{eqnarray}
d_3
=  - \frac{4  \,(2+r_\ell^2)}{(1+r_\ell^2)^3} \, .
\end{eqnarray}
This matches the corrected result of Ref.~\cite{Tantalo:2016vxk}, which removes a typo, discovered with the help of this cross-check, from a previous version.

  \section{Numerical results}
  \label{sec:num}
  In this section we discuss numerically the FVEs derived in~\cref{sec:se,sec:kl2} and estimate the size of the structure-dependent effects. 

\subsection{Self-energy}
\begin{figure}
\centering
    \includegraphics{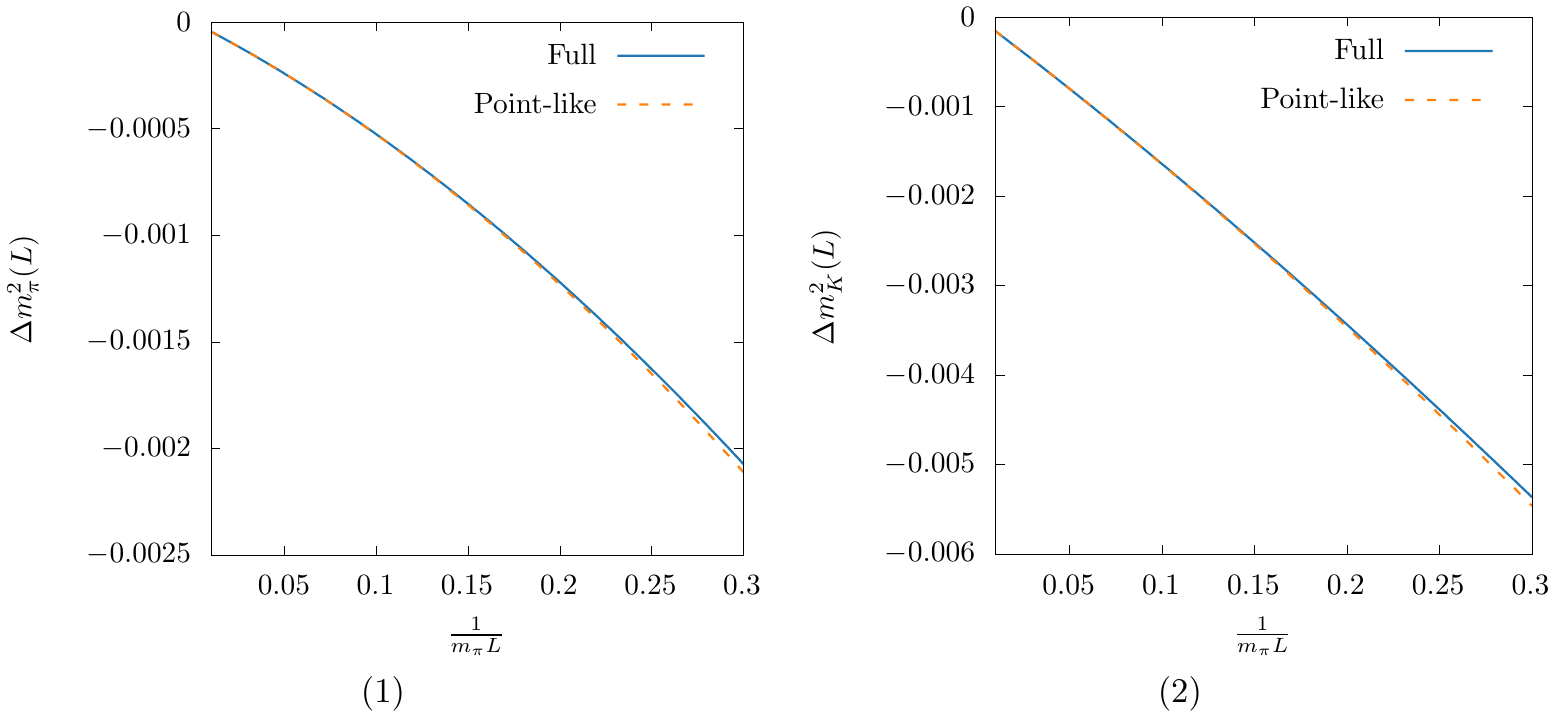}
    \caption{The finite-size scaling in the self-energies of (1) pions and (2) kaons, respectively. The two curves show the point-like result as well as the structure-dependent (SD) one.}
    \label{fig:selfenergies}
\end{figure}

Here we consider the FVEs in~\cref{eq:fveffmass} for the masses of both pions and kaons. However, since at present we have no numerical estimate for the branch-cut contribution $\mathcal{C}$, we have here explicitly put it to zero. As explained in Sec.~\ref{sec:se}, the branch-cut contribution is symptomatic of the non-locality of $\qedl$ and will require more investigation in the future. The
numerical values needed are the meson masses $m_{P}$ and the charge radii $\left\langle r_{P}^2 \right\rangle = 6\, F'(0)$. The charge
radius is a structure-dependent quantity and can e.g.~be measured in
experiments~\cite{10.1093/ptep/ptaa104}, calculated with the help of dispersion theory, see
e.g.~Refs.~\cite{Ananthanarayan:2017efc,Colangelo:2018mtw}, or computed on the
lattice~\cite{Aoki:2019cca}. Here we use the following experimental values from the PDG~\cite{10.1093/ptep/ptaa104}, namely
\begin{align}\label{eq:sevalues}
	m_{\pi^{-}} & = 0.13957039 (18)\,  \textrm{GeV}
\, , 
\; \; \; 
&	\left\langle r_{\pi}^2 \right\rangle  &= 11.19 (0.15)\,  \textrm{GeV}^{-2} 
	\, ,
	\nonumber \\
	m_{K^{-}}&= 0.493677(16) \,  \textrm{GeV}
\, , 
\; \; \; 
&	\left\langle r_{K}^2 \right\rangle  &= 8.08 (1.13)\,  \textrm{GeV}^{-2} 
\, .
\end{align}
In~\cref{fig:selfenergies} we show the FVEs to pion and kaon
self-energies as functions of $(m_{\pi}L)^{-1}$ using only the experimental central values above and values for the finite-size coefficients obtained as in Appendix~\ref{sec:cj}. 
We notice that adding the structure-dependent $1/L^3$-term (with $\mathcal{C}=0$) only generates percent-level deviations from the result through order $1/L^2$. These effects are expected to be of the same order of magnitude as the neglected exponential effects $\bigo(e^{-m_\pi L})$ in typical lattice calculations with $m_\pi L\simeq 4$.

\subsection{Leptonic decays}
Here we numerically study the FVEs derived in~\cref{sec:kl2} for the leptonic decay rates of pions and kaons in the muon channel, i.e.~$P^{-}\rightarrow \mu ^{-}\bar{\nu} _{\mu}$. We make a cross-check of our point-like results with the previous calculation in Ref.~\cite{Lubicz:2016xro}, and in addition compare the relative sizes of $Y^{(2)}(L)$ and $Y^{(1)}(L)$ for pions and kaons.
As an example, we choose the velocity orientation of the lepton to be $\hat{\mathbf{v}}_{\ell} = (1,1,1)/\sqrt{3}$, but this does not affect the overall conclusions.

\subsubsection{Pion decays}
We here compute the FVEs on pion decays using the data from the PDG~\cite{10.1093/ptep/ptaa104} in~\cref{eq:sevalues} together with
\begin{align}
	& 
 m_{\mu} = 0.1056583745(031) \, \textrm{GeV}
	\, , 
\; \; \; 
	f_{\pi^{-}}=0.1307(37) \,  \textrm{GeV}
	\, ,
\; \; \; 
	F_A^{\pi} = 0.0119(1)
\, .
\end{align}
The value of the form factor $F_A^\pi$ is taken from experimental measurements, and it is in good agreement with ChPT and lattice values~\cite{Cirigliano:2011ny,Desiderio:2020oej}. Uncertainties on these quantities are sufficiently small to be safely neglected here.

\begin{figure}
    \centering
    \includegraphics{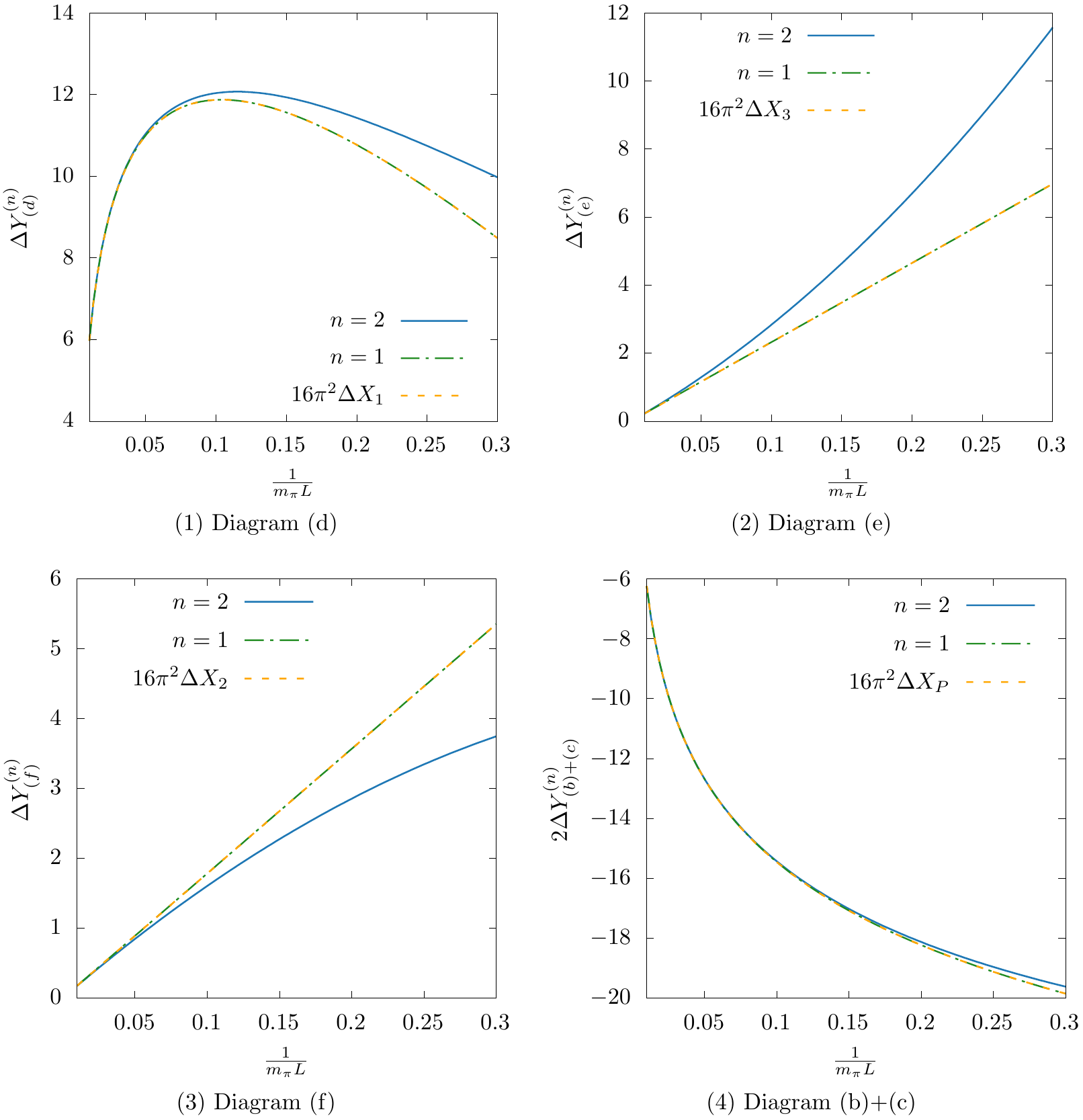}
    \caption{The FV scaling of the indicated diagrams for pions, this in comparison to the purely point-like $\Delta X_{i}$ defined in Ref.~\cite{Lubicz:2016xro}. Included is also the full $1/L^2$-contribution derived herein. Note that $f_n=z_n=0$ here.  }
    \label{fig:comparerm123}
\end{figure}

We first perform a cross-check by comparing our results to Ref.~\cite{Lubicz:2016xro}. In particular, we start by comparing the finite-size scaling in $L$ of the quantities $\Delta X_{i}$ in Eqs.~(\ref{eq:DeltaX1})--(\ref{eq:DeltaXP}) with our $\Delta Y_{(i)}^{(1)}$ using the matching in Eqs.~(\ref{eq:DeltaX1Check})--(\ref{eq:DeltaXPCheck}) and setting $f_n=z_n=0$. The results are reported in~\cref{fig:comparerm123}, and show a complete agreement.
One can notice that the $\bigo(1/L^2)$ term gives sizeable contributions already at $m_{\pi}L \sim 4$. 

\begin{figure}
    \centering
    \includegraphics{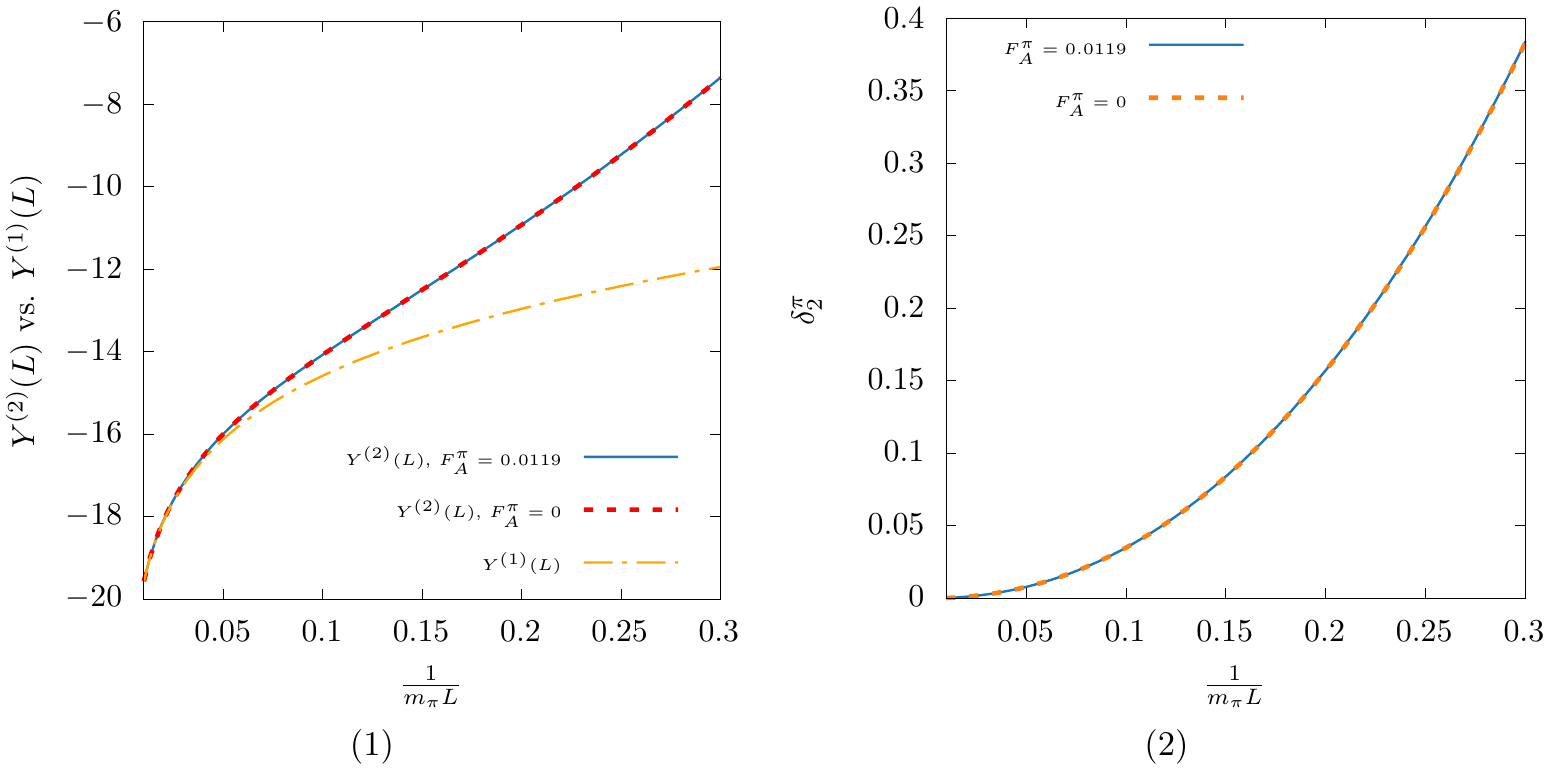}
    \caption{A comparison of the $1/L^2$-corrections to those through order $1/L$ for pions: (1) The structure-dependent function $Y^{(2)}(L)$ compared to the purely point-like $Y^{(1)}(L)$. (2) The relative correction $\delta _{2}^{\pi}$.}\label{fig:piony2}
\end{figure}

In~\cref{fig:piony2}(1) we plot $Y^{(2)}(L)$ and $Y^{(1)}(L)$.
In addition, we include the point-like limit $Y^{(2)}_{\textrm{pt}}(L)$ setting $F_{A}^{\pi} = 0$, and notice that the structure-dependent contribution at $\mathcal{O}(1/L^2)$ is negligible with respect to the point-like one. In total, there is a large effect from the $1/L^2$ contributions already at $m_{\pi}L\sim 4$. In~\cref{fig:piony2}(2) we look at the relative size of the $1/L^2$ correction to that at order $1/L$, defined in terms of the measure
\begin{align}
\delta _{2}^{P} =\left|  \frac{Y^{(2)}(L)-Y^{(1)}(L) }{Y^{(1)}(L)} \right|  \, .
\end{align}
It is clear that the terms at $1/L^2$ are essential already for moderately sized $m_{\pi}L$. 

\subsubsection{Kaon decays}
We next consider kaon decays. The additional data taken from the PDG~\cite{10.1093/ptep/ptaa104} is 
\begin{align}
	& 
	f_{K^{-}}=0.1550(19) \,  \textrm{GeV}
\, ,
\; \;
	F_{A,\, \mathrm{ChPT}}^K = 0.034
\, ,
\nonumber \\
& 
| F_{A, \, \mathrm{Exp}}^K + F_{V, \, \mathrm{Exp}}^K |  =0.165(13)
\, ,
\; \; 
| F_{A, \, \mathrm{Exp}}^K - F_{V, \, \mathrm{Exp}}^K |  = -0.153(33) \, .
\end{align}
The value $F_{A,\, \mathrm{ChPT}}^K$ is the ChPT prediction at order $\mathcal{O}(p^6)$. The two combinations $F_{A, \, \mathrm{Exp}}^K \pm F_{V, \, \mathrm{Exp}}^K$ are instead the PDG averages of measurements in muon experiments, and solving for  $F_{A, \, \mathrm{Exp}}^K $ we find
\begin{align}
F_{A, \, \mathrm{Exp}}^K  = 0.0060(177) \, . 
\end{align}
Note that this disagrees with the ChPT prediction~\cite{10.1093/ptep/ptaa104}. The form factors $F_{A }^K$ and $F_{V}^K$ have also recently been calculated for the first time on the lattice~\cite{Desiderio:2020oej}, and the result for $F_{A} ^K$ is
\begin{align}
F_{A, \, \mathrm{Latt }} ^K=0.0370(88) \, .
\end{align}
Again there is a discrepancy between theory and experiment. This is thoroughly discussed in Ref.~\cite{Frezzotti:2020bfa}, with the conclusion that future experimental and theoretical efforts are needed to study the apparent tension. However, although there are higher order corrections to the ChPT prediction, the practical prospects of improving the value from ChPT are very limited, mainly due to the lack of knowledge of the many low-energy constants at order $\mathcal{O}(p^8)$~\cite{Bijnens:2018lez,Hermansson-Truedsson:2020rtj,Dai:2020cpk,Graf:2020yxt}. In the following, we study $Y^{(2)}(L)$ using all the three values for $F_{A} ^K$ quoted above and compare the respective impacts on the FVEs. 

\begin{figure}
    \centering
    \includegraphics{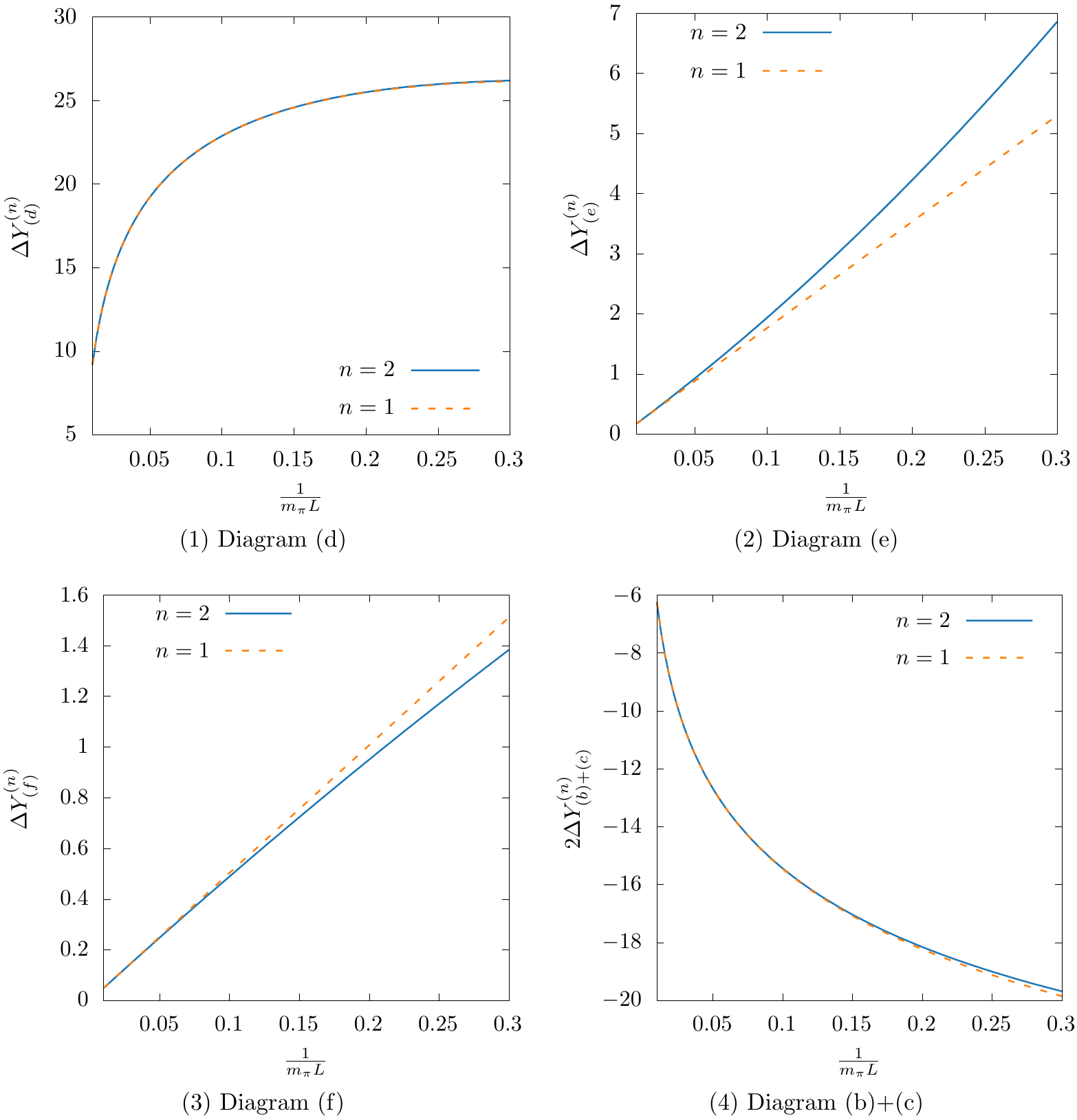}
    \caption{The FV scaling of the indicated diagrams for kaons. }
    \label{fig:kaondiagrams}
\end{figure}

We start by considering the contributions from the various diagrams in~\cref{fig:kaondiagrams}, using here $F_{A,\, \mathrm{ChPT}}^K $. Note that we set $z_n=f_n=0$ since these are are unknown, unphysical quantities that cancel in the end in the sum of diagrams. We see that for diagrams (d) and (b)+(c), the logarithmic terms are completely dominating, whereas for diagrams (e) and (f) the $1/L^2$ terms are sizeable.

\begin{figure}
        \centering
    \includegraphics{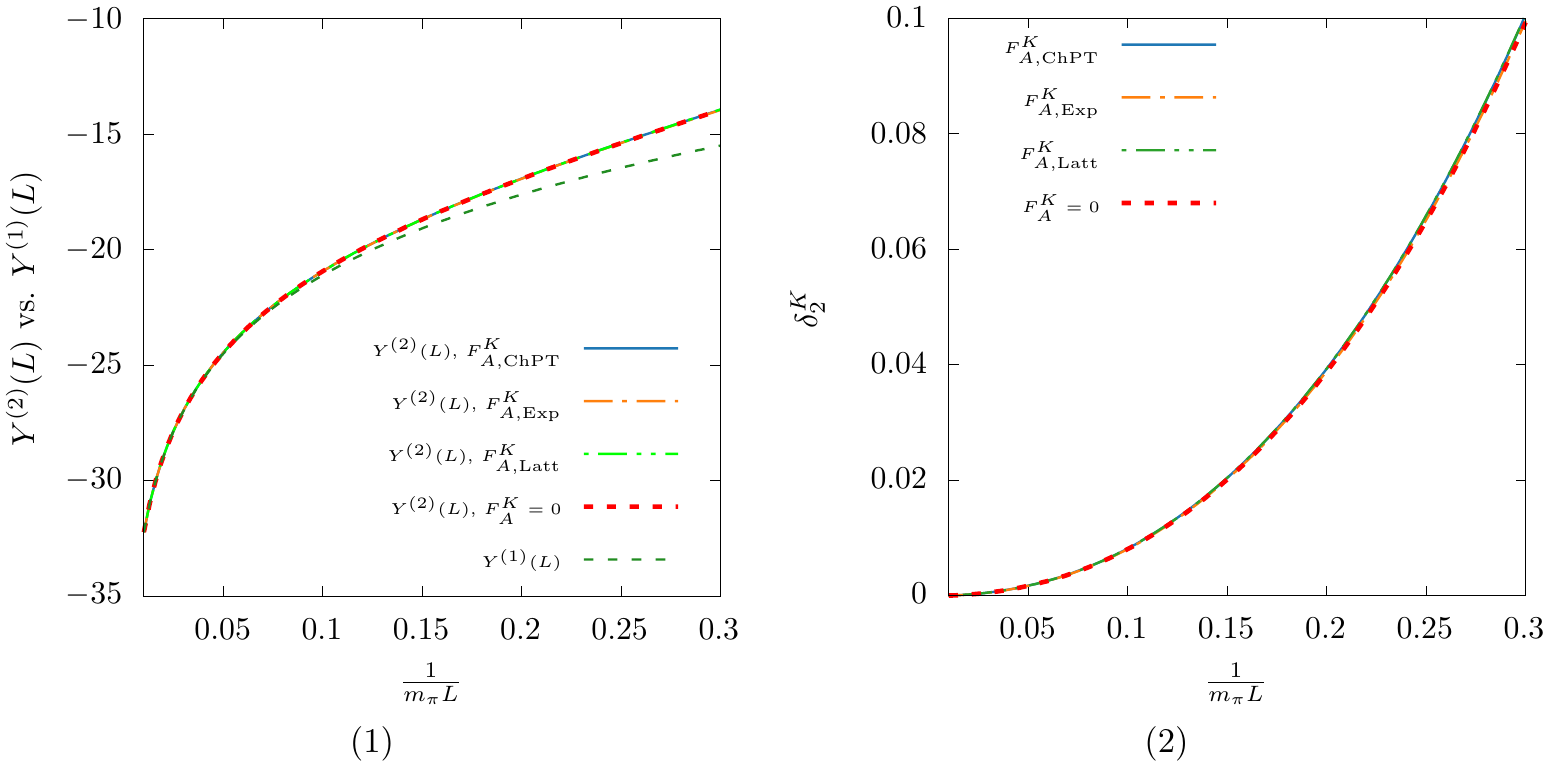}
   \caption{A comparison of the $1/L^2$ corrections to those at $1/L$ for kaons: (1) The structure-dependent function $Y^{(2)}(L)$ compared to the purely point-like $Y^{(1)}(L)$. For $Y^{(2)}(L)$ we have used values for $F_{A}^{K}$ from ChPT, experiments and the lattice as well as put it to zero. (2) The relative correction $\delta _{2}^{K}$, again for four different values of $F_{A}^{K}$. }\label{fig:kaony2}
\end{figure}

The total FVEs through order $1/L^2$ are shown in Fig.~\ref{fig:kaony2}(1) where we compare $Y^{(2)}(L)$ to the point-like $Y^{(1)}(L)$.  Here we use the central values of $F_{A,\, \mathrm{ChPT}}^K$, $F_{A,\, \mathrm{Exp}}^K$ and $F_{A,\, \mathrm{Latt}}^K$ as well as $F_{A}^K = 0$. Just as for pions, we see that setting $F_{A}^K = 0$ does not change the result and nor does a variation within errors for $F_{A,\, \mathrm{ChPT}}^K$, $F_{A,\, \mathrm{Exp}}^K$ and $F_{A,\, \mathrm{Latt}}^K$. Comparing to the decays of pions, the effect of the $1/L^2$ correction is here milder. 
The relative size of the $1/L^2$ term in $Y^{(2)}(L)$ to $Y^{(1)}(L)$ is shown in Fig.~\ref{fig:kaony2}(2) in terms of $\delta _{2}^{K}$, again for different values of $F_{A}^K$. The size of the $F_{A}^K$--dependent term in $Y^{(2)}(L)$ is found to be around the per-cent level for $m_{\pi}L\sim 2$ and it decreases for smaller $m_{\pi}L$. 

  \section{Conclusion}
  \label{sec:conclusions}
In this work we have developed a relativistic and model-independent approach for the determination of electromagnetic finite-size effects. In particular, the method presented here can go beyond the point-like approximation, which until now has proven to be a major stumbling block for more complicated observables such as leptonic decay rates. The defining strategy of the present approach is to decompose the scattering kernel of interest into irreducible vertex functions depending only on on-shell form factors. Similar methods were used in Refs.~\citep{Borsanyi:2014jba,Lucini:2015hfa,Lubicz:2016xro} to demonstrate the universality of the two leading orders in the $1/L$ expansion of EM FVEs for scalar and fermion masses and the leptonic decay width. In Ref.~\cite{Lucini:2015hfa} the authors also considered the higher-order structure-dependent contributions to masses, in the QED$_{\mathcal{C}}$ formulation.

In the expressions presented in this article, the structure dependence enters via physical quantities such as electromagnetic charge radii, polarisabilities and other form factors generally measurable in experiments or on the lattice. We also identify a branch-cut-induced $\bigo(1/L^3)$ effect for the pseudoscalar mass, that can be expressed as a physical spectral integral of the pseudoscalar's Compton amplitude. This contribution is generated directly by the non-locality of $\qedl$ and analogous terms are expected to contribute to any observable at $\bigo(1/L^3)$. In the case of the pseudoscalar mass, the Compton amplitude contribution appears together with a second term at $\bigo(1/L^3)$, proportional to the charge radius of the hadron that exactly matches previous determinations using non-relativistic effective field theories~\citep{Davoudi:2014qua,Lee:2015rua}. 
Regarding leptonic decays, we find that the leading $\bigo(1/L^2)$ structure-dependent effect is proportional to the constant $F_A$ from real radiative decays which can be determined directly in lattice QCD calculations as demonstrated in Ref.~\citep{Desiderio:2020oej} and also experimentally.

In Sec.~\ref{sec:num}, we have estimated the sizes of these corrections using experimental and phenomenological inputs for the structure dependence and typical lattice volumes. In the case of the pseudoscalar mass, the radius correction is found to be mild for light pseudoscalars. However, for leptonic decays, the structure-dependent effects are found to be non-negligible and will play an important role in determinations of CKM matrix elements in lattice QCD+QED with sub-percent accuracy. The finite-size effects presented here are generally expressed in terms of finite-volume coefficients $c_j$, depending on the velocities involved in a given amplitude, which are similar to generalized zeta functions used in finite-volume scattering. We have extended the definition of these coefficients from Ref.~\citep{Davoudi:2018qpl} to infrared-divergent finite-volume sums, and provided a numerically efficient algorithm to evaluate them.

Beyond pseudoscalar masses and leptonic decay rates, our method is general and systematic and can be applied to more complicated observables. The method is also particularly well-suited for autmation using a computer algebra system. We have illustrated this point by releasing a \textsc{Mathematica} notebook~\cite{klfv-zenodo} containing most of the analytic results presented here. Possible future applications include the self-energy of baryons, radiative corrections to pseudoscalar meson semi-leptonic decay rates, and corrections to multi-hadron scattering.

  \begin{acknowledgments}
     A.P.~would like to specially thank Martin Savage for his precious input in
     the early stages of this project. A.P.~additionally thanks Zohreh Davoudi as well as members of the
     RBC-UKQCD collaboration for useful discussions. A.P.~would
     like to thank both the Institute of Nuclear Theory
     at the University of Washington and the Albert Einstein Center for
     Fundamental Physics at the University of Bern for their warm hospitality during visits that played a crucial role for the completion of this work. Similarly, N.~H.--T.~wishes to thank
     the Higgs Centre for Theoretical Physics at the University of Edinburgh for
     hospitality during visits to work on this project. Important parts of
     this work have been completed during these visits. 
     M.T.H.~would like to thank Tim Harris for
     useful discussions and all authors would like to warmly thank Nazario Tantalo
     for his critical read of the manuscript before its first release.
     M.D.C., M.T.H., and A.P.~are supported in part by UK STFC grant ST/P000630/1. Additionally M.T.H.~is
     supported by UKRI Future Leader Fellowship MR/T019956/1. A.P.~additionally
     received funding from the European Research Council (ERC) under the
     European Union's Horizon 2020 research and innovation programme under grant
     agreements No 757646 \& 813942. N.~H.--T.~is funded by the Albert Einstein
     Center for Fundamental Physics at the University of Bern.
  \end{acknowledgments}
  \appendix
  \section{Finite-volume coefficients}
  \label{sec:cj}
  What we define as finite-volume coefficients is a class of special cases of
  $\gamma_{j,k}(\{\psp\};\xi)$ in~\cref{eq:gammajk} which appear frequently while
  computing finite-size effect for typical Feynman integrands. The IR regulator $\xi = \lambda L /(2\pi)$ with $\lambda$ being a photon mass. More specifically,
  these coefficients are a special case of~\cref{eq:gammajk} with
  \begin{equation}\label{eq:ddef}
    f_k(\hat{n}_{\xi},\{\psp\})=d(\hat{\nsp}_\xi;\{\vel\})=\prod_{\{\vel\} }\frac{1}
    {1-\vel\cdot\hat{\nsp}_\xi}\,,
  \end{equation}
  where $\{\vel\}$ is the set of velocities associated with the external momenta $\{\psp\}$,
  and $\hat{\nsp}_\xi$ is the spatial part of $\hat{n}_\xi$ defined in~\cref{eq:defnxi}
  \begin{equation}
    \hat{\nsp}_\xi=\frac{\nsp}{\omega_{\xi}(\nsp)}=\frac{|\nsp|}{\omega_{\xi}(\nsp)}\hat{\nsp}\,.
  \end{equation}
  We denote these coefficients $b_j(\{\vel\};\xi)$
  \begin{equation}
    b_j(\{\vel\};\xi)=\Delta_{\nsp}'\left[\frac{d(\hat{\nsp}_\xi;\{\vel\})}{\omega_{\xi}(\nsp)^{j}}
    \right]\,,
  \end{equation}
  and we additionally define $c_j(\{\vel\})$ to be the finite part in the $\xi\to 0$ limit.
  This last definition is ambiguous in the case of $\log(\xi)$ divergences, and we make in this
  section an explicit choice for it. In the special case where the set $\{\vel\}$ is defined by
  $k$ copies of the same velocity $\vel$, we denote the associated coefficients
  $b_{j,k}(\vel;\xi)$ and $c_{j,k}(\vel)$.
  \subsection{Infrared-finite coefficients}
  These are the coefficients with $j<3$. As discussed in~\cref{sec:sum}
  these coefficients can be evaluated directly at $\xi=0$, giving
  \begin{equation}
    b_j(\{\vel\};\xi=0)=c_j(\{\vel\})=
    \Delta_{\nsp}'\left[\frac{d(\hat{\nsp};\{\vel\})}{|\nsp|^{j}}
    \right]\,.
  \end{equation}
  These are the coefficients discussed in detail in Ref.~\cite{Davoudi:2018qpl}.
  \subsection{Coefficients with power infrared divergences}
  These are the coefficients with $j>3$. Here the finite part $c_j(\{\vel\})$
  is simply given by the finite sum in~\cref{eq:defgammabar}
  \begin{equation}
    c_j(\{\vel\})=\sump_{\nsp\in\Z^3}\frac{d(\hat{\nsp};\{\vel\})}{|\nsp|^{j}}\,,
  \end{equation}
  which will need to be evaluated numerically, and one also needs to compute the integral in~\cref{eq:defphijk}
  \begin{equation}
    \phi_j(\{\vel\})=\int_0^{+\infty}\diff n\,\int_{S^2}\diff^2\hat{\nsp}\, 
    \frac{n^2}{(1+n^2)^{\frac{j}{2}}}\, 
    d\left(\frac{n}{\sqrt{1+n^2}}\hat{\nsp};\{\vel\}\right)
    \,.
  \end{equation}
  Let us look explicitly at the case where $\{\vel\}$ is containing
  $k$ copies of the same velocity $\vel$. Under this assumption and using the definition of $d(\hat{\nsp};\{\vel\})$ in~\cref{eq:ddef} the
  integral above becomes
  \begin{equation}
    \phi_{j,k}(\vel)=\int_0^{+\infty}\diff n\,\int_{S^2}\diff^2\hat{\nsp}\, 
    \frac{n^2}{(1+n^2)^{\frac{j}{2}}}
    \frac{1}{(1-\frac{n}{\sqrt{1+n^2}}\hat{\nsp}\cdot\vel)^k}
    \, .
  \end{equation}
 Here we may freely rotate so that $\vel$ lies along the $z$-axis of the spherical coordinates to obtain
  \begin{equation}
    \phi_{j,k}(\vel)=2\pi\int_0^{+\infty}\diff n\,\int_{-1}^1\diff c\,
    \frac{n^2}{(1+n^2)^{\frac{j}{2}}}
    \frac{1}{(1-\frac{n}{\sqrt{1+n^2}}c|\vel|)^k}\,.
    \label{eq:phijk}
  \end{equation}
  Let us follow with the change of variables $x=\frac{n}{\sqrt{1+n^2}}$
  \begin{equation}
    \phi_{j,k}(\vel)=2\pi\int_0^{1}\diff x\,\int_{-1}^1\diff c\,
    \frac{x^2(1-x^2)^{\frac{j-5}{2}}}{(1-xc|\vel|)^k}\,,
  \end{equation}
  which can be explicitly evaluated to give
  \begin{equation}
    \phi_{j,k}(\vel)=\pi^{\frac{3}{2}}\frac{\Gamma\left(\frac{j-3}{2}\right)}{\Gamma\left(\frac{j}{2}\right)}
    \,{}_2F_1\left(\frac{k}{2},\frac{k+1}{2};\frac{j}{2};\vel^2\right)\,,
  \end{equation}
  where ${}_2F_1$ is a hypergeometric function defined in the usual way. Putting
  everything together,
  \begin{equation}
    b_{j,k}(\vel;\xi)=c_{j,k}(\vel)-
    \frac{\pi^{\frac{3}{2}}}{\xi^{j-3}}\frac{\Gamma\left(\frac{j-3}{2}\right)}{\Gamma\left(\frac{j}{2}\right)}
    \,{}_2F_1\left(\frac{k}{2},\frac{k+1}{2};\frac{j}{2};\vel^2\right)\,.
  \end{equation}
  \subsection{Coefficients with logarithmic infrared divergences}
  These are the coefficients with $j=3$. Reusing the form~\cref{eq:gamma3jform},
  we define $b_3(\{\vel\})$ as
  \begin{equation}
    b_3(\{\vel\};\xi)=c_3(\{\vel\})+4\pi A(\{\vel\})\log(\xi)-B(\{\vel\})\,,
  \end{equation}
  with
  \begin{align}
    A(\{\vel\})&=\frac{1}{4\pi}\int_{S^2}\diff^2\hat{\nsp}\,d(\hat{\nsp};\{\vel\})\,, \\ 
    B(\{\vel\})&=\int_0^{+\infty}\diff n\,\int_{S^2}\diff^2\hat{\nsp}\, 
    \frac{n^2\left[d\left(\frac{n}{\sqrt{1+n^2}}\hat{\nsp};\{\vel\}\right)-d(\hat{\nsp};\{\vel\})\right]}{(1+n^2)^{\frac{3}{2}}}-[1-\log(2)]A(\{\vel\})
    \,,\\
\label{eq:c3def}
    c_3(\{\vel\})&=\lim_{R\to+\infty}\left[\sump_{|\nsp|<R}\frac{d(\hat{\nsp};\{\vel\})}{|\nsp|^{3}}-4\pi A(\{\vel\})\log(R)\right]\,.
  \end{align}
  In the case where $\{\vel\}$ is containing
  $k$ copies of the same velocity $\vel$, one can evaluate explicitly $A_k(\vel)=A(\{\vel\})$ and $B_k(\vel)=B(\{\vel\})$.
  Let us start by $A_k(\vel)$
  \begin{equation}
    A_k(\vel)=\frac{1}{4\pi}\int_{S^2}\diff^2\hat{\nsp}\,\frac{1}{(1-\hat{\nsp}\cdot\vel)^k}
    =\frac{1}{2|\vel|(k-1)}
    \left[
    \left(\frac{1}{1-|\vel|}\right)^{k-1}-
    \left(\frac{1}{1+|\vel|}\right)^{k-1}\right]\,,
  \end{equation}
  which in the $k\to 1$ limit takes the form
  \begin{equation}
  \label{eq:A1v}
    A_1(\vel)=\frac{\arctanh(|\vel|)}{|\vel|}\,.
  \end{equation}
  Again letting $\hat{\vel}$ be along the z-axis one finds that the constant term $B_k(\vel)$ is given by
  \begin{equation}
    B_k(\vel)=2\pi\int_0^{+\infty}\diff n\,\int_{-1}^{1}\diff c\, 
    \frac{n^2\left[(1-\frac{n}{\sqrt{1+n^2}}c|\vel|)^{-k}-(1-c|\vel|)^{-k}\right]}{(1+n^2)^{\frac{3}{2}}}-[1-\log(2)]4\pi A_k(\vel)\, .
  \end{equation}
  The change of variables $x=\frac{n}{\sqrt{1+n^2}}$ can be used again to
  obtain
  \begin{equation}
    B_k(\vel)=2\pi\int_0^{1}\diff x\,\int_{-1}^{1}\diff c\,
    \frac{x^2[(1-c|\vel|x)^{-k}-(1-c|\vel|)^{-k}]}{1-x^2}-[1-\log(2)]4\pi A_k(\vel)\,,
  \end{equation}
  which can be explicitly evaluated 
 \begin{align}
B_k(\vel)= 
& \, 
\frac{\pi}{(1-k) |\vel|}
\Bigg[ \left((1-|\vel|)^{1-k}-(1+|\vel|)^{1-k}\right)
   \left(H_{k-1}+\frac{2 k}{1-k}+\log (2 |\vel|)\right)
   \\ \nonumber
   &\, 
   -(1-|\vel|)^{1-k} \log
   (1-|\vel|)+(1+|\vel|)^{1-k} \log (1+|\vel|)
     \\ \nonumber
   &\, 
   +e^{-i \pi  k} (1-|\vel|)^{1-k} \, 
   \mathrm{B} \left( \frac{|\vel|-1}{2 |\vel|}; k,1-k\right) 
   \\ \nonumber
   &\, 
   +(1+|\vel|)^{1-k} \left(i \pi -e^{-i \pi  k}
   \, \mathrm{B} \left( \frac{1+|\vel|}{2 |\vel|}; k,1-k\right) \right)
   \Bigg]
   \, ,
  \end{align}
  where $\mathrm{B}(z; a,b)$ is the incomplete $\beta$--function
  \begin{align}
      \mathrm{B}(z; a,b) = \int _{0}^{z} \diff u \, u^{a-1}(1-u)^{b-1} \, ,
  \end{align}
  and $H_{k} = \sum _{n=1}^{k}1/n$ is the $k$-th harmonic number. Note that the imaginary terms related to the branch-cut of the $\beta$--functions always cancel in the end in numerical evaluations. In the $k\to 1$ limit the last equation becomes
  \begin{equation}
    B_1(\vel)=\frac{\pi}{|\vel|}\left[  \mathrm{Li}_2\left(\frac{2 |\vel|}{|\vel|-1}\right)-  \mathrm{Li}_2\left(\frac{2
    |\vel|}{|\vel|+1}\right)+4\log(2) |\vel |  A_{1}(\vel ) 
    \right]\,,
  \end{equation}
  where $\mathrm{Li}_2(x)$ is the dilogarithm function. Note that no imaginary terms appear here. In the zero-momentum limit of $B_1(\vel)$ one finds
  \begin{align}
      \lim_{|\vel | \rightarrow 0} B_1(\vel) = -4\pi (1-\log 2) \, .
  \end{align}

\subsection{Numerical evaluation}
In Ref.~\citep{Davoudi:2018qpl}, a method was developed to compute the $c_j$ coefficients for $j<3$ by evaluating sums with a doubly exponential rate of convergence. In this section we show that this technique generalizes to $j\geq 3$ coefficients. We define the acceleration function
\begin{equation}
  f(\nsp)=1-\left(\tanh\{\sinh[|\nsp|
  d(\hat{\nsp};\{\vel \} )^{\frac{1}{j+2}}]\}\right)^{j+2}\,.
\end{equation}
For $j<3$, as demonstrated in Ref.~\citep{Davoudi:2018qpl}, one has the relationship
\begin{equation}
  c_j(\{\vel \})=
  \sump_{\nsp}\frac{f(\eta\nsp)}{|\nsp|^j}d(\hat{\nsp};\{\vel \})-
  4\pi\eta^{j-3}R_jA_{\frac{5}{j+2}}(\{\vel \})\,,
  \label{eq:jlt3num}
\end{equation}
up to corrections which vanish exponentially for $\eta\to 0$, and where
\begin{equation}
  R_j=\int_0^{+\infty}\diff r\,\frac{1-\tanh[\sinh(r)]^{j+2}}{r^{j-2}}\,.
  \label{eq:rj}
\end{equation}
\cref{eq:jlt3num} is very efficient to evaluate numerically $c_j$ at high precision. The sum converges with a double exponential rate and the integral $R_j$ is trivial to evaluate through standard quadrature methods. 

Let us consider $j=3$ in detail. The definition of $ c_{3}(\{\vel \})$ in\cref{eq:c3def} can be rewritten as
\begin{equation}
  c_{3}(\{\vel \}) =
  \lim_{R\to\infty}\left[
  \sump_{|\nsp|<R}
  \frac{d(\hat{\nsp}; \{\vel \}) \left[ 1-f(\eta \mathbf{n})+f(\eta \mathbf{n})\right]}{|\nsp|^3}
  -
  4\pi A_1(\{\vel \} )\log(R)\right]
  \,.
\end{equation}
We can further separate this expression into
\begin{equation}
  c_{3}(\{\vel \}) =
  \sump_{|\nsp|}\frac{d(\hat{\nsp}; \{\vel \}) f(\eta \mathbf{n}) }{|\nsp|^3}+ \lim_{R\to\infty}\left[
  \sump_{|\nsp|<R}
  \frac{d(\hat{\nsp}; \{\vel \}) \left[ 1-f(\eta \mathbf{n})\right]}{|\nsp|^3}
  - 
  4\pi A_1(\{\vel \} )\log(R)
  \right]
  \,.
\end{equation}
The first term on the right-hand side is separately UV-finite so the sum is left unconstrained. Next observe that the properties of $f(\eta \mathbf{n})$ allow us to exchange the UV-regulated sum with $1-f(\eta \mathbf{n})$ in the numerator for a UV-regulated and IR-finite integral up to exponentially small corrections. We thus obtain
\begin{equation}
  c_{3}(\{\vel \}) =
  \sump_{|\nsp|}\frac{d(\hat{\nsp}; \{\vel \})  f(\eta \mathbf{n})}{|\nsp|^3}+ \lim_{R\to\infty}\left[
  \int _{|\nsp|<R}
  \frac{d(\hat{\nsp}; \{\vel \}) \left[ 1-f(\eta \mathbf{n})\right]}{|\nsp|^3}
  - 
  4\pi A_1(\{\vel \} )\log(R)
  \right]
  \,.
\end{equation}
Finally, since the limit over $R$ is arbitrary we may switch $R\rightarrow R/\eta$ and perform a change of variables in the integral to give
\begin{align}
  c_{3}(\{\vel \}) 
  =
  \, &
  \sump_{|\nsp|}\frac{d(\hat{\nsp}; \{\vel \}) f(\eta \mathbf{n})}{|\nsp|^3}
  +4\pi A_1(\{\vel \} )\log(\eta)
  \nonumber \\ &
  + \lim_{R/\eta\to\infty}\left[
  \int _{|\nsp|<R}
  \frac{d(\hat{\nsp}; \{\vel \}) \left[ 1-f( \mathbf{n})\right]}{|\nsp|^3}
  - 
  4\pi A_1(\{\vel \} )\log(R)
  \right]
  \nonumber \\ 
  = \, &
  \sump_{|\nsp|}\frac{d(\hat{\nsp}; \{\vel \})  f(\eta \mathbf{n})}{|\nsp|^3}
  +4\pi A_1(\{\vel \} )\log(\eta) + Q_{3}(\{\vel \})
  \,.
\end{align}
We here defined 
\begin{align}\label{eq:Q3}
    Q_{3}(\{\vel \}) = \lim_{R/\eta\to\infty}\left[
  \int _{|\nsp|<R}
  \frac{d(\hat{\nsp}; \{\vel \}) \left[ 1-f( \mathbf{n})\right]}{|\nsp|^3}
  - 
  4\pi A_1(\{\vel \} )\log(R)
  \right]\, ,
\end{align}
and the expression again holds up to exponential corrections in $\eta$. 

For $j>3$, the same reasoning as above leads to the formula
\begin{equation}
  c_j(\vel_1,\dots,\vel_N)=
  \sump_{\nsp}\frac{f(\eta\nsp)}{|\nsp|^j}d(\hat{\nsp};\vel_1,\dots,\vel_N)
  +4\pi\eta^{j-3}\bar{R}_jA_{\frac{5}{j+2}}(\vel_1,\dots,\vel_N)\,,
\end{equation}
up to exponential corrections in $\eta$, and with
\begin{equation}
  \bar{R}_j=\int_0^{+\infty}\diff r\,\frac{\tanh[\sinh(r)]^{j+2}}{r^{j-2}}\,.
  \label{eq:rbarj}
\end{equation}

\begin{figure}[p]
  \centering
  \includegraphics{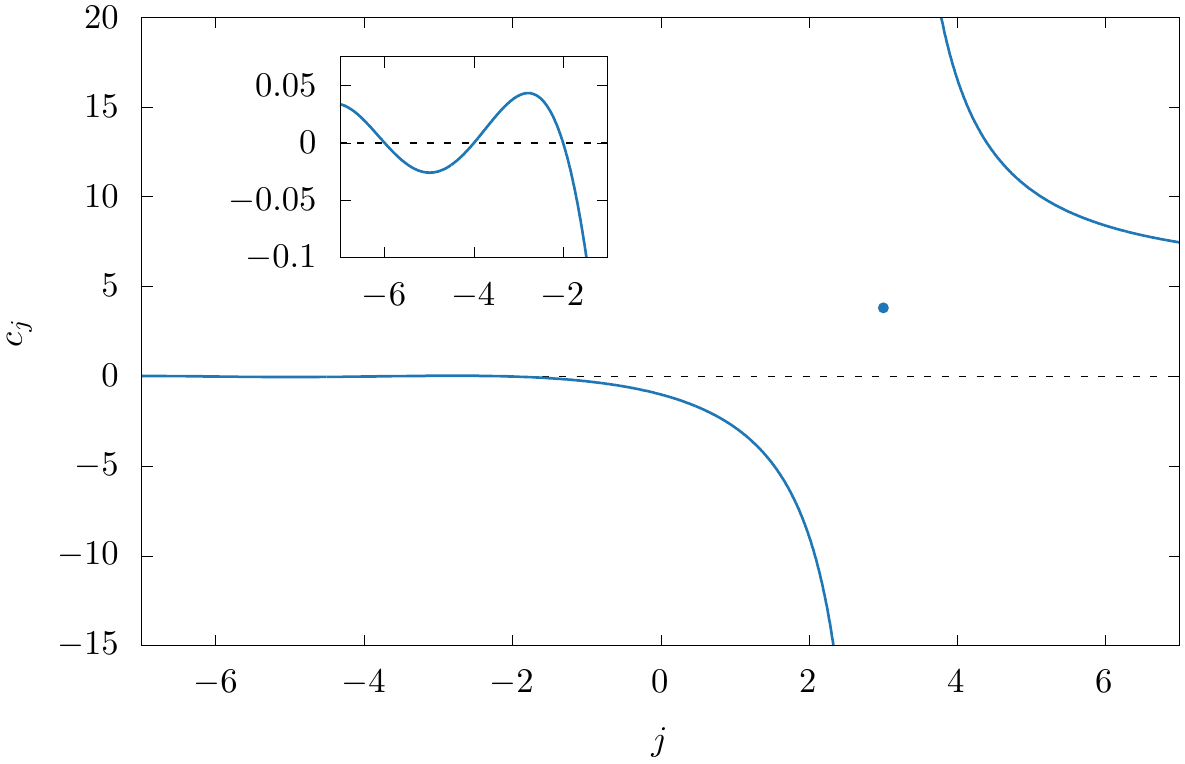}
  \caption{The rest-frame finite-volume coefficients $c_j$ as a function of
  $j$. The isolated dot is $c_3$ and the inset panel is a zoom on the 
  small oscillations in the $-7\leq j\leq -1$ region.}
  \label{fig:cj}
\end{figure}

Using the method described above, we plot in~\cref{fig:cj} the values of the rest-frame coefficients $c_j$ for $-7\leq j\leq 7$. The singularity at $j=3$ is clearly visible. We give in~\cref{tab:cj} explicit values of some of the rest frame coefficients and the constants in~\cref{eq:rj,eq:rbarj,eq:Q3} at zero velocity.
\begin{table}[p]
  \begin{tabular}{|c|c|c|c|}
    \hline
    $j$  & $c_j$      & $R_j$      & $\bar{R}_j$ \\ \hline
    $-5$ & $-0.02587$ & $-1.45833$ &             \\
    $-3$ & $0.04118$  & $-0.25447$ &             \\
    $-1$ & $-0.26660$ & $0.17191$  &             \\
    $0$  & $-1$       & $0.32289$  &             \\
    $1$  & $-2.83730$ & $0.52702$  &             \\
    $2$  & $-8.91363$ & $1.04744$  &             \\
    $3$  & $3.82192$  &            &             \\
    $4$  & $16.53232$ &            & $0.93684$   \\
    $5$  & $10.37752$ &            & $0.43505$   \\ \hline
    \end{tabular}
  \caption{Values of selected zero-velocity finite-volume coefficients and the integrals~\cref{eq:rj,eq:rbarj}. We also find $Q_{3}(0)= -0.730289$ from~\cref{eq:Q3}. }
  \label{tab:cj}
\end{table}

  \clearpage
  \bibliography{klfv}

\begin{thebibliography}{43}%
\makeatletter
\providecommand \@ifxundefined [1]{%
 \@ifx{#1\undefined}
}%
\providecommand \@ifnum [1]{%
 \ifnum #1\expandafter \@firstoftwo
 \else \expandafter \@secondoftwo
 \fi
}%
\providecommand \@ifx [1]{%
 \ifx #1\expandafter \@firstoftwo
 \else \expandafter \@secondoftwo
 \fi
}%
\providecommand \natexlab [1]{#1}%
\providecommand \enquote  [1]{``#1''}%
\providecommand \bibnamefont  [1]{#1}%
\providecommand \bibfnamefont [1]{#1}%
\providecommand \citenamefont [1]{#1}%
\providecommand \href@noop [0]{\@secondoftwo}%
\providecommand \href [0]{\begingroup \@sanitize@url \@href}%
\providecommand \@href[1]{\@@startlink{#1}\@@href}%
\providecommand \@@href[1]{\endgroup#1\@@endlink}%
\providecommand \@sanitize@url [0]{\catcode `\\12\catcode `\$12\catcode
  `\&12\catcode `\#12\catcode `\^12\catcode `\_12\catcode `\%12\relax}%
\providecommand \@@startlink[1]{}%
\providecommand \@@endlink[0]{}%
\providecommand \url  [0]{\begingroup\@sanitize@url \@url }%
\providecommand \@url [1]{\endgroup\@href {#1}{\urlprefix }}%
\providecommand \urlprefix  [0]{URL }%
\providecommand \Eprint [0]{\href }%
\providecommand \doibase [0]{https://doi.org/}%
\providecommand \selectlanguage [0]{\@gobble}%
\providecommand \bibinfo  [0]{\@secondoftwo}%
\providecommand \bibfield  [0]{\@secondoftwo}%
\providecommand \translation [1]{[#1]}%
\providecommand \BibitemOpen [0]{}%
\providecommand \bibitemStop [0]{}%
\providecommand \bibitemNoStop [0]{.\EOS\space}%
\providecommand \EOS [0]{\spacefactor3000\relax}%
\providecommand \BibitemShut  [1]{\csname bibitem#1\endcsname}%
\let\auto@bib@innerbib\@empty
\bibitem [{\citenamefont {Aoyama}\ \emph {et~al.}(2020)\citenamefont {Aoyama}
  \emph {et~al.}}]{Aoyama:2020ynm}%
  \BibitemOpen
  \bibfield  {author} {\bibinfo {author} {\bibfnamefont {T.}~\bibnamefont
  {Aoyama}} \emph {et~al.},\ }\bibfield  {title} {\bibinfo {title} {{The
  anomalous magnetic moment of the muon in the Standard Model}},\ }\href
  {https://doi.org/10.1016/j.physrep.2020.07.006} {\bibfield  {journal}
  {\bibinfo  {journal} {Phys. Rept.}\ }\textbf {\bibinfo {volume} {887}},\
  \bibinfo {pages} {1} (\bibinfo {year} {2020})},\ \Eprint
  {https://arxiv.org/abs/2006.04822} {arXiv:2006.04822 [hep-ph]} \BibitemShut
  {NoStop}%
\bibitem [{\citenamefont {Aoki}\ \emph {et~al.}(2020)\citenamefont {Aoki} \emph
  {et~al.}}]{Aoki:2019cca}%
  \BibitemOpen
  \bibfield  {author} {\bibinfo {author} {\bibfnamefont {S.}~\bibnamefont
  {Aoki}} \emph {et~al.} (\bibinfo {collaboration} {Flavour Lattice Averaging
  Group}),\ }\bibfield  {title} {\bibinfo {title} {{FLAG Review 2019: Flavour
  Lattice Averaging Group (FLAG)}},\ }\href
  {https://doi.org/10.1140/epjc/s10052-019-7354-7} {\bibfield  {journal}
  {\bibinfo  {journal} {Eur. Phys. J. C}\ }\textbf {\bibinfo {volume} {80}},\
  \bibinfo {pages} {113} (\bibinfo {year} {2020})},\ \Eprint
  {https://arxiv.org/abs/1902.08191} {arXiv:1902.08191 [hep-lat]} \BibitemShut
  {NoStop}%
\bibitem [{\citenamefont {Carrasco}\ \emph {et~al.}(2015)\citenamefont
  {Carrasco}, \citenamefont {Lubicz}, \citenamefont {Martinelli}, \citenamefont
  {Sachrajda}, \citenamefont {Tantalo}, \citenamefont {Tarantino},\ and\
  \citenamefont {Testa}}]{Carrasco:2015xwa}%
  \BibitemOpen
  \bibfield  {author} {\bibinfo {author} {\bibfnamefont {N.}~\bibnamefont
  {Carrasco}}, \bibinfo {author} {\bibfnamefont {V.}~\bibnamefont {Lubicz}},
  \bibinfo {author} {\bibfnamefont {G.}~\bibnamefont {Martinelli}}, \bibinfo
  {author} {\bibfnamefont {C.}~\bibnamefont {Sachrajda}}, \bibinfo {author}
  {\bibfnamefont {N.}~\bibnamefont {Tantalo}}, \bibinfo {author} {\bibfnamefont
  {C.}~\bibnamefont {Tarantino}},\ and\ \bibinfo {author} {\bibfnamefont
  {M.}~\bibnamefont {Testa}},\ }\bibfield  {title} {\bibinfo {title} {{QED
  Corrections to Hadronic Processes in Lattice QCD}},\ }\href
  {https://doi.org/10.1103/PhysRevD.91.074506} {\bibfield  {journal} {\bibinfo
  {journal} {Phys. Rev. D}\ }\textbf {\bibinfo {volume} {91}},\ \bibinfo
  {pages} {074506} (\bibinfo {year} {2015})},\ \Eprint
  {https://arxiv.org/abs/1502.00257} {arXiv:1502.00257 [hep-lat]} \BibitemShut
  {NoStop}%
\bibitem [{\citenamefont {Di~Carlo}\ \emph {et~al.}(2019)\citenamefont
  {Di~Carlo}, \citenamefont {Giusti}, \citenamefont {Lubicz}, \citenamefont
  {Martinelli}, \citenamefont {Sachrajda}, \citenamefont {Sanfilippo},
  \citenamefont {Simula},\ and\ \citenamefont {Tantalo}}]{DiCarlo:2019thl}%
  \BibitemOpen
  \bibfield  {author} {\bibinfo {author} {\bibfnamefont {M.}~\bibnamefont
  {Di~Carlo}}, \bibinfo {author} {\bibfnamefont {D.}~\bibnamefont {Giusti}},
  \bibinfo {author} {\bibfnamefont {V.}~\bibnamefont {Lubicz}}, \bibinfo
  {author} {\bibfnamefont {G.}~\bibnamefont {Martinelli}}, \bibinfo {author}
  {\bibfnamefont {C.}~\bibnamefont {Sachrajda}}, \bibinfo {author}
  {\bibfnamefont {F.}~\bibnamefont {Sanfilippo}}, \bibinfo {author}
  {\bibfnamefont {S.}~\bibnamefont {Simula}},\ and\ \bibinfo {author}
  {\bibfnamefont {N.}~\bibnamefont {Tantalo}},\ }\bibfield  {title} {\bibinfo
  {title} {{Light-meson leptonic decay rates in lattice QCD+QED}},\ }\href
  {https://doi.org/10.1103/PhysRevD.100.034514} {\bibfield  {journal} {\bibinfo
   {journal} {Phys. Rev. D}\ }\textbf {\bibinfo {volume} {100}},\ \bibinfo
  {pages} {034514} (\bibinfo {year} {2019})},\ \Eprint
  {https://arxiv.org/abs/1904.08731} {arXiv:1904.08731 [hep-lat]} \BibitemShut
  {NoStop}%
\bibitem [{\citenamefont {Desiderio}\ \emph {et~al.}(2021)\citenamefont
  {Desiderio} \emph {et~al.}}]{Desiderio:2020oej}%
  \BibitemOpen
  \bibfield  {author} {\bibinfo {author} {\bibfnamefont {A.}~\bibnamefont
  {Desiderio}} \emph {et~al.},\ }\bibfield  {title} {\bibinfo {title} {{First
  lattice calculation of radiative leptonic decay rates of pseudoscalar
  mesons}},\ }\href {https://doi.org/10.1103/PhysRevD.103.014502} {\bibfield
  {journal} {\bibinfo  {journal} {Phys. Rev. D}\ }\textbf {\bibinfo {volume}
  {103}},\ \bibinfo {pages} {014502} (\bibinfo {year} {2021})},\ \Eprint
  {https://arxiv.org/abs/2006.05358} {arXiv:2006.05358 [hep-lat]} \BibitemShut
  {NoStop}%
\bibitem [{\citenamefont {Frezzotti}\ \emph {et~al.}(2021)\citenamefont
  {Frezzotti}, \citenamefont {Garofalo}, \citenamefont {Lubicz}, \citenamefont
  {Martinelli}, \citenamefont {Sachrajda}, \citenamefont {Sanfilippo},
  \citenamefont {Simula},\ and\ \citenamefont {Tantalo}}]{Frezzotti:2020bfa}%
  \BibitemOpen
  \bibfield  {author} {\bibinfo {author} {\bibfnamefont {R.}~\bibnamefont
  {Frezzotti}}, \bibinfo {author} {\bibfnamefont {M.}~\bibnamefont {Garofalo}},
  \bibinfo {author} {\bibfnamefont {V.}~\bibnamefont {Lubicz}}, \bibinfo
  {author} {\bibfnamefont {G.}~\bibnamefont {Martinelli}}, \bibinfo {author}
  {\bibfnamefont {C.~T.}\ \bibnamefont {Sachrajda}}, \bibinfo {author}
  {\bibfnamefont {F.}~\bibnamefont {Sanfilippo}}, \bibinfo {author}
  {\bibfnamefont {S.}~\bibnamefont {Simula}},\ and\ \bibinfo {author}
  {\bibfnamefont {N.}~\bibnamefont {Tantalo}},\ }\bibfield  {title} {\bibinfo
  {title} {{Comparison of lattice QCD+QED predictions for radiative leptonic
  decays of light mesons with experimental data}},\ }\href
  {https://doi.org/10.1103/PhysRevD.103.053005} {\bibfield  {journal} {\bibinfo
   {journal} {Phys. Rev. D}\ }\textbf {\bibinfo {volume} {103}},\ \bibinfo
  {pages} {053005} (\bibinfo {year} {2021})},\ \Eprint
  {https://arxiv.org/abs/2012.02120} {arXiv:2012.02120 [hep-ph]} \BibitemShut
  {NoStop}%
\bibitem [{\citenamefont {Borsanyi}\ \emph {et~al.}(2015)\citenamefont
  {Borsanyi} \emph {et~al.}}]{Borsanyi:2014jba}%
  \BibitemOpen
  \bibfield  {author} {\bibinfo {author} {\bibfnamefont {S.}~\bibnamefont
  {Borsanyi}} \emph {et~al.},\ }\bibfield  {title} {\bibinfo {title} {{Ab
  initio calculation of the neutron-proton mass difference}},\ }\href
  {https://doi.org/10.1126/science.1257050} {\bibfield  {journal} {\bibinfo
  {journal} {Science}\ }\textbf {\bibinfo {volume} {347}},\ \bibinfo {pages}
  {1452} (\bibinfo {year} {2015})},\ \Eprint {https://arxiv.org/abs/1406.4088}
  {arXiv:1406.4088 [hep-lat]} \BibitemShut {NoStop}%
\bibitem [{\citenamefont {Davoudi}\ and\ \citenamefont
  {Savage}(2014)}]{Davoudi:2014qua}%
  \BibitemOpen
  \bibfield  {author} {\bibinfo {author} {\bibfnamefont {Z.}~\bibnamefont
  {Davoudi}}\ and\ \bibinfo {author} {\bibfnamefont {M.~J.}\ \bibnamefont
  {Savage}},\ }\bibfield  {title} {\bibinfo {title} {{Finite-Volume
  Electromagnetic Corrections to the Masses of Mesons, Baryons and Nuclei}},\
  }\href {https://doi.org/10.1103/PhysRevD.90.054503} {\bibfield  {journal}
  {\bibinfo  {journal} {Phys. Rev.}\ }\textbf {\bibinfo {volume} {D90}},\
  \bibinfo {pages} {054503} (\bibinfo {year} {2014})},\ \Eprint
  {https://arxiv.org/abs/1402.6741} {arXiv:1402.6741 [hep-lat]} \BibitemShut
  {NoStop}%
\bibitem [{\citenamefont {Lubicz}\ \emph {et~al.}(2017)\citenamefont {Lubicz},
  \citenamefont {Martinelli}, \citenamefont {Sachrajda}, \citenamefont
  {Sanfilippo}, \citenamefont {Simula},\ and\ \citenamefont
  {Tantalo}}]{Lubicz:2016xro}%
  \BibitemOpen
  \bibfield  {author} {\bibinfo {author} {\bibfnamefont {V.}~\bibnamefont
  {Lubicz}}, \bibinfo {author} {\bibfnamefont {G.}~\bibnamefont {Martinelli}},
  \bibinfo {author} {\bibfnamefont {C.~T.}\ \bibnamefont {Sachrajda}}, \bibinfo
  {author} {\bibfnamefont {F.}~\bibnamefont {Sanfilippo}}, \bibinfo {author}
  {\bibfnamefont {S.}~\bibnamefont {Simula}},\ and\ \bibinfo {author}
  {\bibfnamefont {N.}~\bibnamefont {Tantalo}},\ }\bibfield  {title} {\bibinfo
  {title} {{Finite-Volume QED Corrections to Decay Amplitudes in Lattice
  QCD}},\ }\href {https://doi.org/10.1103/PhysRevD.95.034504} {\bibfield
  {journal} {\bibinfo  {journal} {Phys. Rev.}\ }\textbf {\bibinfo {volume}
  {D95}},\ \bibinfo {pages} {034504} (\bibinfo {year} {2017})},\ \Eprint
  {https://arxiv.org/abs/1611.08497} {arXiv:1611.08497 [hep-lat]} \BibitemShut
  {NoStop}%
\bibitem [{\citenamefont {Davoudi}\ \emph {et~al.}(2019)\citenamefont
  {Davoudi}, \citenamefont {Harrison}, \citenamefont {Jüttner}, \citenamefont
  {Portelli},\ and\ \citenamefont {Savage}}]{Davoudi:2018qpl}%
  \BibitemOpen
  \bibfield  {author} {\bibinfo {author} {\bibfnamefont {Z.}~\bibnamefont
  {Davoudi}}, \bibinfo {author} {\bibfnamefont {J.}~\bibnamefont {Harrison}},
  \bibinfo {author} {\bibfnamefont {A.}~\bibnamefont {Jüttner}}, \bibinfo
  {author} {\bibfnamefont {A.}~\bibnamefont {Portelli}},\ and\ \bibinfo
  {author} {\bibfnamefont {M.~J.}\ \bibnamefont {Savage}},\ }\bibfield  {title}
  {\bibinfo {title} {{Theoretical aspects of quantum electrodynamics in a
  finite volume with periodic boundary conditions}},\ }\href
  {https://doi.org/10.1103/PhysRevD.99.034510} {\bibfield  {journal} {\bibinfo
  {journal} {Phys. Rev.}\ }\textbf {\bibinfo {volume} {D99}},\ \bibinfo {pages}
  {034510} (\bibinfo {year} {2019})},\ \Eprint
  {https://arxiv.org/abs/1810.05923} {arXiv:1810.05923 [hep-lat]} \BibitemShut
  {NoStop}%
\bibitem [{\citenamefont {Bijnens}\ \emph
  {et~al.}(2019{\natexlab{a}})\citenamefont {Bijnens}, \citenamefont
  {Harrison}, \citenamefont {Hermansson-Truedsson}, \citenamefont {Janowski},
  \citenamefont {Jüttner},\ and\ \citenamefont {Portelli}}]{Bijnens:2019ejw}%
  \BibitemOpen
  \bibfield  {author} {\bibinfo {author} {\bibfnamefont {J.}~\bibnamefont
  {Bijnens}}, \bibinfo {author} {\bibfnamefont {J.}~\bibnamefont {Harrison}},
  \bibinfo {author} {\bibfnamefont {N.}~\bibnamefont {Hermansson-Truedsson}},
  \bibinfo {author} {\bibfnamefont {T.}~\bibnamefont {Janowski}}, \bibinfo
  {author} {\bibfnamefont {A.}~\bibnamefont {Jüttner}},\ and\ \bibinfo
  {author} {\bibfnamefont {A.}~\bibnamefont {Portelli}},\ }\bibfield  {title}
  {\bibinfo {title} {{Electromagnetic finite-size effects to the hadronic
  vacuum polarization}},\ }\href {https://doi.org/10.1103/PhysRevD.100.014508}
  {\bibfield  {journal} {\bibinfo  {journal} {Phys. Rev. D}\ }\textbf {\bibinfo
  {volume} {100}},\ \bibinfo {pages} {014508} (\bibinfo {year}
  {2019}{\natexlab{a}})},\ \Eprint {https://arxiv.org/abs/1903.10591}
  {arXiv:1903.10591 [hep-lat]} \BibitemShut {NoStop}%
\bibitem [{\citenamefont {Hayakawa}\ and\ \citenamefont
  {Uno}(2008)}]{Hayakawa:2008an}%
  \BibitemOpen
  \bibfield  {author} {\bibinfo {author} {\bibfnamefont {M.}~\bibnamefont
  {Hayakawa}}\ and\ \bibinfo {author} {\bibfnamefont {S.}~\bibnamefont {Uno}},\
  }\bibfield  {title} {\bibinfo {title} {{QED in finite volume and finite size
  scaling effect on electromagnetic properties of hadrons}},\ }\href
  {https://doi.org/10.1143/PTP.120.413} {\bibfield  {journal} {\bibinfo
  {journal} {Prog. Theor. Phys.}\ }\textbf {\bibinfo {volume} {120}},\ \bibinfo
  {pages} {413} (\bibinfo {year} {2008})},\ \Eprint
  {https://arxiv.org/abs/0804.2044} {arXiv:0804.2044 [hep-ph]} \BibitemShut
  {NoStop}%
\bibitem [{\citenamefont {Wiese}(1992)}]{WIESE199245}%
  \BibitemOpen
  \bibfield  {author} {\bibinfo {author} {\bibfnamefont {U.-J.}\ \bibnamefont
  {Wiese}},\ }\bibfield  {title} {\bibinfo {title} {C- and {G}-periodic {QCD}
  at finite temperature},\ }\href
  {https://doi.org/https://doi.org/10.1016/0550-3213(92)90333-7} {\bibfield
  {journal} {\bibinfo  {journal} {Nuclear Physics B}\ }\textbf {\bibinfo
  {volume} {375}},\ \bibinfo {pages} {45 } (\bibinfo {year}
  {1992})}\BibitemShut {NoStop}%
\bibitem [{\citenamefont {Kronfeld}\ and\ \citenamefont
  {Wiese}(1993)}]{Kronfeld:1992ae}%
  \BibitemOpen
  \bibfield  {author} {\bibinfo {author} {\bibfnamefont {A.~S.}\ \bibnamefont
  {Kronfeld}}\ and\ \bibinfo {author} {\bibfnamefont {U.}~\bibnamefont
  {Wiese}},\ }\bibfield  {title} {\bibinfo {title} {{{SU(N)} gauge theories
  with {C} periodic boundary conditions. 2. Small volume dynamics}},\ }\href
  {https://doi.org/10.1016/0550-3213(93)90302-6} {\bibfield  {journal}
  {\bibinfo  {journal} {Nucl. Phys. B}\ }\textbf {\bibinfo {volume} {401}},\
  \bibinfo {pages} {190} (\bibinfo {year} {1993})},\ \Eprint
  {https://arxiv.org/abs/hep-lat/9210008} {arXiv:hep-lat/9210008} \BibitemShut
  {NoStop}%
\bibitem [{\citenamefont {Kronfeld}\ and\ \citenamefont
  {Wiese}(1991)}]{KRONFELD1991521}%
  \BibitemOpen
  \bibfield  {author} {\bibinfo {author} {\bibfnamefont {A.}~\bibnamefont
  {Kronfeld}}\ and\ \bibinfo {author} {\bibfnamefont {U.-J.}\ \bibnamefont
  {Wiese}},\ }\bibfield  {title} {\bibinfo {title} {{SU(N) gauge theories with
  C-periodic boundary conditions (I). Topological structure}},\ }\href
  {https://doi.org/https://doi.org/10.1016/0550-3213(91)90479-H} {\bibfield
  {journal} {\bibinfo  {journal} {Nuclear Physics B}\ }\textbf {\bibinfo
  {volume} {357}},\ \bibinfo {pages} {521 } (\bibinfo {year}
  {1991})}\BibitemShut {NoStop}%
\bibitem [{\citenamefont {Polley}(1993)}]{Polley:1993bn}%
  \BibitemOpen
  \bibfield  {author} {\bibinfo {author} {\bibfnamefont {L.}~\bibnamefont
  {Polley}},\ }\bibfield  {title} {\bibinfo {title} {{Boundaries for $SU(3)_{C}
  \times U(1)_{el}$ lattice gauge theory with a chemical potential}},\ }\href
  {https://doi.org/10.1007/BF01555844} {\bibfield  {journal} {\bibinfo
  {journal} {Z. Phys. C}\ }\textbf {\bibinfo {volume} {59}},\ \bibinfo {pages}
  {105} (\bibinfo {year} {1993})}\BibitemShut {NoStop}%
\bibitem [{\citenamefont {Lucini}\ \emph {et~al.}(2016)\citenamefont {Lucini},
  \citenamefont {Patella}, \citenamefont {Ramos},\ and\ \citenamefont
  {Tantalo}}]{Lucini:2015hfa}%
  \BibitemOpen
  \bibfield  {author} {\bibinfo {author} {\bibfnamefont {B.}~\bibnamefont
  {Lucini}}, \bibinfo {author} {\bibfnamefont {A.}~\bibnamefont {Patella}},
  \bibinfo {author} {\bibfnamefont {A.}~\bibnamefont {Ramos}},\ and\ \bibinfo
  {author} {\bibfnamefont {N.}~\bibnamefont {Tantalo}},\ }\bibfield  {title}
  {\bibinfo {title} {{Charged hadrons in local finite-volume QED+QCD with
  C$^{\star }$ boundary conditions}},\ }\href
  {https://doi.org/10.1007/JHEP02(2016)076} {\bibfield  {journal} {\bibinfo
  {journal} {JHEP}\ }\textbf {\bibinfo {volume} {02}},\ \bibinfo {pages}
  {076}},\ \Eprint {https://arxiv.org/abs/1509.01636} {arXiv:1509.01636
  [hep-th]} \BibitemShut {NoStop}%
\bibitem [{\citenamefont {Endres}\ \emph {et~al.}(2016)\citenamefont {Endres},
  \citenamefont {Shindler}, \citenamefont {Tiburzi},\ and\ \citenamefont
  {Walker-Loud}}]{Endres:2015gda}%
  \BibitemOpen
  \bibfield  {author} {\bibinfo {author} {\bibfnamefont {M.~G.}\ \bibnamefont
  {Endres}}, \bibinfo {author} {\bibfnamefont {A.}~\bibnamefont {Shindler}},
  \bibinfo {author} {\bibfnamefont {B.~C.}\ \bibnamefont {Tiburzi}},\ and\
  \bibinfo {author} {\bibfnamefont {A.}~\bibnamefont {Walker-Loud}},\
  }\bibfield  {title} {\bibinfo {title} {{Massive photons: an infrared
  regularization scheme for lattice QCD+QED}},\ }\href
  {https://doi.org/10.1103/PhysRevLett.117.072002} {\bibfield  {journal}
  {\bibinfo  {journal} {Phys. Rev. Lett.}\ }\textbf {\bibinfo {volume} {117}},\
  \bibinfo {pages} {072002} (\bibinfo {year} {2016})},\ \Eprint
  {https://arxiv.org/abs/1507.08916} {arXiv:1507.08916 [hep-lat]} \BibitemShut
  {NoStop}%
\bibitem [{\citenamefont {Bussone}\ \emph {et~al.}(2018)\citenamefont
  {Bussone}, \citenamefont {Della~Morte},\ and\ \citenamefont
  {Janowski}}]{Bussone:2017xkb}%
  \BibitemOpen
  \bibfield  {author} {\bibinfo {author} {\bibfnamefont {A.}~\bibnamefont
  {Bussone}}, \bibinfo {author} {\bibfnamefont {M.}~\bibnamefont
  {Della~Morte}},\ and\ \bibinfo {author} {\bibfnamefont {T.}~\bibnamefont
  {Janowski}},\ }\bibfield  {title} {\bibinfo {title} {{Electromagnetic
  corrections to the hadronic vacuum polarization of the photon within
  QED$_{\rm L}$ and QED$_{\rm M}$}},\ }\href
  {https://doi.org/10.1051/epjconf/201817506005} {\bibfield  {journal}
  {\bibinfo  {journal} {EPJ Web Conf.}\ }\textbf {\bibinfo {volume} {175}},\
  \bibinfo {pages} {06005} (\bibinfo {year} {2018})},\ \Eprint
  {https://arxiv.org/abs/1710.06024} {arXiv:1710.06024 [hep-lat]} \BibitemShut
  {NoStop}%
\bibitem [{\citenamefont {Duncan}\ \emph {et~al.}(1996)\citenamefont {Duncan},
  \citenamefont {Eichten},\ and\ \citenamefont {Thacker}}]{Duncan:1996xy}%
  \BibitemOpen
  \bibfield  {author} {\bibinfo {author} {\bibfnamefont {A.}~\bibnamefont
  {Duncan}}, \bibinfo {author} {\bibfnamefont {E.}~\bibnamefont {Eichten}},\
  and\ \bibinfo {author} {\bibfnamefont {H.}~\bibnamefont {Thacker}},\
  }\bibfield  {title} {\bibinfo {title} {{Electromagnetic splittings and light
  quark masses in lattice QCD}},\ }\href
  {https://doi.org/10.1103/PhysRevLett.76.3894} {\bibfield  {journal} {\bibinfo
   {journal} {Phys. Rev. Lett.}\ }\textbf {\bibinfo {volume} {76}},\ \bibinfo
  {pages} {3894} (\bibinfo {year} {1996})},\ \Eprint
  {https://arxiv.org/abs/hep-lat/9602005} {arXiv:hep-lat/9602005} \BibitemShut
  {NoStop}%
\bibitem [{\citenamefont {Duncan}\ \emph {et~al.}(1997)\citenamefont {Duncan},
  \citenamefont {Eichten},\ and\ \citenamefont {Thacker}}]{Duncan:1996be}%
  \BibitemOpen
  \bibfield  {author} {\bibinfo {author} {\bibfnamefont {A.}~\bibnamefont
  {Duncan}}, \bibinfo {author} {\bibfnamefont {E.}~\bibnamefont {Eichten}},\
  and\ \bibinfo {author} {\bibfnamefont {H.}~\bibnamefont {Thacker}},\
  }\bibfield  {title} {\bibinfo {title} {{Electromagnetic structure of light
  baryons in lattice QCD}},\ }\href
  {https://doi.org/10.1016/S0370-2693(97)00850-2} {\bibfield  {journal}
  {\bibinfo  {journal} {Phys. Lett. B}\ }\textbf {\bibinfo {volume} {409}},\
  \bibinfo {pages} {387} (\bibinfo {year} {1997})},\ \Eprint
  {https://arxiv.org/abs/hep-lat/9607032} {arXiv:hep-lat/9607032} \BibitemShut
  {NoStop}%
\bibitem [{\citenamefont {Feng}\ and\ \citenamefont
  {Jin}(2019)}]{Feng:2018qpx}%
  \BibitemOpen
  \bibfield  {author} {\bibinfo {author} {\bibfnamefont {X.}~\bibnamefont
  {Feng}}\ and\ \bibinfo {author} {\bibfnamefont {L.}~\bibnamefont {Jin}},\
  }\bibfield  {title} {\bibinfo {title} {{QED self energies from lattice QCD
  without power-law finite-volume errors}},\ }\href
  {https://doi.org/10.1103/PhysRevD.100.094509} {\bibfield  {journal} {\bibinfo
   {journal} {Phys. Rev. D}\ }\textbf {\bibinfo {volume} {100}},\ \bibinfo
  {pages} {094509} (\bibinfo {year} {2019})},\ \Eprint
  {https://arxiv.org/abs/1812.09817} {arXiv:1812.09817 [hep-lat]} \BibitemShut
  {NoStop}%
\bibitem [{\citenamefont {Christ}\ \emph {et~al.}(2020)\citenamefont {Christ},
  \citenamefont {Feng}, \citenamefont {Lu-Chang},\ and\ \citenamefont
  {Sachrajda}}]{Christ:2020jlp}%
  \BibitemOpen
  \bibfield  {author} {\bibinfo {author} {\bibfnamefont {N.~H.}\ \bibnamefont
  {Christ}}, \bibinfo {author} {\bibfnamefont {X.}~\bibnamefont {Feng}},
  \bibinfo {author} {\bibfnamefont {J.}~\bibnamefont {Lu-Chang}},\ and\
  \bibinfo {author} {\bibfnamefont {C.~T.}\ \bibnamefont {Sachrajda}},\
  }\bibfield  {title} {\bibinfo {title} {{Electromagnetic corrections to
  leptonic pion decay from lattice QCD using infinite-volume reconstruction
  method}},\ }\href {https://doi.org/10.22323/1.363.0259} {\bibfield  {journal}
  {\bibinfo  {journal} {PoS}\ }\textbf {\bibinfo {volume} {LATTICE2019}},\
  \bibinfo {pages} {259} (\bibinfo {year} {2020})}\BibitemShut {NoStop}%
\bibitem [{\citenamefont {Feng}\ \emph {et~al.}(2021)\citenamefont {Feng},
  \citenamefont {Jin},\ and\ \citenamefont {Riberdy}}]{Feng:2021zek}%
  \BibitemOpen
  \bibfield  {author} {\bibinfo {author} {\bibfnamefont {X.}~\bibnamefont
  {Feng}}, \bibinfo {author} {\bibfnamefont {L.}~\bibnamefont {Jin}},\ and\
  \bibinfo {author} {\bibfnamefont {M.~J.}\ \bibnamefont {Riberdy}},\
  }\bibfield  {title} {\bibinfo {title} {{Lattice QCD calculation of the pion
  mass splitting}},\ }\href@noop {} {\bibfield  {journal} {\bibinfo  {journal}
  {arXiv}\ } (\bibinfo {year} {2021})},\ \Eprint
  {https://arxiv.org/abs/2108.05311} {arXiv:2108.05311 [hep-lat]} \BibitemShut
  {NoStop}%
\bibitem [{\citenamefont {Tantalo}\ \emph {et~al.}(2016)\citenamefont
  {Tantalo}, \citenamefont {Lubicz}, \citenamefont {Martinelli}, \citenamefont
  {Sachrajda}, \citenamefont {Sanfilippo},\ and\ \citenamefont
  {Simula}}]{Tantalo:2016vxk}%
  \BibitemOpen
  \bibfield  {author} {\bibinfo {author} {\bibfnamefont {N.}~\bibnamefont
  {Tantalo}}, \bibinfo {author} {\bibfnamefont {V.}~\bibnamefont {Lubicz}},
  \bibinfo {author} {\bibfnamefont {G.}~\bibnamefont {Martinelli}}, \bibinfo
  {author} {\bibfnamefont {C.~T.}\ \bibnamefont {Sachrajda}}, \bibinfo {author}
  {\bibfnamefont {F.}~\bibnamefont {Sanfilippo}},\ and\ \bibinfo {author}
  {\bibfnamefont {S.}~\bibnamefont {Simula}},\ }\bibfield  {title} {\bibinfo
  {title} {{Electromagnetic corrections to leptonic decay rates of charged
  pseudoscalar mesons: finite-volume effects}},\ }\href@noop {} {\bibfield
  {journal} {\bibinfo  {journal} {PoS}\ }\textbf {\bibinfo {volume}
  {LATTICE2016}} (\bibinfo {year} {2016})},\ \Eprint
  {https://arxiv.org/abs/1612.00199} {arXiv:1612.00199 [hep-lat]} \BibitemShut
  {NoStop}%
\bibitem [{\citenamefont {Luscher}(1986{\natexlab{a}})}]{Luscher:1986pf}%
  \BibitemOpen
  \bibfield  {author} {\bibinfo {author} {\bibfnamefont {M.}~\bibnamefont
  {Luscher}},\ }\bibfield  {title} {\bibinfo {title} {{Volume Dependence of the
  Energy Spectrum in Massive Quantum Field Theories. 2. Scattering States}},\
  }\href {https://doi.org/10.1007/BF01211097} {\bibfield  {journal} {\bibinfo
  {journal} {Commun. Math. Phys.}\ }\textbf {\bibinfo {volume} {105}},\
  \bibinfo {pages} {153} (\bibinfo {year} {1986}{\natexlab{a}})}\BibitemShut
  {NoStop}%
\bibitem [{\citenamefont {Kim}\ \emph {et~al.}(2005)\citenamefont {Kim},
  \citenamefont {Sachrajda},\ and\ \citenamefont {Sharpe}}]{Kim:2005gf}%
  \BibitemOpen
  \bibfield  {author} {\bibinfo {author} {\bibfnamefont {C.~h.}\ \bibnamefont
  {Kim}}, \bibinfo {author} {\bibfnamefont {C.~T.}\ \bibnamefont {Sachrajda}},\
  and\ \bibinfo {author} {\bibfnamefont {S.~R.}\ \bibnamefont {Sharpe}},\
  }\bibfield  {title} {\bibinfo {title} {{Finite-volume effects for two-hadron
  states in moving frames}},\ }\href
  {https://doi.org/10.1016/j.nuclphysb.2005.08.029} {\bibfield  {journal}
  {\bibinfo  {journal} {Nucl. Phys.}\ }\textbf {\bibinfo {volume} {B727}},\
  \bibinfo {pages} {218} (\bibinfo {year} {2005})},\ \Eprint
  {https://arxiv.org/abs/hep-lat/0507006} {arXiv:hep-lat/0507006 [hep-lat]}
  \BibitemShut {NoStop}%
\bibitem [{\citenamefont {Hansen}\ and\ \citenamefont
  {Sharpe}(2015)}]{Hansen:2015zga}%
  \BibitemOpen
  \bibfield  {author} {\bibinfo {author} {\bibfnamefont {M.~T.}\ \bibnamefont
  {Hansen}}\ and\ \bibinfo {author} {\bibfnamefont {S.~R.}\ \bibnamefont
  {Sharpe}},\ }\bibfield  {title} {\bibinfo {title} {{Expressing the
  three-particle finite-volume spectrum in terms of the three-to-three
  scattering amplitude}},\ }\href {https://doi.org/10.1103/PhysRevD.92.114509}
  {\bibfield  {journal} {\bibinfo  {journal} {Phys. Rev.}\ }\textbf {\bibinfo
  {volume} {D92}},\ \bibinfo {pages} {114509} (\bibinfo {year} {2015})},\
  \Eprint {https://arxiv.org/abs/1504.04248} {arXiv:1504.04248 [hep-lat]}
  \BibitemShut {NoStop}%
\bibitem [{\citenamefont {Di~Carlo}\ \emph {et~al.}(2021)\citenamefont
  {Di~Carlo}, \citenamefont {Hansen}, \citenamefont {Hermansson-Truedsson},\
  and\ \citenamefont {Portelli}}]{klfv-zenodo}%
  \BibitemOpen
  \bibfield  {author} {\bibinfo {author} {\bibfnamefont {M.}~\bibnamefont
  {Di~Carlo}}, \bibinfo {author} {\bibfnamefont {M.~T.}\ \bibnamefont
  {Hansen}}, \bibinfo {author} {\bibfnamefont {N.}~\bibnamefont
  {Hermansson-Truedsson}},\ and\ \bibinfo {author} {\bibfnamefont
  {A.}~\bibnamefont {Portelli}},\ }\bibfield  {title} {\bibinfo {title}
  {Relativistic, model-independent determination of electromagnetic finite-size
  effects beyond the point-like approximation}\ }\href
  {https://doi.org/10.5281/zenodo.5607232} {10.5281/zenodo.5607232} (\bibinfo
  {year} {2021})\BibitemShut {NoStop}%
\bibitem [{\citenamefont {Luscher}(1986{\natexlab{b}})}]{Luscher:1985dn}%
  \BibitemOpen
  \bibfield  {author} {\bibinfo {author} {\bibfnamefont {M.}~\bibnamefont
  {Luscher}},\ }\bibfield  {title} {\bibinfo {title} {{Volume Dependence of the
  Energy Spectrum in Massive Quantum Field Theories. 1. Stable Particle
  States}},\ }\href {https://doi.org/10.1007/BF01211589} {\bibfield  {journal}
  {\bibinfo  {journal} {Commun. Math. Phys.}\ }\textbf {\bibinfo {volume}
  {104}},\ \bibinfo {pages} {177} (\bibinfo {year}
  {1986}{\natexlab{b}})}\BibitemShut {NoStop}%
\bibitem [{\citenamefont {Rudy}\ \emph {et~al.}(1994)\citenamefont {Rudy},
  \citenamefont {Fearing},\ and\ \citenamefont {Scherer}}]{Rudy:1994qb}%
  \BibitemOpen
  \bibfield  {author} {\bibinfo {author} {\bibfnamefont {T.~E.}\ \bibnamefont
  {Rudy}}, \bibinfo {author} {\bibfnamefont {H.~W.}\ \bibnamefont {Fearing}},\
  and\ \bibinfo {author} {\bibfnamefont {S.}~\bibnamefont {Scherer}},\
  }\bibfield  {title} {\bibinfo {title} {{The off-shell electromagnetic
  form-factors of pions and kaons in chiral perturbation theory}},\ }\href
  {https://doi.org/10.1103/PhysRevC.50.447} {\bibfield  {journal} {\bibinfo
  {journal} {Phys. Rev.}\ }\textbf {\bibinfo {volume} {C50}},\ \bibinfo {pages}
  {447} (\bibinfo {year} {1994})},\ \Eprint
  {https://arxiv.org/abs/hep-ph/9401302} {arXiv:hep-ph/9401302 [hep-ph]}
  \BibitemShut {NoStop}%
\bibitem [{\citenamefont {Fearing}\ and\ \citenamefont
  {Scherer}(1998)}]{Fearing:1996gs}%
  \BibitemOpen
  \bibfield  {author} {\bibinfo {author} {\bibfnamefont {H.~W.}\ \bibnamefont
  {Fearing}}\ and\ \bibinfo {author} {\bibfnamefont {S.}~\bibnamefont
  {Scherer}},\ }\bibfield  {title} {\bibinfo {title} {{Virtual Compton
  scattering off spin zero particles at low-energies}},\ }\href
  {https://doi.org/10.1007/s006010050067} {\bibfield  {journal} {\bibinfo
  {journal} {Few Body Syst.}\ }\textbf {\bibinfo {volume} {23}},\ \bibinfo
  {pages} {111} (\bibinfo {year} {1998})},\ \Eprint
  {https://arxiv.org/abs/nucl-th/9607056} {arXiv:nucl-th/9607056 [nucl-th]}
  \BibitemShut {NoStop}%
\bibitem [{\citenamefont {Sirlin}(1980)}]{Sirlin:1980nh}%
  \BibitemOpen
  \bibfield  {author} {\bibinfo {author} {\bibfnamefont {A.}~\bibnamefont
  {Sirlin}},\ }\bibfield  {title} {\bibinfo {title} {{Radiative Corrections in
  the $SU(2)_L \times U(1)$ Theory: A Simple Renormalization Framework}},\
  }\href {https://doi.org/10.1103/PhysRevD.22.971} {\bibfield  {journal}
  {\bibinfo  {journal} {Phys. Rev. D}\ }\textbf {\bibinfo {volume} {22}},\
  \bibinfo {pages} {971} (\bibinfo {year} {1980})}\BibitemShut {NoStop}%
\bibitem [{\citenamefont {Bijnens}\ \emph {et~al.}(1993)\citenamefont
  {Bijnens}, \citenamefont {Ecker},\ and\ \citenamefont
  {Gasser}}]{Bijnens:1992en}%
  \BibitemOpen
  \bibfield  {author} {\bibinfo {author} {\bibfnamefont {J.}~\bibnamefont
  {Bijnens}}, \bibinfo {author} {\bibfnamefont {G.}~\bibnamefont {Ecker}},\
  and\ \bibinfo {author} {\bibfnamefont {J.}~\bibnamefont {Gasser}},\
  }\bibfield  {title} {\bibinfo {title} {{Radiative semileptonic kaon
  decays}},\ }\href {https://doi.org/10.1016/0550-3213(93)90259-R} {\bibfield
  {journal} {\bibinfo  {journal} {Nucl. Phys.}\ }\textbf {\bibinfo {volume}
  {B396}},\ \bibinfo {pages} {81} (\bibinfo {year} {1993})},\ \Eprint
  {https://arxiv.org/abs/hep-ph/9209261} {arXiv:hep-ph/9209261 [hep-ph]}
  \BibitemShut {NoStop}%
\bibitem [{\citenamefont {Cirigliano}\ \emph {et~al.}(2012)\citenamefont
  {Cirigliano}, \citenamefont {Ecker}, \citenamefont {Neufeld}, \citenamefont
  {Pich},\ and\ \citenamefont {Portoles}}]{Cirigliano:2011ny}%
  \BibitemOpen
  \bibfield  {author} {\bibinfo {author} {\bibfnamefont {V.}~\bibnamefont
  {Cirigliano}}, \bibinfo {author} {\bibfnamefont {G.}~\bibnamefont {Ecker}},
  \bibinfo {author} {\bibfnamefont {H.}~\bibnamefont {Neufeld}}, \bibinfo
  {author} {\bibfnamefont {A.}~\bibnamefont {Pich}},\ and\ \bibinfo {author}
  {\bibfnamefont {J.}~\bibnamefont {Portoles}},\ }\bibfield  {title} {\bibinfo
  {title} {{Kaon Decays in the Standard Model}},\ }\href
  {https://doi.org/10.1103/RevModPhys.84.399} {\bibfield  {journal} {\bibinfo
  {journal} {Rev. Mod. Phys.}\ }\textbf {\bibinfo {volume} {84}},\ \bibinfo
  {pages} {399} (\bibinfo {year} {2012})},\ \Eprint
  {https://arxiv.org/abs/1107.6001} {arXiv:1107.6001 [hep-ph]} \BibitemShut
  {NoStop}%
\bibitem [{\citenamefont {Group}\ \emph {et~al.}(2020)\citenamefont {Group},
  \citenamefont {Zyla} \emph {et~al.}}]{10.1093/ptep/ptaa104}%
  \BibitemOpen
  \bibfield  {author} {\bibinfo {author} {\bibfnamefont {P.~D.}\ \bibnamefont
  {Group}}, \bibinfo {author} {\bibfnamefont {P.~A.}\ \bibnamefont {Zyla}},
  \emph {et~al.},\ }\bibfield  {title} {\bibinfo {title} {{Review of Particle
  Physics}},\ }\bibfield  {journal} {\bibinfo  {journal} {Progress of
  Theoretical and Experimental Physics}\ }\textbf {\bibinfo {volume} {2020}},\
  \href {https://doi.org/10.1093/ptep/ptaa104} {10.1093/ptep/ptaa104} (\bibinfo
  {year} {2020}),\ \bibinfo {note} {083C01},\ \Eprint
  {https://arxiv.org/abs/https://academic.oup.com/ptep/article-pdf/2020/8/083C01/33653179/ptaa104.pdf}
  {https://academic.oup.com/ptep/article-pdf/2020/8/083C01/33653179/ptaa104.pdf}
  \BibitemShut {NoStop}%
\bibitem [{\citenamefont {Ananthanarayan}\ \emph {et~al.}(2017)\citenamefont
  {Ananthanarayan}, \citenamefont {Caprini},\ and\ \citenamefont
  {Das}}]{Ananthanarayan:2017efc}%
  \BibitemOpen
  \bibfield  {author} {\bibinfo {author} {\bibfnamefont {B.}~\bibnamefont
  {Ananthanarayan}}, \bibinfo {author} {\bibfnamefont {I.}~\bibnamefont
  {Caprini}},\ and\ \bibinfo {author} {\bibfnamefont {D.}~\bibnamefont {Das}},\
  }\bibfield  {title} {\bibinfo {title} {{Electromagnetic charge radius of the
  pion at high precision}},\ }\href
  {https://doi.org/10.1103/PhysRevLett.119.132002} {\bibfield  {journal}
  {\bibinfo  {journal} {Phys. Rev. Lett.}\ }\textbf {\bibinfo {volume} {119}},\
  \bibinfo {pages} {132002} (\bibinfo {year} {2017})},\ \Eprint
  {https://arxiv.org/abs/1706.04020} {arXiv:1706.04020 [hep-ph]} \BibitemShut
  {NoStop}%
\bibitem [{\citenamefont {Colangelo}\ \emph {et~al.}(2019)\citenamefont
  {Colangelo}, \citenamefont {Hoferichter},\ and\ \citenamefont
  {Stoffer}}]{Colangelo:2018mtw}%
  \BibitemOpen
  \bibfield  {author} {\bibinfo {author} {\bibfnamefont {G.}~\bibnamefont
  {Colangelo}}, \bibinfo {author} {\bibfnamefont {M.}~\bibnamefont
  {Hoferichter}},\ and\ \bibinfo {author} {\bibfnamefont {P.}~\bibnamefont
  {Stoffer}},\ }\bibfield  {title} {\bibinfo {title} {{Two-pion contribution to
  hadronic vacuum polarization}},\ }\href
  {https://doi.org/10.1007/JHEP02(2019)006} {\bibfield  {journal} {\bibinfo
  {journal} {JHEP}\ }\textbf {\bibinfo {volume} {02}},\ \bibinfo {pages}
  {006}},\ \Eprint {https://arxiv.org/abs/1810.00007} {arXiv:1810.00007
  [hep-ph]} \BibitemShut {NoStop}%
\bibitem [{\citenamefont {Bijnens}\ \emph
  {et~al.}(2019{\natexlab{b}})\citenamefont {Bijnens}, \citenamefont
  {Hermansson-Truedsson},\ and\ \citenamefont {Wang}}]{Bijnens:2018lez}%
  \BibitemOpen
  \bibfield  {author} {\bibinfo {author} {\bibfnamefont {J.}~\bibnamefont
  {Bijnens}}, \bibinfo {author} {\bibfnamefont {N.}~\bibnamefont
  {Hermansson-Truedsson}},\ and\ \bibinfo {author} {\bibfnamefont
  {S.}~\bibnamefont {Wang}},\ }\bibfield  {title} {\bibinfo {title} {{The order
  p$^{8}$ mesonic chiral Lagrangian}},\ }\href
  {https://doi.org/10.1007/JHEP01(2019)102} {\bibfield  {journal} {\bibinfo
  {journal} {JHEP}\ }\textbf {\bibinfo {volume} {01}},\ \bibinfo {pages}
  {102}},\ \Eprint {https://arxiv.org/abs/1810.06834} {arXiv:1810.06834
  [hep-ph]} \BibitemShut {NoStop}%
\bibitem [{\citenamefont
  {Hermansson-Truedsson}(2020)}]{Hermansson-Truedsson:2020rtj}%
  \BibitemOpen
  \bibfield  {author} {\bibinfo {author} {\bibfnamefont {N.}~\bibnamefont
  {Hermansson-Truedsson}},\ }\bibfield  {title} {\bibinfo {title} {{Chiral
  Perturbation Theory at NNNLO}},\ }\href {https://doi.org/10.3390/sym12081262}
  {\bibfield  {journal} {\bibinfo  {journal} {Symmetry}\ }\textbf {\bibinfo
  {volume} {12}},\ \bibinfo {pages} {1262} (\bibinfo {year} {2020})},\ \Eprint
  {https://arxiv.org/abs/2006.01430} {arXiv:2006.01430 [hep-ph]} \BibitemShut
  {NoStop}%
\bibitem [{\citenamefont {Dai}\ \emph {et~al.}(2020)\citenamefont {Dai},
  \citenamefont {Low}, \citenamefont {Mehen},\ and\ \citenamefont
  {Mohapatra}}]{Dai:2020cpk}%
  \BibitemOpen
  \bibfield  {author} {\bibinfo {author} {\bibfnamefont {L.}~\bibnamefont
  {Dai}}, \bibinfo {author} {\bibfnamefont {I.}~\bibnamefont {Low}}, \bibinfo
  {author} {\bibfnamefont {T.}~\bibnamefont {Mehen}},\ and\ \bibinfo {author}
  {\bibfnamefont {A.}~\bibnamefont {Mohapatra}},\ }\bibfield  {title} {\bibinfo
  {title} {{Operator Counting and Soft Blocks in Chiral Perturbation Theory}},\
  }\href {https://doi.org/10.1103/PhysRevD.102.116011} {\bibfield  {journal}
  {\bibinfo  {journal} {Phys. Rev. D}\ }\textbf {\bibinfo {volume} {102}},\
  \bibinfo {pages} {116011} (\bibinfo {year} {2020})},\ \Eprint
  {https://arxiv.org/abs/2009.01819} {arXiv:2009.01819 [hep-ph]} \BibitemShut
  {NoStop}%
\bibitem [{\citenamefont {Graf}\ \emph {et~al.}(2021)\citenamefont {Graf},
  \citenamefont {Henning}, \citenamefont {Lu}, \citenamefont {Melia},\ and\
  \citenamefont {Murayama}}]{Graf:2020yxt}%
  \BibitemOpen
  \bibfield  {author} {\bibinfo {author} {\bibfnamefont {L.}~\bibnamefont
  {Graf}}, \bibinfo {author} {\bibfnamefont {B.}~\bibnamefont {Henning}},
  \bibinfo {author} {\bibfnamefont {X.}~\bibnamefont {Lu}}, \bibinfo {author}
  {\bibfnamefont {T.}~\bibnamefont {Melia}},\ and\ \bibinfo {author}
  {\bibfnamefont {H.}~\bibnamefont {Murayama}},\ }\bibfield  {title} {\bibinfo
  {title} {{2, 12, 117, 1959, 45171, 1170086, \textellipsis{}: a Hilbert series
  for the QCD chiral Lagrangian}},\ }\href
  {https://doi.org/10.1007/JHEP01(2021)142} {\bibfield  {journal} {\bibinfo
  {journal} {JHEP}\ }\textbf {\bibinfo {volume} {01}},\ \bibinfo {pages}
  {142}},\ \Eprint {https://arxiv.org/abs/2009.01239} {arXiv:2009.01239
  [hep-ph]} \BibitemShut {NoStop}%
\bibitem [{\citenamefont {Lee}\ and\ \citenamefont
  {Tiburzi}(2016)}]{Lee:2015rua}%
  \BibitemOpen
  \bibfield  {author} {\bibinfo {author} {\bibfnamefont {J.-W.}\ \bibnamefont
  {Lee}}\ and\ \bibinfo {author} {\bibfnamefont {B.~C.}\ \bibnamefont
  {Tiburzi}},\ }\bibfield  {title} {\bibinfo {title} {{Finite Volume
  Corrections to the Electromagnetic Mass of Composite Particles}},\ }\href
  {https://doi.org/10.1103/PhysRevD.93.034012} {\bibfield  {journal} {\bibinfo
  {journal} {Phys. Rev. D}\ }\textbf {\bibinfo {volume} {93}},\ \bibinfo
  {pages} {034012} (\bibinfo {year} {2016})},\ \Eprint
  {https://arxiv.org/abs/1508.04165} {arXiv:1508.04165 [hep-lat]} \BibitemShut
  {NoStop}%
\end{thebibliography}%
\end{document}